\documentclass{aastex62}

\usepackage{hyperref}
\usepackage{mathtools}
\newcommand{\ratio}{{\langle B_{t}^{2} \rangle}/{\langle B_{0}^{2} \rangle}}
\newcommand{\dispfunct}{1 - \langle\cos[\Delta \phi(\ell)] \rangle}

\graphicspath{{./}{figures/}}

\received{2019}
\revised{2019}
\accepted{\today}
\submitjournal{ApJ}

\shorttitle{OMC-1 B-field maps}
\shortauthors{Guerra et al.}
\begin{document}

\title{Maps of Magnetic Field Strength in the OMC-1 using HAWC+ FIR Polarimetric data}

\correspondingauthor{J. A. Guerra}
\email{jordan.guerraaguilera@villanova.edu}

\author[0000-0001-8819-9648]{Jordan A. Guerra}
\affil{Department of Physics, Villanova University, 800 E. Lancaster Ave., Villanova, PA 19085, USA}

\author[0000-0003-0016-0533]{David T. Chuss}
\affil{Department of Physics, Villanova University, 800 E. Lancaster Ave., Villanova, PA 19085, USA}

\author{C. Darren Dowell}
\affil{NASA Jet Propulsion Laboratory, California Institute of Technology, 4800 Oak Grove Drive, Pasadena, CA 91109, USA}

\author[0000-0003-4420-8674]{Martin Houde}
\affil{Department of Physics and Astronomy, University of Western Ontario, 1151 Richmond Street, London, ON N6A 3K7, Canada}

\author[0000-0003-3503-3446]{Joseph M. Michail}
\affil{Department of Astrophysics and Planetary Science, 800 E. Lancaster Ave., Villanova University, Villanova, PA 19085, USA}
\affil{Department of Physics, Villanova University, 800 E. Lancaster Ave., Villanova, PA 19085, USA}
\affil{Center for Interdisciplinary Exploration and Research in Astrophysics (CIERA) and Department of Physics and Astronomy, Northwestern University, 1800 Sherman Ave., Evanston, IL 60201, USA}

\author[0000-0001-5389-5635]{Javad Siah}
\affil{Department of Physics, Villanova University, 800 E. Lancaster Ave., Villanova, PA 19085, USA}

\author[0000-0002-7567-4451]{Edward J. Wollack}
\affil{NASA Goddard Space Flight Center, Greenbelt, MD 20771, USA}

%\author{Joseph Michail}
%\affiliation{Villanova}
%\affiliation{AAS Journals Associate Editor-in-Chief}
%\nocollaboration

%\author{Javad Siah}
%\affiliation{Villanova}

%\collaboration{(HAWC+ Science Team)}

%\altaffiliation{Creator of AASTeX v6.2}
%\collaboration{(LaTeX collaboration)}

%\affiliation{AAS Director of Publishing}
%\affiliation{American Astronomical Society \\
%2000 Florida Ave., NW, Suite 300 \\
%Washington, DC 20009-1231, USA}

%\author{Jeff Lewandowski}
%\affiliation{IOP Senior Publisher for the AAS Journals}
%\affiliation{IOP Publishing, Washington, DC 20005}



\begin{abstract}

Far-infrared (FIR) dust polarimetry enables the study of interstellar magnetic fields via tracing of the polarized emission from dust grains that are partially aligned with the direction of the field. The advent of high quality polarimetric data has permitted the use of statistical methods to extract both the direction and magnitude of the magnetic field. In this work, the Davis-Chandrasekhar-Fermi technique is used to make maps of the plane-of-sky (POS) component of the magnetic field in the Orion Molecular Cloud (OMC-1) by combining polarization maps at 53, 89, 154 and 214 \micron\ from HAWC+/SOFIA with maps of density and velocity dispersion. In addition, maps of the local dispersion of polarization angles are used in conjuction with Zeeman measurements to estimate a map of the strength of the line-of-sight (LOS) component of the field. Combining these maps, information about the three-dimensional magnetic field configuration (integrated along the line-of-sight) is inferred over the OMC-1 region. POS magnetic field strengths of up to 2 mG are observed near the BN/KL object, while the OMC-1 bar shows strengths of up to a few hundred $\mu$G. 
These estimates of the magnetic field components are used to produce maps of the mass-to-magnetic flux ratio ($M/\Phi$) -- a metric for probing the conditions for star formation in molecular clouds -- and determine regions of sub- and super-criticality in OMC-1. Such maps can provide invaluable input and comparison to MHD simulations of star formation processes in filamentary structures of molecular clouds.

\end{abstract}



\section{Introduction} \label{sec:intro}

Magnetized turbulence is believed to play an important role in regulating the star formation activity in the interstellar medium (ISM) over a wide range of scales. The free-electron density in the ISM is sufficiently high that magnetic field lines are frozen into the gas, allowing gravitational collapse parallel to the field lines. Across magnetic field lines, the collapse can modify the geometry by compressing the field lines to create regions of enhanced magnetic field strength. Simultaneously, the collapse in this direction is slowed by increase of magnetic pressure. Therefore, whether a molecular cloud will form filaments, dense cores, and protostars depends on (among other factors) the interplay between the mass $M$ of a region and its magnetic flux $\Phi$. The two relevant regimes correspond to $M/\Phi < (M/\Phi)_{\rm c}$ (subcritical) or $M/\Phi > (M/\Phi)_{\rm c}$ (supercritical), where $(M/\Phi)_{\rm c} = 1/(2\pi\sqrt{G})$ is the critical value of mass-to-magnetic flux ratio \citep{Crutcher2019}, with $G$ being the gravitational constant. If a region is supercritical, the magnetic field is insufficient to halt the collapse of the cloud, and stars will eventually form. On the other hand, if the cloud is subcritical, the magnetic pressure will prevent gravitational collapse \citep{Mouschovias1976}. More specifically, magnetohydrodynamical (MHD) numerical simulations have shown that the gravitational collapse and the star formation rate (SFR) greatly depend on physical parameters such as the virial parameter, $\alpha_{\rm vir}\equiv 2E_{\rm kin}/|E_{\rm grav}|$, the sonic mach number, ($M_{a}$), the ratio of gas pressure to magnetic pressure, $\beta \equiv p/(B^{2}/8\pi)$, and the turbulence forcing parameter \citep{Federrath2012,Price2008}. Many of these parameters depend on mass and magnetic field strength. There is also growing interest and capability for deciphering the three-dimensional magnetic field.  \citet{Tahani2019} utilize a method based on Faraday rotation measurements to probe the three-dimensional geometry of Orion-A and \citet{Chen2019} use statistical properties of the polarization magnitude to infer the angle of inclination of the magnetic field in Vela-C. Such techniques are likely to improve the fidelity of comparisons of observations with models.

The Orion Molecular Cloud (OMC) complex is the nearest ($\sim$ 380 pc; \cite{Kounkel2017}) region in which massive star formation is occurring. In particular, the OMC-1 region contains molecular gas and dust in the form of a ridge roughly oriented north-south. Inside this molecular ridge lies the Becklin-Neugebauer (BN) object -- a massive young stellar object -- and the Kleinmann-Low (KL) nebula. Previous studies using sub-millimeter \citep{Tang2010} and far-infrared (FIR) polarization measurements \citep{Schleuning1998,Vallee1999,Houde2004,Ward-Thompson2017}, have revealed that the OMC-1 region exhibits the ``hourglass''-shaped magnetic field expected from the MHD considerations described above -- that is, a fairly uniform magnetic field oriented approximately northwest-southeast that displays a pinch orthogonal to this direction. Recently, \citet{Chuss2019} using multi-wavelength IR observations from the {\it High-resolution Airborne Wideband Camera} (HAWC+; \citet{Harper2018}) on board the {\it Stratospheric Observatory For Infrared Astronomy} (SOFIA; \citet{SOFIA}) confirmed the general hourglass shape of the magnetic field on large scales. At shorter wavelengths (53 and 89 $\micron$) deviations from this geometry are observed approximately perpendicular to the north-south direction near the BN/KL location. In addition, at the same location, low polarization fractions are observed, perhaps signaling that the magnetic field is predominantly along the line-of-sight direction. \citet{Chuss2019} also studied the magnetized turbulence through the structure of the dispersion of the polarization vectors and determined that three distinct parts of OMC-1 (the dense region around BN/KL, the HII region ionized by the radiation from the Trapezium cluster, and the photo-dissociation region of the OMC-1 bar) indeed display different properties. These authors infer a plane-of-sky magnetic field strength of $\sim$ 1 mG for the BN/KL region, and approximately 1/3 of that value was measured for the bar and Trapezium HII regions. These authors have found the bar region to be highly turbulent as indicated by the large dispersion in polarization angle. The HII region was found to possess lower dispersion, which was interpreted to correspond to lower turbulence, while the magnetic field in the high-density BN/KL region was found to have approximately equal contributions from organized and turbulent components. These results indicate that there are spatial variations in the magnetized turbulence throughout the OMC-1 complex. Naturally, these spatial variations are of importance for the ongoing star formation in the region.

Ultimately, understanding the star formation process depends on the ability to connect new high quality data sets with increasingly sophisticated magnetohydrodynamics models \citep[see {\it e.g.}][]{Federrath2016}. This paper takes a step towards this goal by utilizing the order-of-magnitude increase in the number of HAWC+/SOFIA polarization vectors over previous FIR polarimetry data sets to create maps of the magnetic field strength across the region. This work extends the classical Davis-Chandrasekhar-Fermi technique for estimating the strength of the plane-of-sky (POS) component of the magnetic field using the variation of polarization direction. Utilizing the methods for separating the large scale field variation from the turbulent component of the polarization direction variation \citep{Hildebrand2009,Houde2009}, the technique is applied on a pixel-by-pixel basis.  Maps of velocity dispersion and column density are then used to estimate the magnitude of the POS component of the magnetic field.  The strength of the line-of-sight (LOS) component of the magnetic field is also addressed. Based on the idea that the local dispersion of the polarization direction is related to the inclination angle of the field \citep{Hensley2019}, an estimate of this component is derived that is constrained to be consistent with Zeeman measurements.  The total estimate of the magnetic field strength over the region is then used to estimate $M/\Phi$ for the region.

Section 2 reviews the HAWC+/SOFIA observations used in this work. Section 3 describes the implementation of the DCF technique along with the calculation of the maps of local dispersion. Section 4 describes the construction of the POS and LOS magnetic field maps and discusses the criticality of the OMC-1 regions based on the derived map of $M/\Phi$. Finally, a summary is presented in Section 5.

\section{Observations} \label{sec:observations}

Multi-wavelength polarimetric observations of OMC-1 have been obtained with the HAWC+ instrument on SOFIA and first reported in \citet{Chuss2019}. Data consist of maps of Stokes parameters $I$, $Q$, and $U$ (and their associated uncertainties) for far-infrared (FIR) continuum emission centered at wavelengths of 53, 89, 154, and 214 \micron, observed with nominal beam sizes of 4.9\arcsec, 7.8\arcsec, 13.6\arcsec, and 18.2\arcsec, respectively \citep{Harper2018}. Maps of polarization angles $\phi$ and polarization fraction $p$ are calculated from the Stokes parameters as $\phi = \frac{1}{2}\arctan{(u/q)}$ and $p = \sqrt{(q^{2} + u^{2})}$. Here, normalized Stokes parameters are defined as $q\equiv Q /I$ and $u\equiv U/I$. In adherence to the IAU standard definition, $\phi$ is measured east of north. The polarization fraction is debiased by $p = \sqrt{p_{m}^{2} - \sigma_{p}^{2}}$, where $\sigma_p$ is the uncertainty in the polarization fraction \citep{Serkowski1974} and $p_{m}$ is the measured polarization. Resulting maps have the resolution $1/4$ of beam size per pixel. For full details on these data sets and the associated reduction details, see \citet{Chuss2019}.

\begin{figure}[!h]
    \centering
    \includegraphics[width=7.0in]{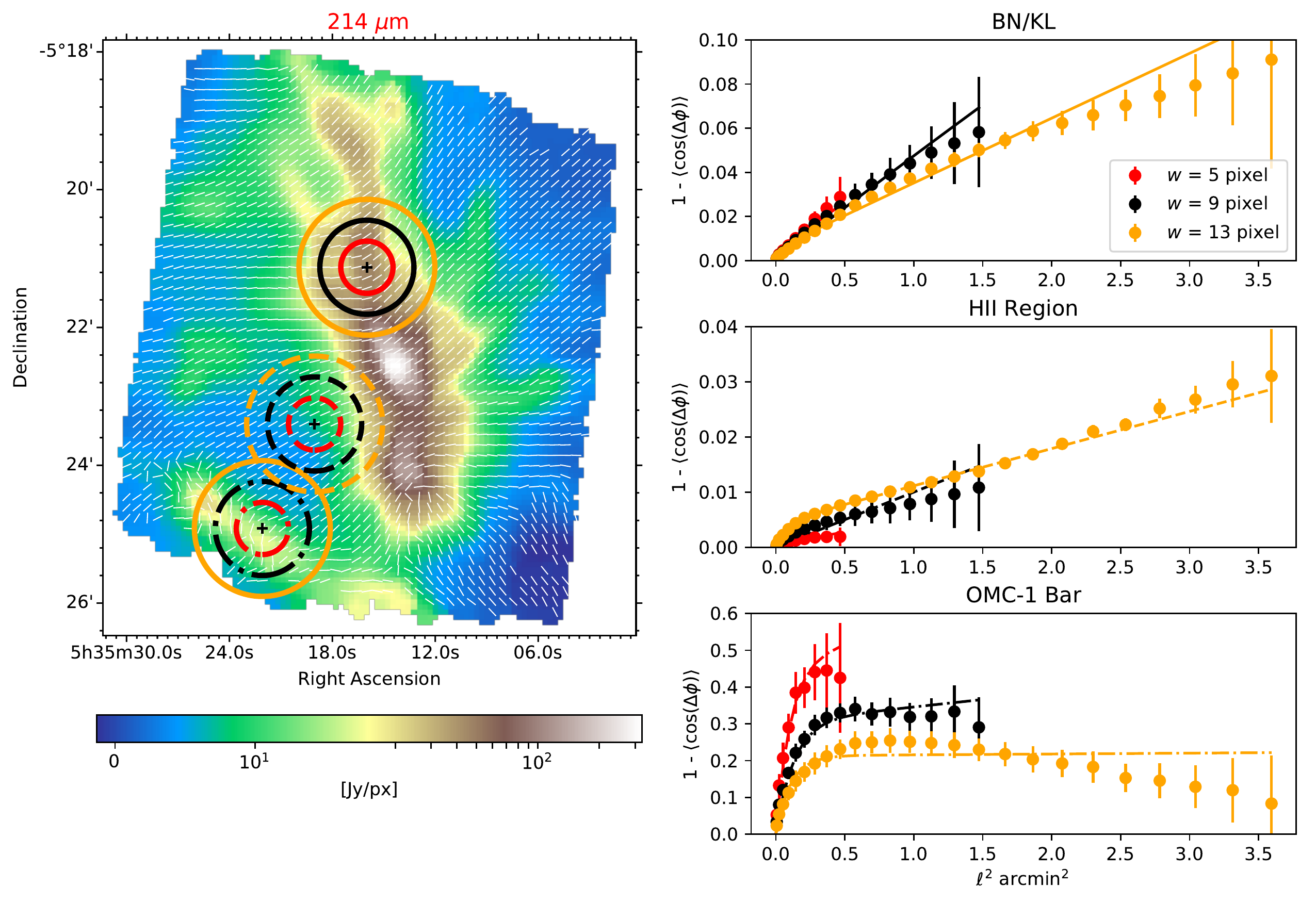}
    \caption{{\it Left:} HAWC+ 214 $\micron$ data for OMC-1. The color-scaled background corresponds to the Stokes I intensity, and the overlaid Nyquist-sampled, constant-length vectors represent the direction of the POS magnetic field direction inferred from the 214 um HAWC+ polarization data (white). Three locations in OMC-1 are identified with crosses: one north of the BN/KL object, one in the HII region, and one in the bar. For each location, three circles are shown representing the kernels used to calculate the dispersion functions shown on the right. {\it Right panels:} Dispersion functions for the three locations in OMC-1 ({\it Left}). Top, middle, and bottom panels correspond to locations near the BN/KL object, HII region, and the bar. Black, red, and dark yellow symbols correspond to kernel sizes of 5, 9, and 13 pixel. Lines (in the same colors) correspond to the fit using the model in  (Eq. \ref{eq:disp_model})}.
    \label{fig:sample_DFs}
\end{figure}

\clearpage

\section{Methods} \label{sec:methods}

In order to construct maps of magnetic field strength from IR polarization vector maps, the variation of polarization angles across the image is studied in two ways. First, for estimating the magnitude of the plane-of-the-sky (POS) magnetic field component, the Davis-Chandrasekhar-Fermi \citep[DCF;][]{Davis1951,Chandrasekhar1953} method is utilized and the dispersion-related parameters are determined based on the work of \citet{Hildebrand2009} and \citet{Houde2009,Houde2013}. In these works, the turbulent magnetic field contribution is obtained by fitting a model to the variation of the polarization vectors as a function of angular separation. In this paper, these methods are applied to estimate the magnitude of the POS component of the magnetic field, $B_{\rm POS}$, for each pixel across the source. For each pixel, dispersion-related parameters are obtained by analysis of a two-point structure function, here referred to as the ``dispersion function'', calculated using data within a circular region centered on the pixel in question. Details are provided below. The size of the circular region used for determining the dispersion function sets the resolution for the maps of the magnetized-turbulence parameters; however, in determining the ultimate resolution of the magnetic field maps, the resolutions of the auxiliary data sets (velocity dispersion and density maps; see Section \ref{sec:H2_vel}) must also be considered.

The spatial variation of the magnetic field is also commonly quantified by the (local) dispersion, $\mathcal{S}$ \citep{Planck2018,Fissel2016,Chuss2019,Hensley2019}. In contrast to the dispersion function, $\mathcal{S}$ has no explicit dependence on spatial scale since it corresponds to the root-mean-squared variation of all polarization directions within a region surrounding a pixel, relative to the mean direction. Again, details are provided below. This quantity has been utilized to probe grain alignment efficiency \citep{Fissel2016, Chuss2019}, but it has also been suggested that $\mathcal{S}$ can provide information about the angle of the field with respect to the plane-of-sky \citep{Falceta2008,Hensley2019}. Maps of this quantity were created and used to produce an estimate of magnitude of the line-of-sight component of the magnetic field, $B_{\mathrm{LOS}}$. In this section, the construction of the dispersion function and the method for obtaining maps of fit parameters are described.  The construction of maps of $\mathcal{S}$ are also detailed.

\subsection{Applying the DCF Technique Across a Polarization Map} \label{sec:disp}

Assuming that the magnetic field in a molecular cloud is composed of ordered, large-scale magnetic component, $\mathbf{B_{0}}$, and a turbulent component, $\mathbf{B_{t}}$, the total magnetic field in the POS is $\mathbf{B}_{\rm POS} = \mathbf{B_{0}} + \mathbf{B_{t}}$.  The DCF expression for $B_{\mathrm{POS}}$ can be then be written as \citep{Houde2009},

\begin{equation}
    B_{\rm POS} \simeq \sqrt{4\pi\rho}\sigma_v\!\!\left[\frac{\langle B_{t}^{2}\rangle}{\langle B_{0}^{2}\rangle}\right]^{-1/2},
    \label{eq:dcf_eq}
\end{equation}

\noindent
where $\rho$ and $\sigma_v$ are the mass density and velocity dispersion of the cloud, and $\ratio$ is the ratio of the LOS-averaged turbulent-to-ordered magnetic energy densities. In Eq. \ref{eq:dcf_eq}, the dispersion of polarization vectors is approximated as $\sigma^{2}_{\phi} \approx \ratio$.

Defining $\Delta\phi(\ell)$ as the angle difference between two points separated by an angle $\ell$ on the sky, \citet{Houde2009} proposed the 2-point dispersion function of polarization vectors difference, $\dispfunct$, and modeled it as the superposition of large-scale field structure and small-scale, beam-integrated turbulence,

\begin{equation}
    \dispfunct = \frac{1}{1+\mathcal{N} \left[\frac{\langle B_{t}^{2}\rangle}{\langle B_{0}^{2}\rangle}\right]^{-1}}\times \left\{1-\exp\left[-\frac{\ell^{2}}{2(\delta^{2} + 2W^{2})}\right]\right\} + a_{2}\ell^{2},
    \label{eq:disp_model}
\end{equation}

\noindent
where $\ell$ is the angular distance between a pair of polarization vectors; $\delta$ and $W$ are the (Gaussian) turbulence correlation length, and telescope beam width (FWHM), respectively.  $\mathcal{N}$ is the number of turbulent cells along the line of sight, where $\mathcal{N}^{-1} = \sqrt{2\pi}\delta^{3}/[(\delta^{2} + 2W^{2})\Delta']$ (with $\Delta^{\prime}$ being the cloud's effective thickness). Here, angle brackets indicate an average over all such pairs of polarization vectors having angular separation $\ell$. This analysis is applied locally at each position (pixel) in a polarization map by applying a two-dimensional normalized circular top hat kernel that defines the region over which $\dispfunct$ is evaluated. The size of this top-hat kernel is characterized by a radius $w$ (measured in pixels). All vectors within a distance $w$ from the position in question are multiplied by unity, while those outside this radius are multiplied by zero.  This symmetric kernel ensures that when calculating $\dispfunct$, no preference is given to a particular direction. In this way, a dispersion function is constructed for each pixel in the map, and corresponding magnetized turbulence parameters can be obtained. 

The dispersion function at each pixel is fitted with Eq. \ref{eq:disp_model}. Using a Monte-Carlo Markov-Chain (MCMC) solver \citep[\textsc{emcee};][]{Foreman-Mackey2013}, the parameters $\delta$, $a_{2}$, and $\ratio$ are determined. The MCMC solver explicitly produces posterior distributions for the first two parameters ($\delta$, $a_{2}$) while posterior distributions of $\ratio$ are obtained from similar distributions for the product $\Delta^{\prime}[\ratio]^{-1}$. The value of $\Delta'$ is determined by calculating the half-width at half-maximum (HWHM) value of the polarized flux ($p\times I$) auto-correlation function over the entire field of view \citep[see][for details]{Houde2009}. 

During the MCMC fitting of dispersion functions, numerical values of all three parameters are constrained to be positive. No upper bounds are imposed for parameters $a_{2}$ and $\Delta^{\prime}\left[\ratio\right]^{-1}$, although $\delta$ has a natural upper bound equal to the diameter of the circular kernel used to calculate the dispersion function. In maps of parameters presented in the following sections, values at each pixel were estimated from posterior distributions with well-defined unimodal shapes that do not fall against the boundaries (see corner plots in Figure \ref{fig:MCMC_corner_plots}, Appendix \ref{sec:appendixB}). The convergence of the MCMC chains was inspected for several sample locations in the maps (BN/KL object and the bar; see Figure \ref{fig:sample_DFs},{\it Left} for reference), and posterior distributions are typically constructed with chains for which maximum likelihood displays $\approx$1\% of variance. Appendix \ref{sec:appendixB} contains further discussion on the quality and convergence of the MCMC results.

\subsection{Optimization of the DCF Kernel Size} \label{sec:calib}

The choice of kernel size depends on several factors. Trivially, the kernel needs to be larger than both the beam size and turbulence coherence scale to be useful.  Kernels that are too small also run the risk that the there will be an insufficient number of pairs of measurements to create a dispersion function with the requisite fidelity. Large kernels potentially degrade the final resolution of the maps, depending on the resolutions of the auxiliary data sets.  In addition, for large kernels, the spatial variation of the large-scale component of the field must be considered. The model in Eq. \ref{eq:disp_model} is valid only for small spatial scales, {\it i.e.}, for the case where $\ell<$ few times the observation's beam size, $W$\citep{Houde2011}. Since the dispersion function is calculated for a circular kernel of radius $w$ (in pixels), the largest physical spatial scale that is included in the dispersion function is $\ell_{max} = 2\times w\times \mathrm{(pixel\,size)}$. This section describes a method for determining an optimal kernel for the DCF studies in which fits to various kernels are done and tested for fidelity over the data set. To ensure that all dispersion function are calculated with the same number of pixels thus statistical properties are kept constant across all bands, it is desirable to use a single kernel size, $w_{opt}$ for all maps.

\begin{figure}[!h]
    \centering
    \includegraphics[width=3.0in]{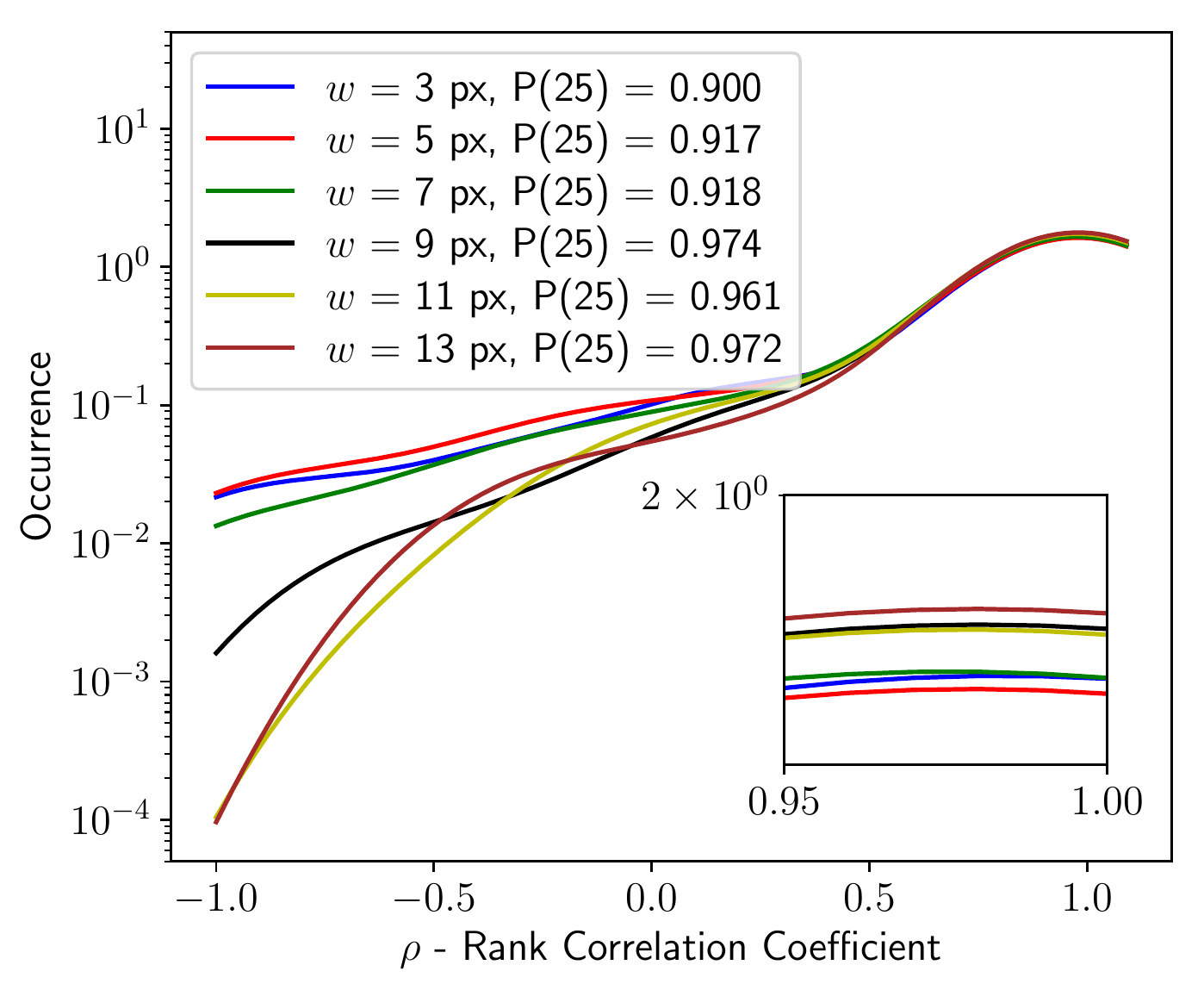}
    \includegraphics[width=3.0in]{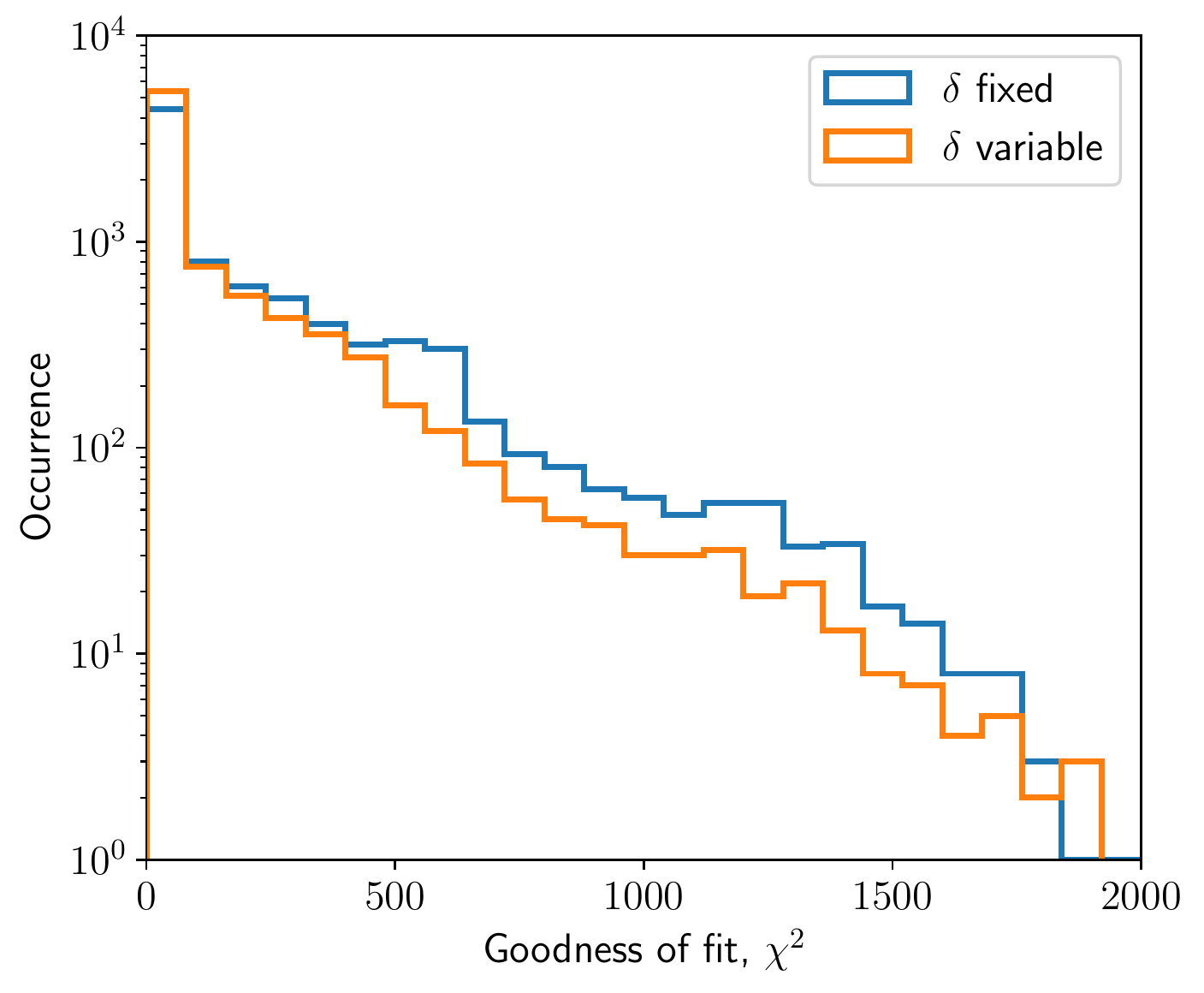}
    \caption{{\it Left:} Smoothed distribution of the rank correlation (Spearman) $\rho$ values for the 214 \micron\ data. Each curve include the values of $\rho$ for every pixel in the map. Rank correlations are calculated for different kernel radii ($w = $3,5,7,9,11,13 px). With larger kernels the $\rho$-values density distribution becomes narrower around its median value (1.0). Kernel size $w = $ 9 pixel provides the distribution of $\rho$ with highest median and smallest width (measured by the 25 percentile, denoted P(25) here). {\it Right:} Histograms comparing the reduced $\chi^{2}$ values (goodness-of-fit parameter) for the two approaches taken to fit dispersion functions: $\delta$ fixed (blue) and $\delta$ variable (orange). Using a $\delta$-variable approach shows fewer occurrences for larger values of $\chi^{2}$ in comparison to the $\delta$-fixed approach. However, performing the $\delta$-fixed MCMC fitting of Eq. \ref{eq:disp_model} allowed to study the covariance between the fitted parameters. See Appendix \ref{sec:appendixA} for details.}
    \label{fig:rho_dist}
\end{figure}

To determine $w_{\rm opt}$ for the analysis, the 214 $\micron$ data are used. First, a dispersion function for each pixel/position is calculated using odd values for $w$ ranging from 3 to 13 pixels.   Figure~\ref{fig:sample_DFs} shows examples of dispersion functions ({\it Right panels}) constructed for sample positions in three physically different regions in OMC-1 (as shown in the Left panel): the BN/KL region and molecular ridge, the intercloud HII region, and the bar. Each panel on the right shows dispersion functions calculated using three kernels of different sizes (see circles in Figure \ref{fig:sample_DFs}). For each $w$ value, dispersion functions are fitted for each map pixel using Eq. \ref{eq:disp_model} and the MCMC algorithm. The quality of the fit is quantified by evaluating the reduced $\chi^{2}$ goodness-of-fit parameter and the non-linear rank correlation (Spearman $\rho$) coefficient between the best-fit version of Eq. \ref{eq:disp_model} and data. Therefore, for each value of $w$, a map of $\rho$ and $\chi^{2}$ are obtained. Figure \ref{fig:rho_dist}({\it Left}) displays the histogram of $\rho$ values for each value of $w$. These distributions are smoothed using Gaussian-kernel density estimation (KDE) for better visualization. Values of $\rho$ range -1 to 1, with the latter corresponding to perfect positive correlation in a slowly-varying, non-linear fashion. The distribution of $\rho$ for with the narrowest shape is for $w = $9 px. This means 75\% of the dispersion functions in this case show $\rho > $0.974. Therefore, $w_{\rm opt} = $9 px is used for data at each wavelength.

\subsection{DCF Parameters Maps}

\begin{figure}[!h]
    \centering
    \includegraphics[width=2.3in]{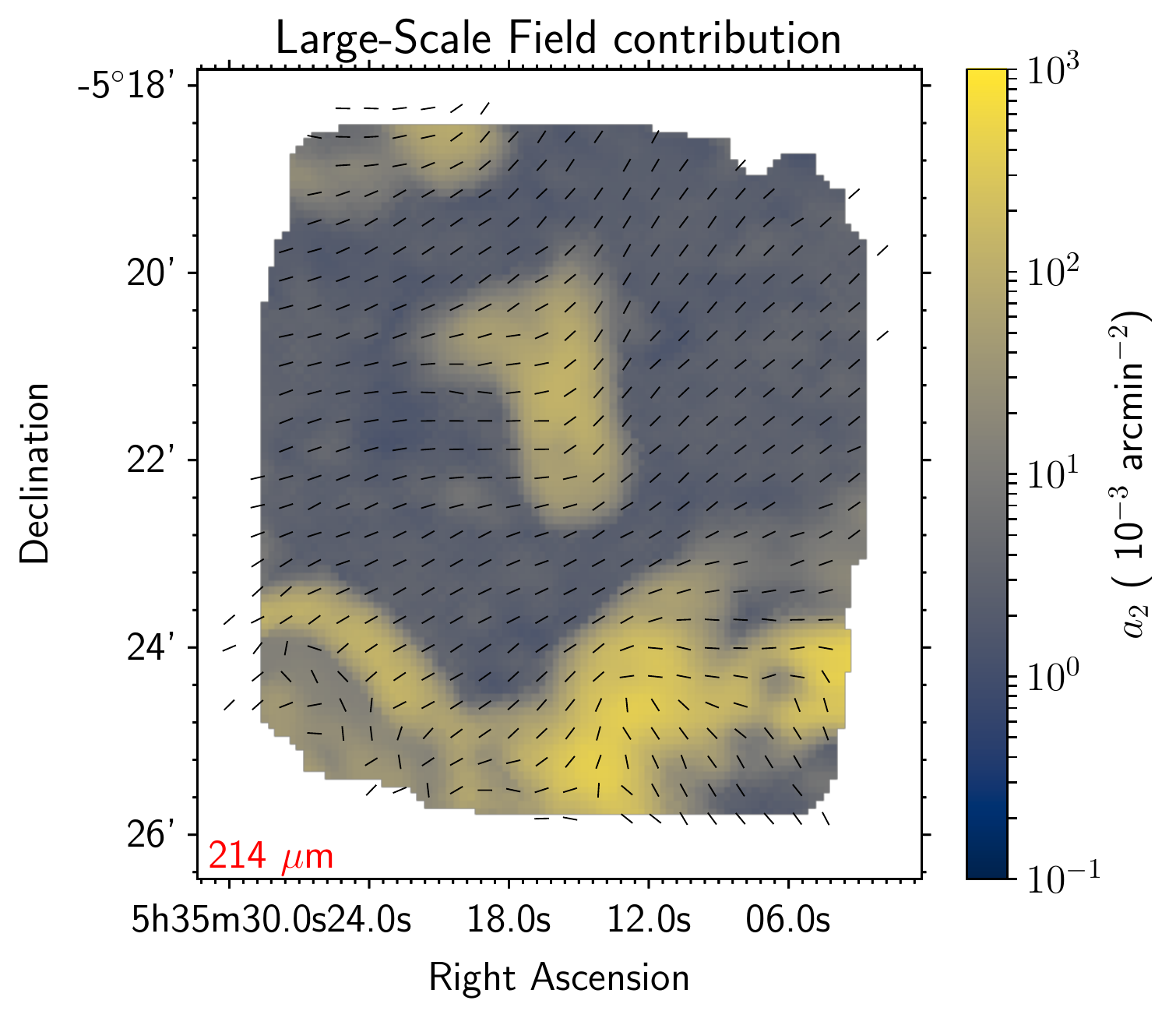}
    \includegraphics[width=2.3in]{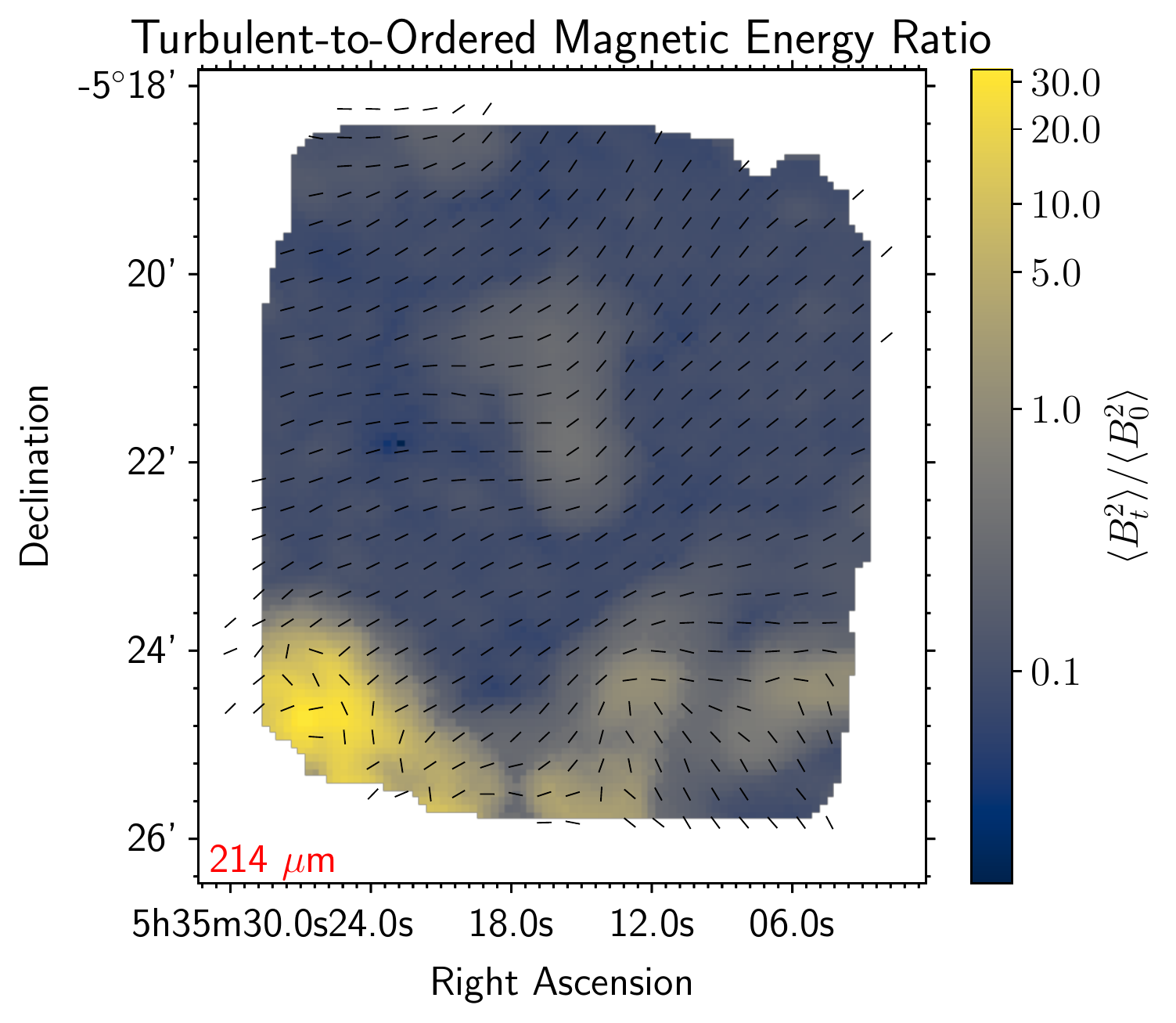}
    \includegraphics[width=2.3in]{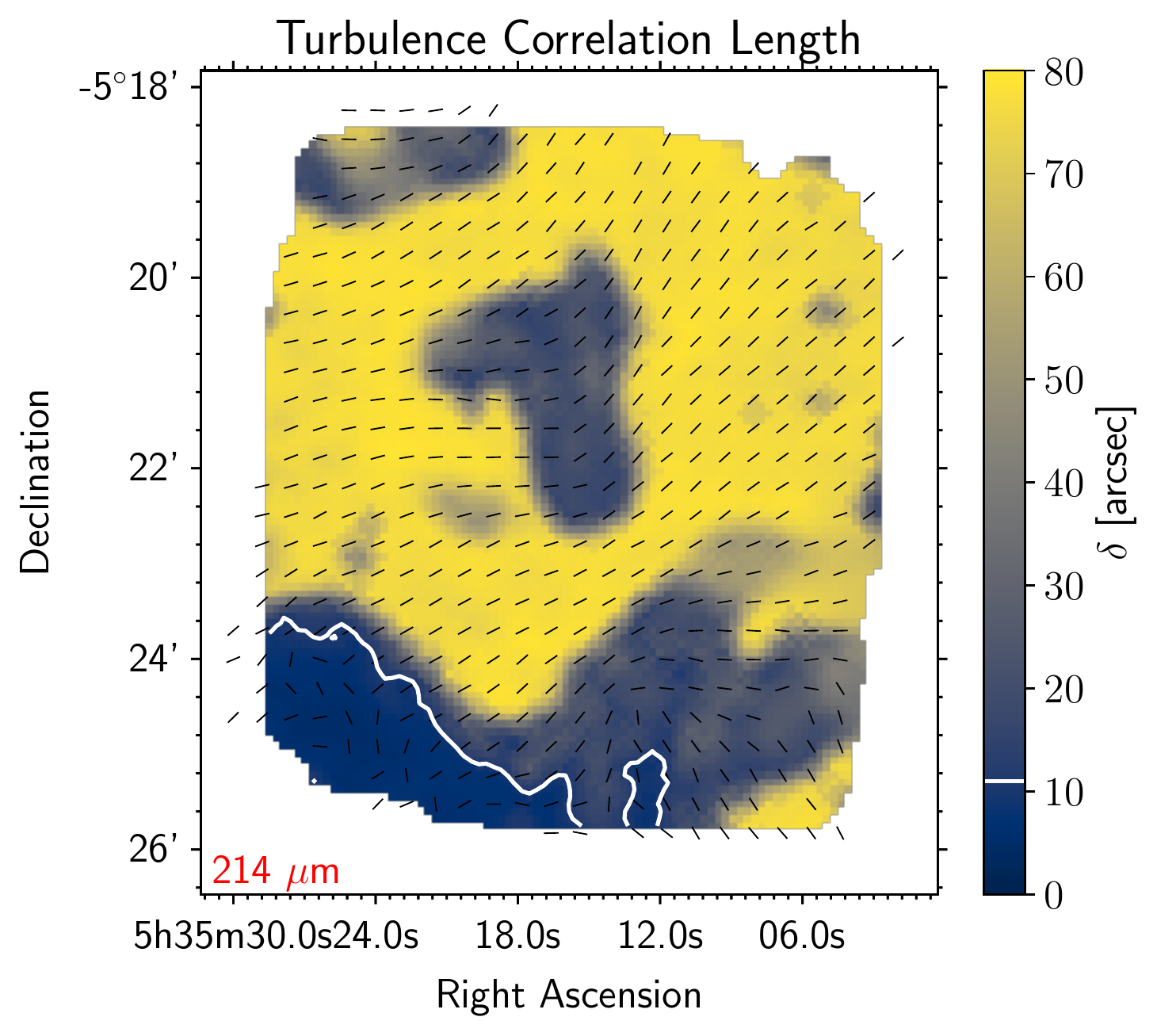}
    \caption{Maps of parameters $a_{2}$ ({\it Left}),  $\ratio$ ({\it Middle}), and $\delta$ ({\it Right}) for 214 $\micron$ data using $w_{\rm opt}$ = 9 pixels. In all three parameters maps a similar structure is observed: large-scale field and turbulence contributions to the dispersion of polarization vectors is larger in a region near and around the BN/KL object and in the OMC-1 bar. In these regions the turbulence's correlation length appears shorter. The white contour in the map of $\delta$ encompass those locations where the turbulence correlation length is equal or shorter than the auto-correlated beam size, $\sqrt{2}W$.}
    \label{fig:param_maps_delta}
\end{figure}

Figure \ref{fig:param_maps_delta} displays the parameter maps calculated for 214 $\micron$ data using $w_{\rm opt}$ = 9 px. To obtain $\ratio$, $\Delta^{\prime}$ needs to be estimated.  As stated in Section \ref{sec:disp}, values of $\Delta^{\prime}$ are estimated to be the width of the autocorrelation function of the polarized intensity. A value of $\Delta^{\prime}_{214} = 91.0$\arcsec\ is obtained for the $\ratio$ map in Figure \ref{fig:param_maps_delta} ({\it Middle}). For the other HAWC+ wavelengths, values of the cloud's effective thickness used are $\Delta^{\prime}_{154} = 81.6$\arcsec, $\Delta^{\prime}_{89} = 80.0$\arcsec, and $\Delta^{\prime}_{53} = 71.4$\arcsec.  Maps of parameters have an angular resolution set by the size of kernel used to calculate the dispersion functions. A FWHM value for each band is defined to be the FWHM of the Gaussian having the same area as the kernel. The result is FWHM$  = 1.88 w_{\rm opt}$. This corresponds to 77.0\arcsec\ for 214~\micron, 57.5\arcsec\ for 154~\micron, 33.0\arcsec\ for 89~\micron, and 20.7\arcsec\ for 53~\micron. The maps in Figure \ref{fig:param_maps_delta} are shown with beam-sampled inferred magnetic field vectors superposed. These maps were first cleaned by removing outliers using the Chauvenet criterion. That is, values that exceed three times the standard deviation within a $3\times 3$ region centered at the pixel are removed and replaced with interpolated values. Less than $\sim10$\% of the pixels in each map are replaced by this process. 

In all three maps a similar spatial structure is seen. First, larger values of the $a_{2}$ and $\ratio$ values appear in locations where the polarization vectors are seen to have larger deviation from uniformity. This occurs near the BN/KL and the bar regions. The largest values of both $a_{2}$ and $\ratio$ are seen in the bar region where the polarization vectors (see Figure~\ref{fig:sample_DFs}) are visibly most random. In the $\delta$-map (Figure \ref{fig:param_maps_delta}, {\it Right}) the turbulence correlation length appears shorter in the same regions where the $a_{2}$ and $\ratio$ parameters are larger. Indeed, it appears that $a_{2}$ and $\delta$ are negatively correlated. Table 4 in \citet{Chuss2019}, where single values of the three parameters are reported for the three regions (BN/KL, bar, and HII), also hints at a negative correlation. This anticorrelation is expected because small values of $\delta$ represent large numbers of turbulent cells in the column of gas and therefore provide a larger contribution to the dispersion. 

In Figure \ref{fig:param_maps_delta} ({\it Right}) a white contour delineates the level at which  $\delta$ is equal to the auto-correlated beam size, $\sqrt{2}W$, for the corresponding HAWC+ wavelength (see color bar). In regions where $\delta$ is lower than this level, the contribution of the gas turbulence cannot be properly resolved by the angular resolution of the polarimetric observations, and the resulting $B_{\rm POS}$ can be overestimated. The majority of such areas are located in the OMC-1 bar. It is worth noting that in the OMC-1 bar the potential effect of reference beam contamination is particularly large for longer HAWC+ wavelengths \citep{Chuss2019}. However the effect of reference beam contamination on dispersion functions needs to be properly studied. Over most of the map, the turbulence scale is resolved.

To check that the structure seen in Figure \ref{fig:param_maps_delta} is not due to covariance among the fitted parameters, the dispersion functions are re-fitted assuming a constant value of the local turbulence correlation length ($\delta$) to apply to each pixel in the map.  This value is set to the global value of correlation length $\delta_{0}$, calculated using the data from the entire map. In this case, only maps of $a_{2}$ and $\ratio$ are constructed. The resulting maps (Appendix \ref{sec:appendixA}) show a similar spatial structure as those in Figure \ref{fig:param_maps_delta}. This provides confidence that the structure in the maps of Figure \ref{fig:param_maps_delta} is not due to a fitting-induced covariance. In addition, by comparing the histograms of the reduced goodness-of-fit $\chi^{2}$ values (Figure \ref{fig:rho_dist}, {\it Right}) obtained for $w_{\rm opt}$ and using the $\delta$-fixed (blue) and $\delta$-variable (orange) approaches, it is clear that in the latter case, lower occurrence is seen in all bins of $\chi^{2}$ but the smallest. This implies that allowing $\delta$ to be determined by the MCMC fit improves the fit in most pixels.

The focus of this work is the construction of magnetic field maps, for which only the $\ratio$ map is used from this analysis. Further interpretation of the other DCF parameter maps is deferred to future work.

\subsection{Local Dispersion Maps}\label{sec:LocDisp}

The local dispersion quantifies the deviations of polarization angles at a given position from the average direction calculated within a circle centered at that position. Although the local dispersion has a variety of definitions, this work utilizes a version similar to that of \citet{Planck2018}. The measured dispersion, $\mathcal{S}_m$, within an angular radius, centered at each pixel can be written as

\begin{equation}
    \mathcal{S}_m \equiv \sqrt{\frac{1}{N}\sum_{i=1}^N\left(~\phi_i - \overline{\phi}~\right)^2},
\end{equation}

\noindent
where $\phi_i$ is the polarization angle of pixel \textit{i}, $\overline{\phi}$ is the average polarization angle of pixels inside the kernel, and $N$ is the number of pixels within the kernel. To calculate the average polarization direction, $\overline{\phi}$, first the error-weighted normalized Stokes parameters, $\overline{q}$ and $\overline{u}$, are calculated, and their associated errors are propagated accordingly. Then, $\overline{\phi} \equiv 0.5$ \texttt{arctan2}$(\overline{u}, \overline{q})$, where \texttt{arctan2} places the polarization angle in the correct quadrant. Since polarization is represented by a pseudo-vector, if $\left|\phi_i - \overline{\phi}\right| > 90^\circ$, the supplement of the angle between pseudo-vectors is taken.
        
To correct for bias in the positive-definite quantity $\mathcal{S}$, \citet{Planck2018} is followed

\begin{equation}
\mathcal{S} = 
\begin{dcases} 
\sqrt{\mathcal{S}_m^{2} - \sigma^{2}_{\mathcal{S}}} & \text{if } \mathcal{S}_m > \sigma_{\mathcal{S}} \\
0 & \text{otherwise }
\end{dcases}
,
\end{equation}

\noindent
where $\sigma_{\mathcal{S}}^{2}$ is found by propagating the uncertainties in both the individual $\phi$'s and the mean value $\overline{\phi}$

\begin{equation}
        \sigma_\mathcal{S}^2 = \frac{1}{N^2\mathcal{S}_m^2}\sum_{i=1}^N\left(~\phi_i - \overline{\phi}~\right)^2 \sigma_{\phi_i}^2
        + \frac{\sigma^2_{\overline{\phi}}}{N^2\mathcal{S}_m^2}\left[\sum_{i=1}^N\left(~\phi_i - \overline{\phi}~\right)\right]^2.
\end{equation}
Here, $\sigma_{\phi_i}$ is the polarization angle error for pixel $i$ and $\sigma_{\overline{\phi}}$ is the error in the mean polarization angle calculated by propagating the errors in the values of ${\phi_i}$ over the region defined by the kernel. 
        
\begin{figure}[h!]
    \centering
    \includegraphics[width=3.5in]{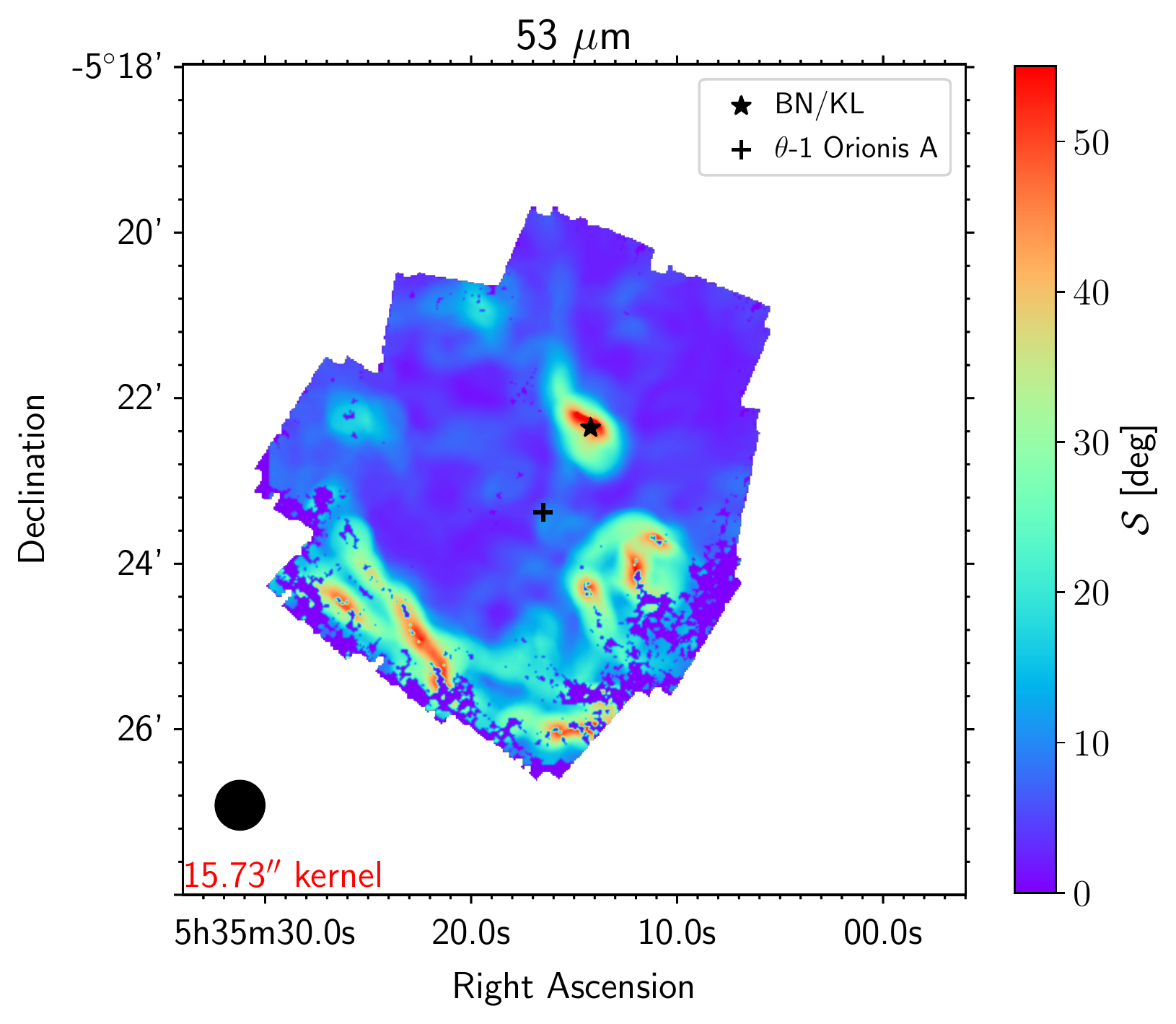}
    \includegraphics[width=3.5in]{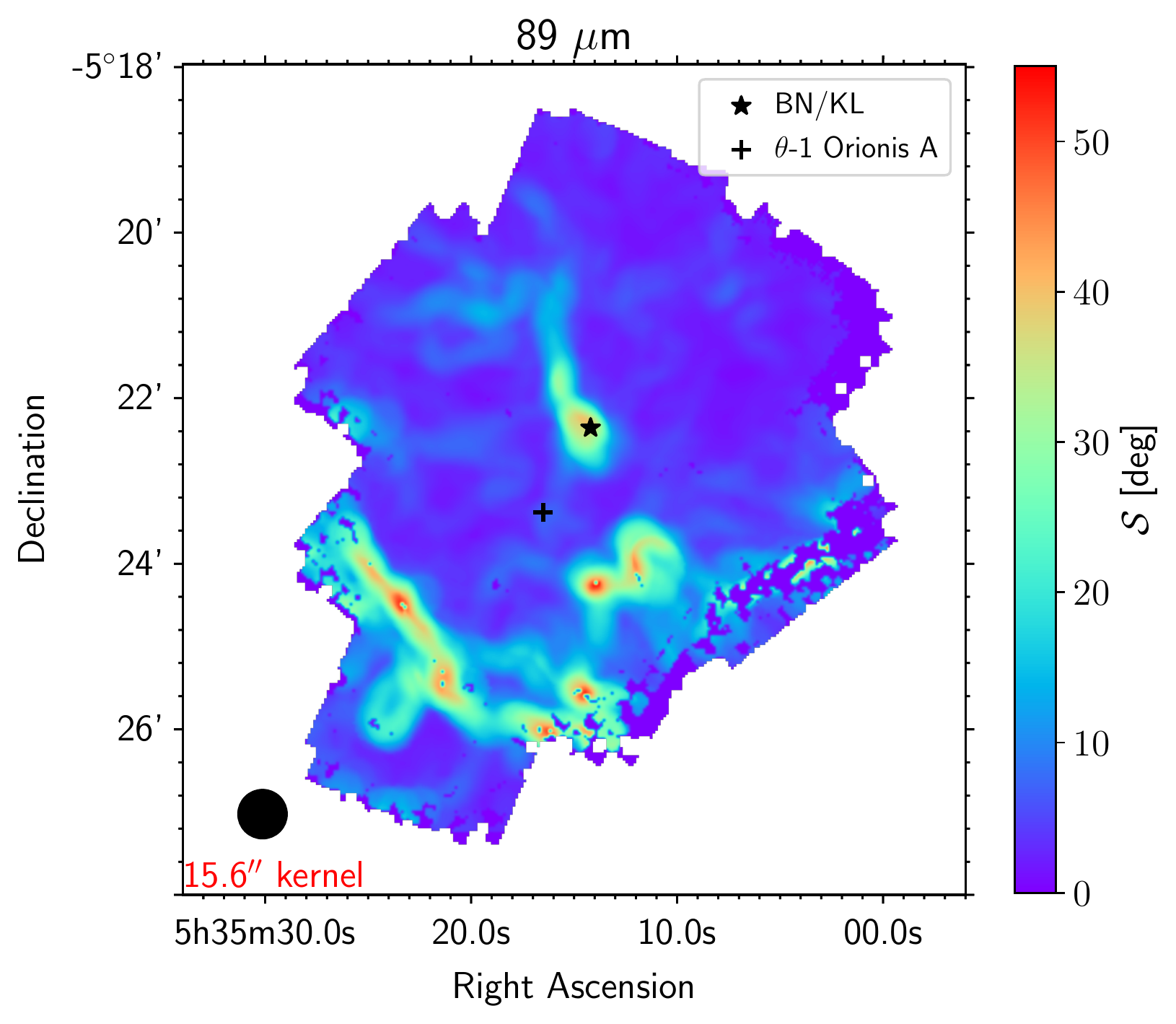}\\
    \includegraphics[width=3.5in]{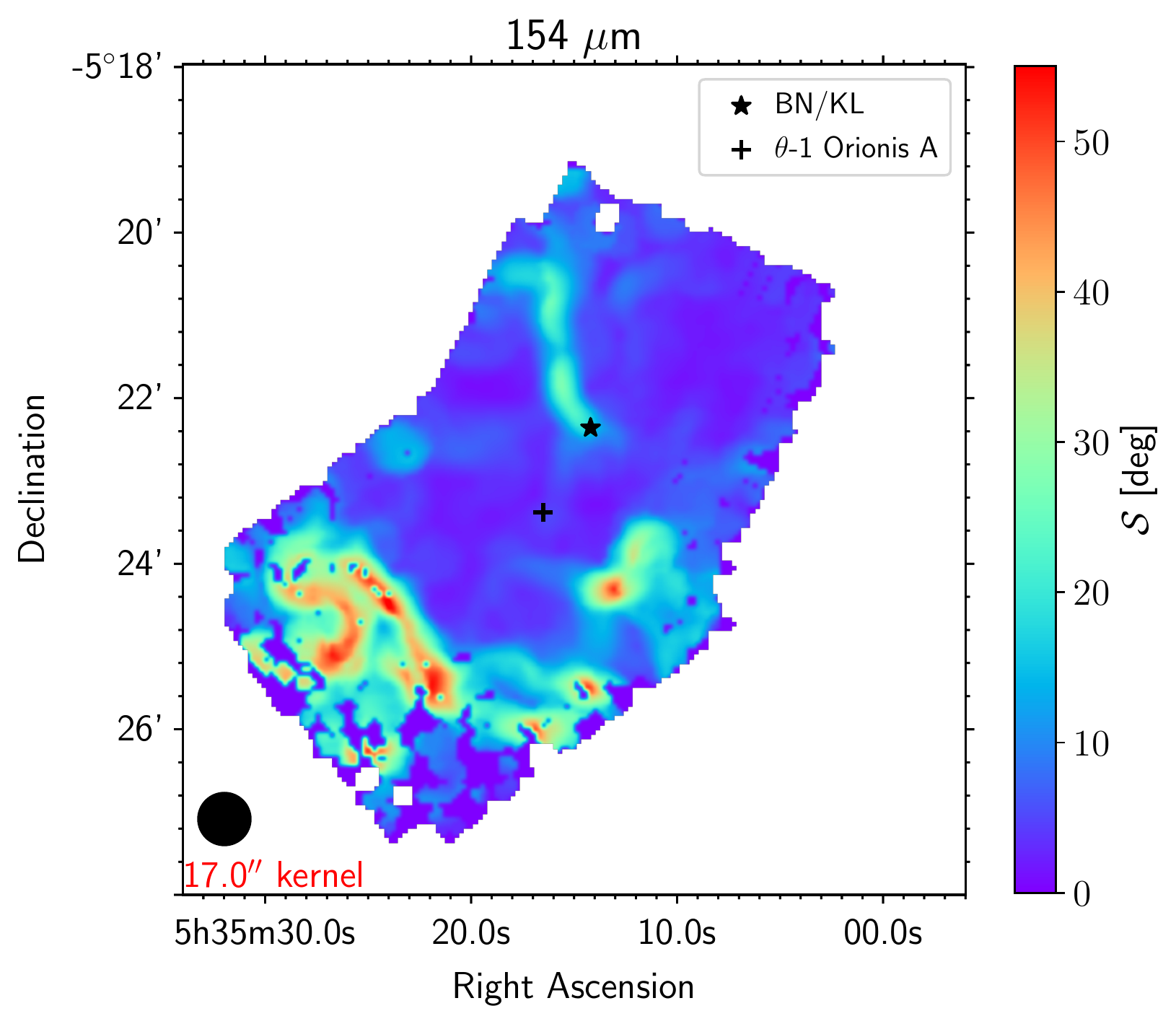}
    \includegraphics[width=3.5in]{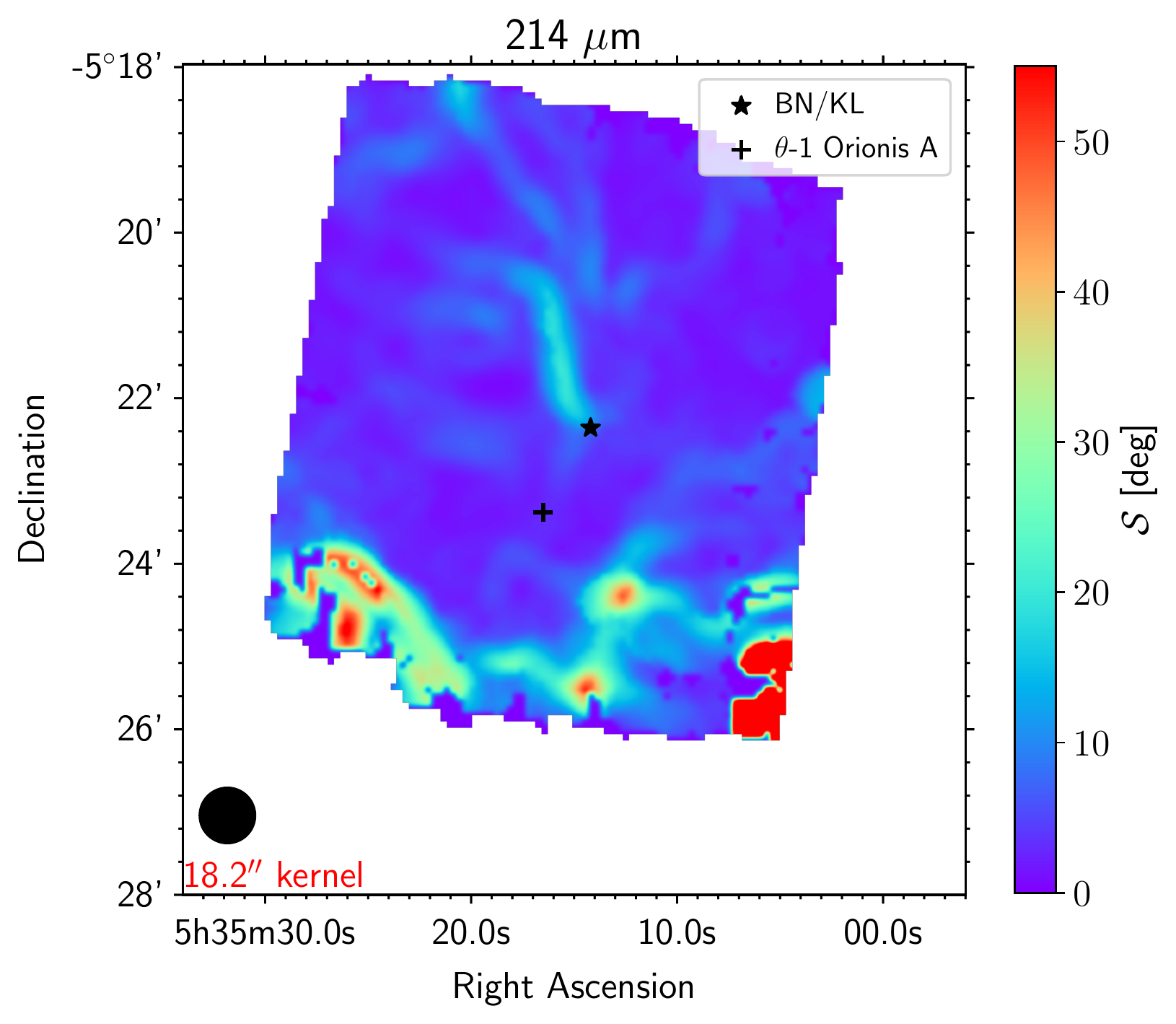}
    \caption{Maps of $\mathcal{S}$ for all four HAWC+ bands are shown. The local dispersion, $\mathcal{S}$, is calculated using circular regions of $\sim$ 15 - 18.2\arcsec~angular diameter. In order to observe more spatial details, the color scale of $\mathcal{S}$ is capped at 55$^{\circ}$. 
    }
    \label{fig:dispersion}
\end{figure}

For calculating $\mathcal{S}$, the radius of the circular kernel was chosen to be an integer number of pixels that approximately corresponds to the angular resolution of the velocity dispersion maps (Section \ref{sec:H2_vel}) which is $\approx$ 16$^{\prime\prime}$ (32$^{\prime\prime}$ FWHM). The precise values of each kernel is shown for each HAWC+ band in the $\mathcal{S}$-maps of Figure \ref{fig:dispersion}. Although, testing revealed that $\mathcal{S}$ in general increases with larger kernel sizes, the growth did not continue much beyond 2 - 3 times the HAWC+ beam size. This choice of kernel size, which is at least twice the beam diameter for each band, guarantees that in all cases $\mathcal{S}$ values are not under-sampling the HAWC+ data.

Figure \ref{fig:dispersion} shows maps of $\mathcal{S}$ across all four HAWC+ wavelengths using an angular radius of $\sim$16\arcsec. The scale of $\mathcal{S}$ values saturates at 55$^{\circ}$. Higher-than-average values of $\mathcal{S}$ seem to appear in elongated structures that are reasonably spatially consistent across all wavelengths. The observation of such elongated structures was first reported by \citet{Planck2018} in the 353-GHz data of Planck and subsequently studied by \citet{Clark2019}. It is also clear that larger-than-average dispersion $\mathcal{S}$ is observed at the same locations ({\it i.e.}, BN/KL, the bar) where the analysis of dispersion functions (previous section) produced large values of $\ratio$, which will correspond to lower POS magnetic field strengths. 

The parameter $\mathcal{S}$ has been shown to be negatively correlated with the polarization fraction; regions with large dispersion display low values of polarization fraction \citep{Chuss2019,Fissel2016,Planck2018}. Several explanations are possible for $\mathcal{S}$-$p$ anticorrelation: 1) highly turbulent fields cancel contributions from different depths along the line of sight; 2) in dense regions of the cloud, grains may be less well-aligned than in cloud envelopes; and 3) the magnetic field is mostly oriented in the LOS direction. According to \citet{Hensley2019}, in a magnetic field mostly oriented in the LOS direction, even small perturbations can induce large changes in $\phi$, leading to large values of $\mathcal{S}$. Therefore, one would expect regions for which the field is predominantly oriented along the LOS to have low $p$ and high $\mathcal{S}$. In order to estimate the LOS component, for which the DCF technique is insensitive, empirical relations between $\mathcal{S}$ and measurement of the LOS magnetic field may be a viable exploratory path.

\section{Results} \label{sec:results}

\subsection{Plane-of-Sky (POS) Component}\label{sec:pos_field}

As is evident from Eq. \ref{eq:dcf_eq}, the spatial variations in all three quantities ($\rho$, $\sigma_{v}$, and $\ratio$) determine the spatial variations in the resulting $B_{\rm POS}$. Previous work has been limited to providing estimates of $B_{\rm POS}$ that were averaged over large regions (or in most cases, over the entire map) because of the lack of spatial information available for one or more of these three quantities. In this work, maps for all three variables are used, and maps of POS magnetic field strength are produced for the first time.

\subsubsection{Column Density and Velocity Dispersion maps}\label{sec:H2_vel}

\begin{figure}[!h]
    \centering
    \includegraphics[width=3.5in]{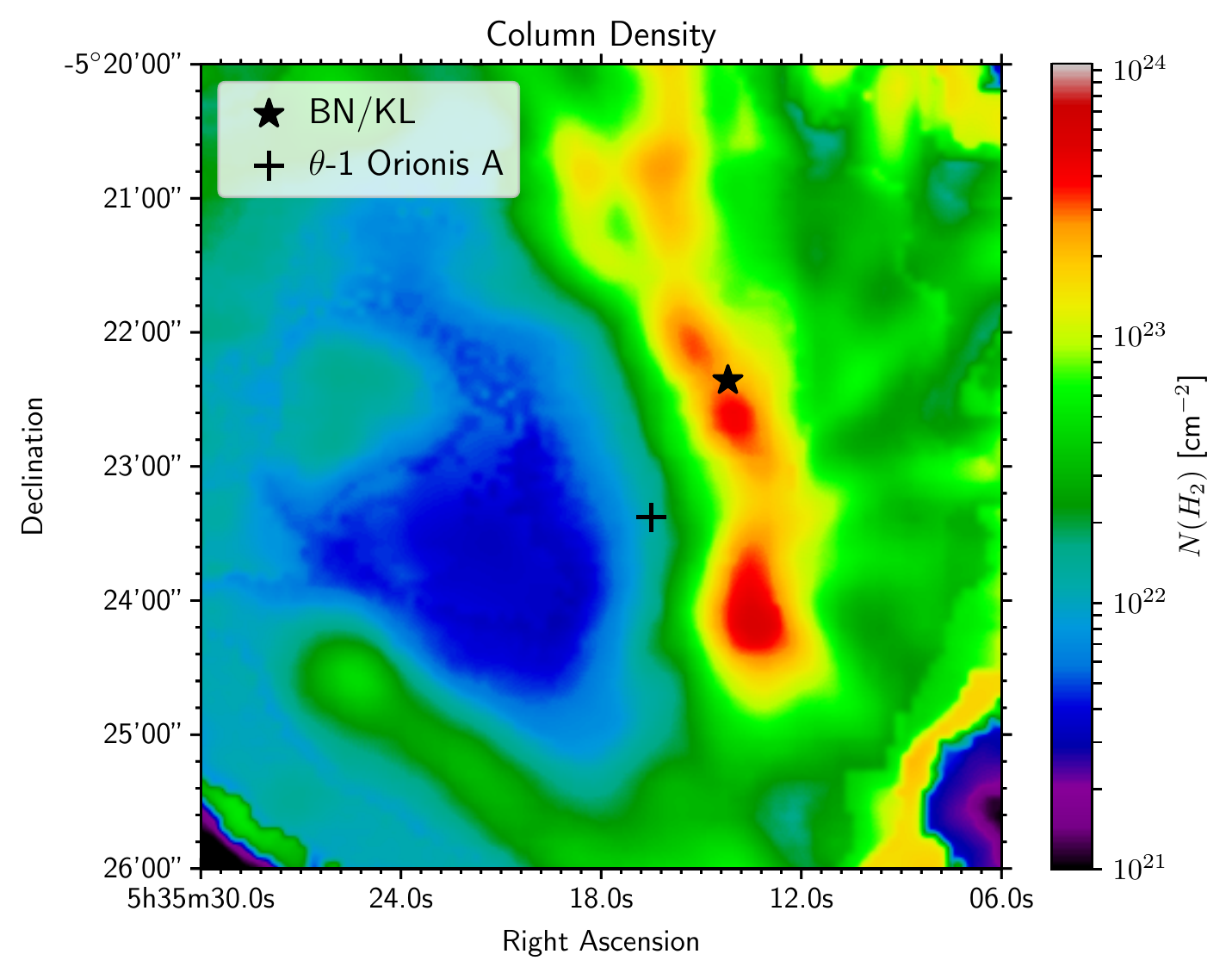}
    \includegraphics[width=3.5in]{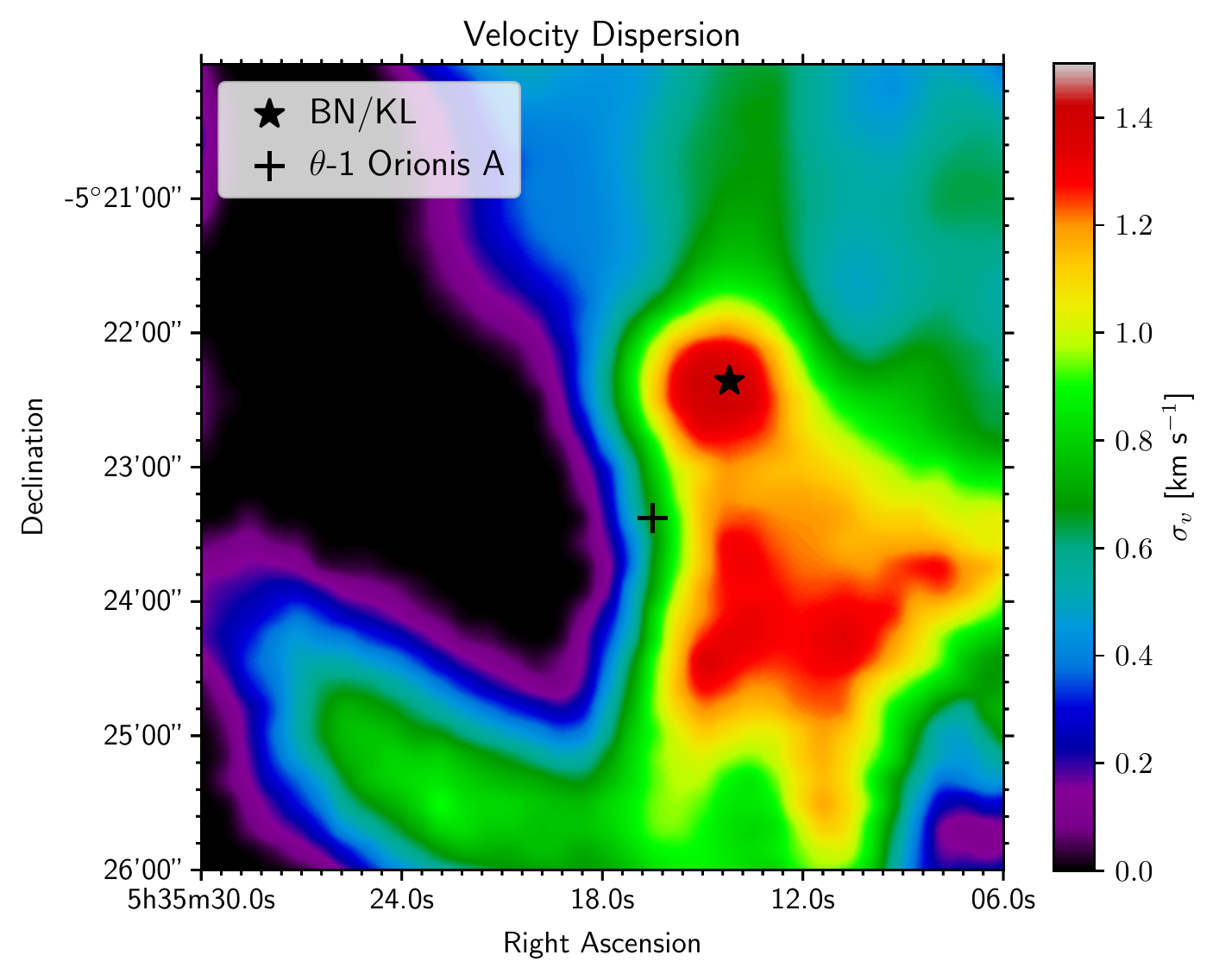}
    \caption{{\it Left:} Column density map for the OMC-1 region obtained thought the fitting of Spectral Energy Density using infrared photometric data from different instruments (for larger spectral coverage), including HAWC+. The angular resolution of this map is 22\arcsec\  \citep{Chuss2019}. {\it Right:} OMC-1 velocity dispersion map determined from the multi-Gaussian fitting of the hyperfine structure of $NH_{3}(1,1)$. These observations were taken with a FWHM beam of 32\arcsec\ \citep{Friesen2017}.}
    \label{fig:den_vel_maps}
\end{figure}

First, a mass density map is obtained using the map of $H_{2}$ column density ($N(H_{2})$; Figure \ref{fig:den_vel_maps}, {\it Left}) from \citet{Chuss2019} and assuming an uniform depth of the cloud of $\sim$10$^{17}$ cm. This $N(H_{2})$ map was obtained by fitting of the spectral energy distribution (SED) of IR emission from OMC-1 in the range 53 $\micron$ - 35 mm, including data from the four HAWC+ bands. In \citet{Chuss2019}, the values of $N(H_{2})$ used to estimated $B_{\rm POS}$ for the three distinct regions (BN/KL, the bar, and the HII area) corresponded to average values from the same $N(H_{2})$ map used here, calculated inside each region. See \citet{Chuss2019} for a discussion of details.

Velocity dispersion values for the OMC-1 region can be obtained from emission line spectra of an appropriate molecular tracer. The ammonia molecule, NH$_{3}$, has been used as a probe of dense clouds and clouds cores \citep[$n > 2\times 10^{3}$ cm$^{-3}$;][]{Friesen2017}. In particular, the emission line from the (1,1) transition of NH$_{3}$ (rest frequency 23694.4955 MHz) has been found to be highly correlated with the dust column density derived from Herschel observations in the OMC-1 region \citep[see Fig.~7 in][]{Friesen2017}, and thus, the coexistence of this molecular tracer with the polarized dust emission is inferred. Most of the OMC-1 region exhibits column density values (estimated from thermal dust SEDs) in excess of 10$^{22}$ cm$^{-2}$. The exception is in the HII region surrounding the Trapezium cluster where the column density drops to approximately half of this value.  Because of the lack of information about the velocity dispersion, the HII region is excluded from the analysis

These $\sigma_{v}$ data\footnote{Available at: \href{https://dataverse.harvard.edu/dataverse/GAS\_Project}{https://dataverse.harvard.edu/dataverse/GAS\_Project}} were obtained with the Green Bank Telescope as part of the Greenbank Ammonia Survey \citep[GAS; ][]{Friesen2017} to map all star-forming regions in the Gould Belt. This survey includes NH$_{3}$(1,1) observations of the Orion-A (North) filament taken with a beam of 32$^{\prime\prime}$ (23 GHz). At the distance of Orion-A, this beam yields a resolution of $\sim0.06$ pc. In order to estimate $\sigma_{v}$, the hyperfine structure of the NH$_{3}(1,1)$ line is modelled for thermal and non-thermal widening with a multi-Gaussian model, assuming that each hyperfine splitting has the same $\sigma_{v}$ value. The resulting map of $\sigma_{v}$ can be seen in the right panel of Figure \ref{fig:den_vel_maps}. Velocity dispersion values range from $\sim$0.8 $\mathrm{km\,s^{-1}}$ in areas like the bar to $\sim$1.5 $\mathrm{km\,s^{-1}}$ near the BN/KL object. 

\subsubsection{Maps of $B_{\rm{POS}}$}

As stated before, maps of $B_{\rm POS}$ are obtained by combining the maps of $\rho$, $\sigma_{v}$, and $\ratio$ according to Equation~\ref{eq:dcf_eq}, once they have been smoothed to a common resolution.  In doing so, the following steps are taken:

\begin{enumerate}

    \item Maps of $\rho$ and $\sigma_{v}$ are re-projected to the pixelization of each HAWC+ data map (which is the same as the pixelation of $\ratio$).
    \item Maps are smoothed by convolving the original maps with a Gaussian kernel of $\sigma = \sqrt{\sigma_{\rm T}^{2} - \sigma^{2}_{\rm O}}$, where the subscripts T and O signal the $\sigma$ value of the target and original resolutions, respectively. The target resolution is set by that of the map with the lowest resolution among all the maps involved in the calculation.
    
\end{enumerate}

For the 214, 154, and 89~\micron\ maps, the angular resolution for the POS magnetic field maps is set by the $\ratio$ maps: 77$^{\prime\prime}$, 58$^{\prime\prime}$, and 33$^{\prime\prime}$, respectively. For the 53~\micron\ map, the resolution is 32$^{\prime\prime}$, limited the resolution of the $\sigma_{v}$ map. Thus, the 53~\micron\ map is unique in that the $\ratio$ map is smoothed to a courser resolution. For the remainder of the bands, this quantity sets the resolution of the $B_{\rm POS}$ map. 

\begin{figure}[!h]
    \centering
    \includegraphics[width=3.5in]{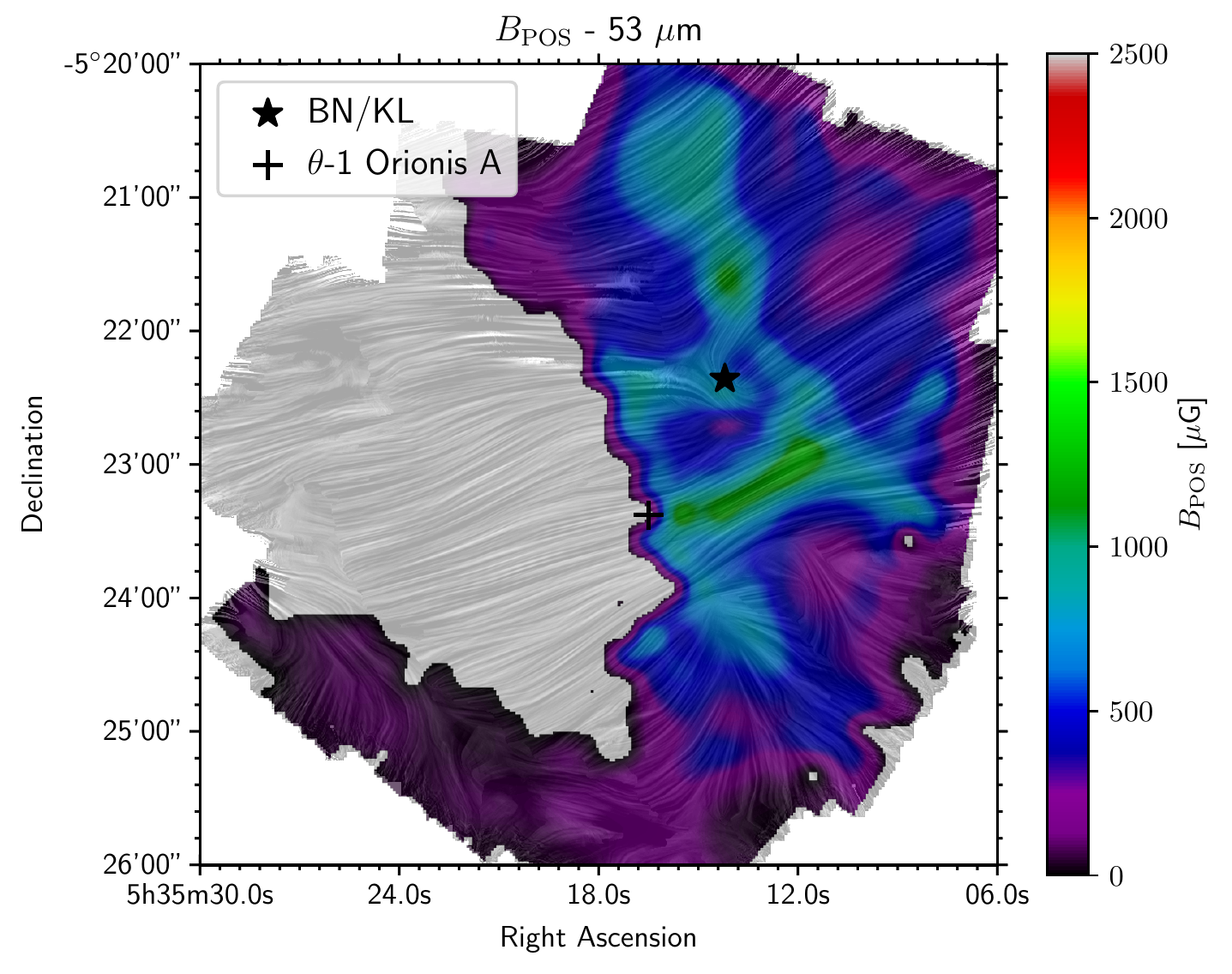}
    \includegraphics[width=3.5in]{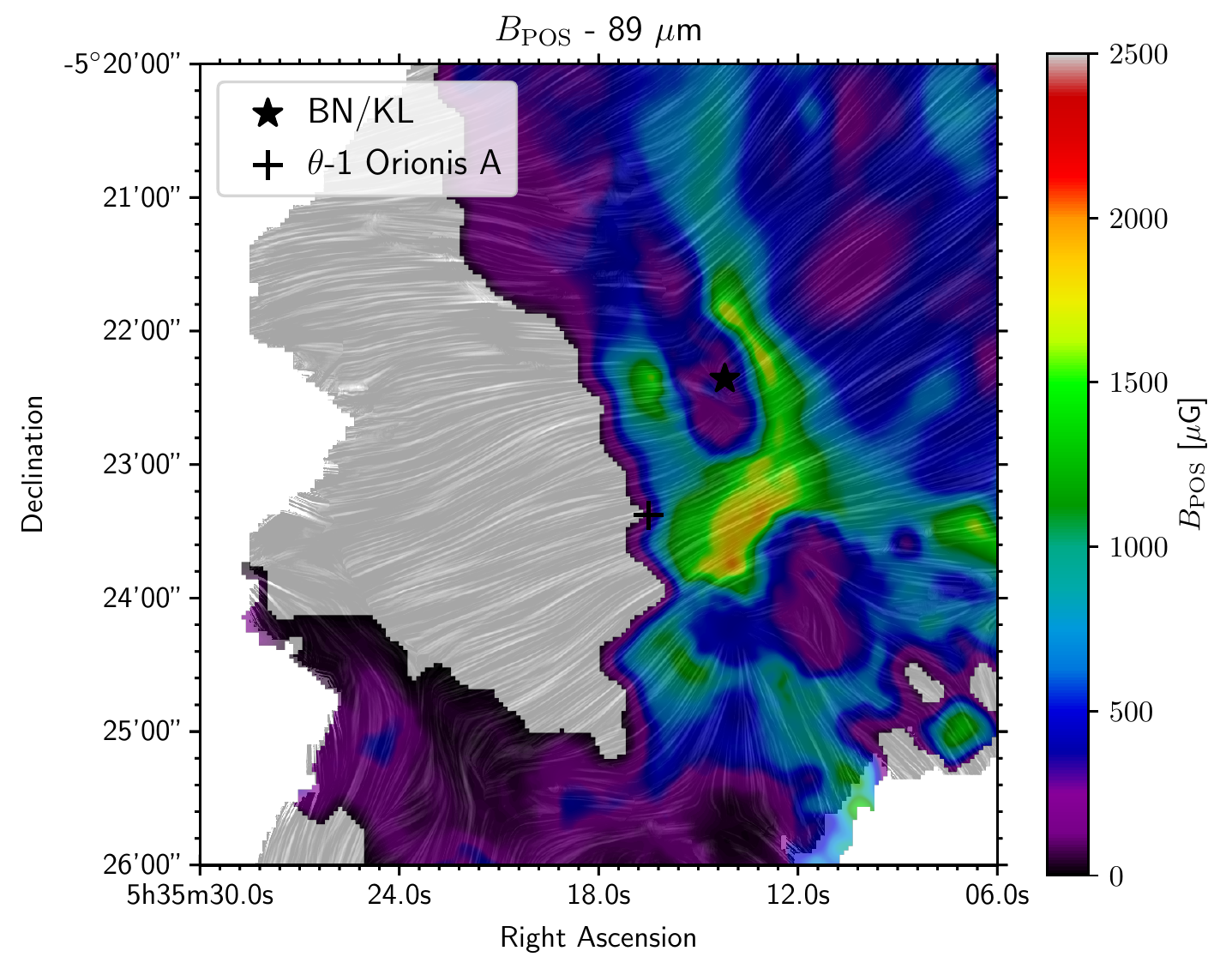}
    \includegraphics[width=3.5in]{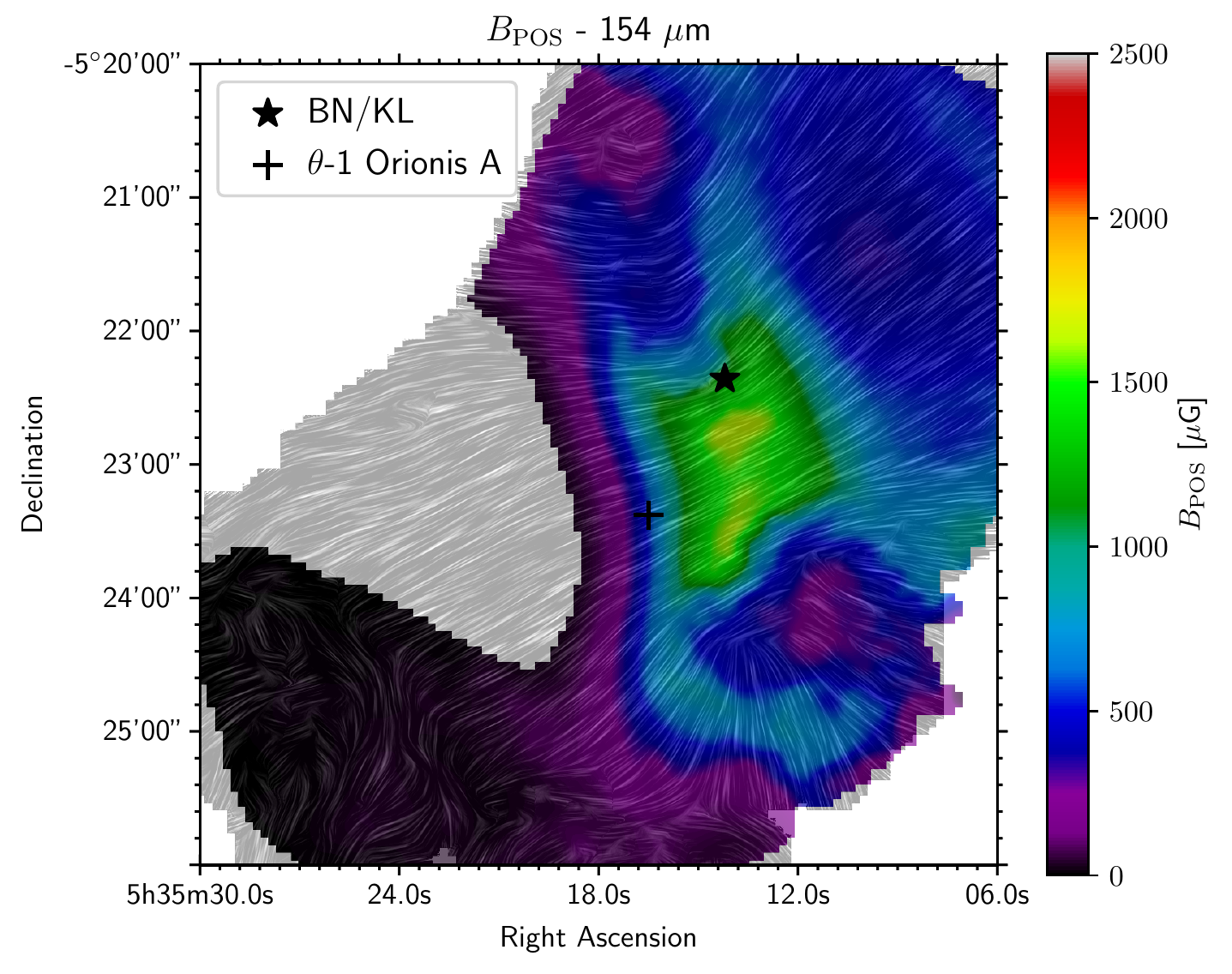}
    \includegraphics[width=3.5in]{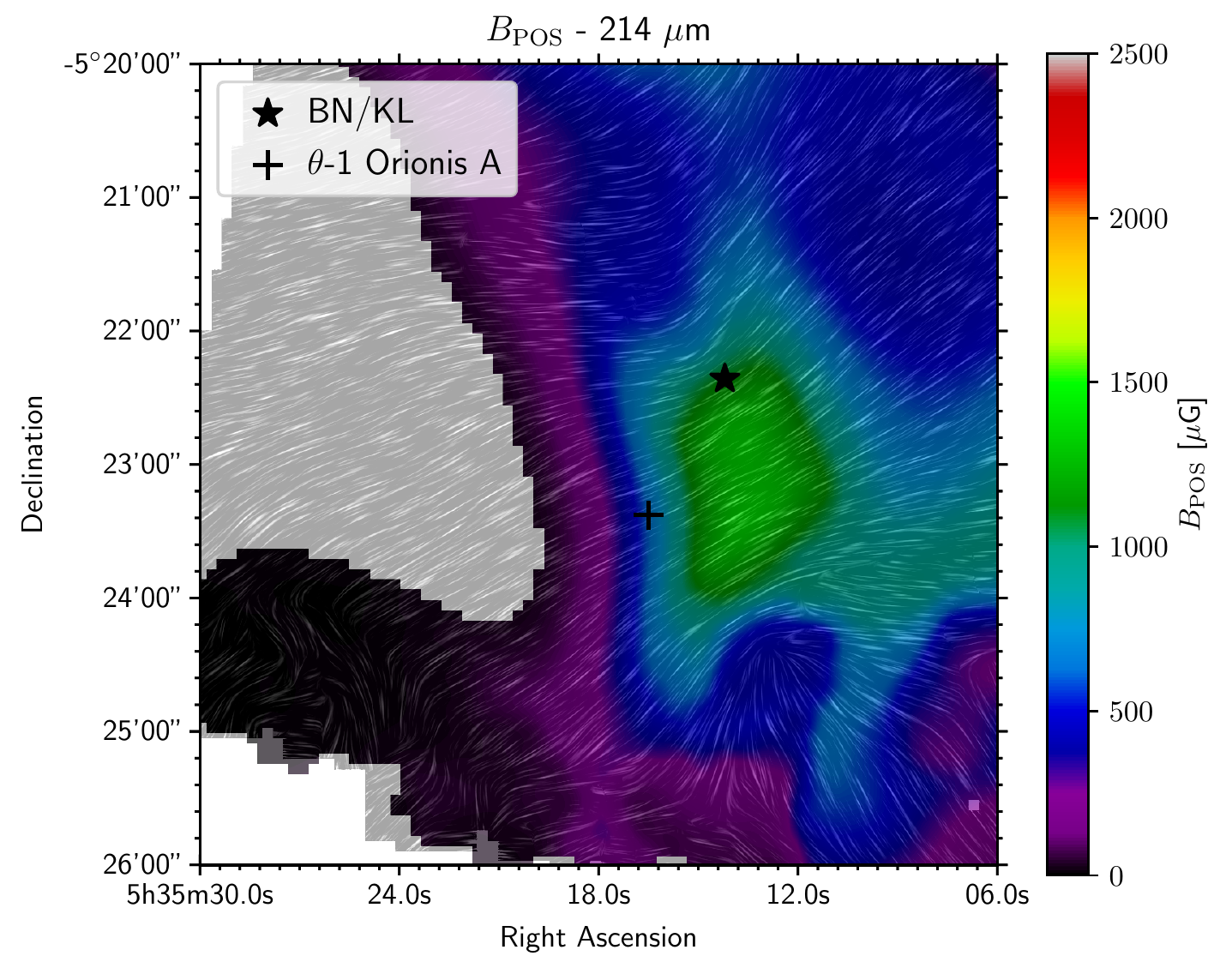}
    \caption{Maps of POS magnetic field strength for 53 $\micron$ (Top left), 89 $\micron$ (Top right), 154 $\micron$ (Bottom left), and 214 $\micron$ (Bottom right). Inferred magnetic field orientation is shown by LIC contours. For reference, the locations of the BN/KL object (star) and the Trapezium cluster (cross) are included as well. The angular resolution in each map is 32$^{\prime\prime}$ (53 $\micron$), 33$^{\prime\prime}$ (89 $\micron$), 58$^{\prime\prime}$ (154 $\micron$), and 77$^{\prime\prime}$ (214 $\micron$).}
    \label{fig:B_maps}
\end{figure}

The resulting maps of the plane-of-the-sky magnetic field are displayed in Figure~\ref{fig:B_maps} for all four HAWC+ wavelengths. The inferred POS magnetic field direction for each map is shown as a line integral contour \citep[LIC; ][]{Cabral1993} overlay. Analyzing all four maps, one finds: 1) $B_{\rm POS}$ values range from $\sim$ 100 $\mu$G to a maximum value of $\sim$2000 $\mu$G; 2) the maximum $B_{\rm POS}$ value in the map increases with increasing angular resolution of the maps from 214 to 89 \micron; 3) for the 53~\micron\ map, the maximum value of $B_{\rm POS}$ is slightly lower,  $\sim$ 1500 $\mu$G; 4) the largest field strengths ($\sim$ 2000 $\mu$G) are consistently observed around and south of the BN/KL object, where both the mass density and velocity dispersion are large and the angular dispersion low; 5) weaker $B_{\rm POS}$ strengths are observed in the bar region, due to a combination of low density and large polarization dispersion.

Differences in the four maps potentially indicate that the four wavelengths are each preferentially sampling a different part of the cloud along the line-of-sight. For example, the short-wavelength (53 and 89 $\micron$) maps clearly show a region where the POS magnetic field strength decreases, that it is not observed in the long wavelengths (154 and 214 $\micron$) maps. This region is approximately centered at the BN/KL-object location (Figure \ref{fig:B_maps}, top panels) and is approximately 3.6 pc $\times$ 4.7 pc in area. This region is coincident with the BN/KL explosion \citep{Bally2017} and the decreased inferred POS field strength may be due to the sensitivity to the explosion morphology at these bands due to their higher resolutions and sensitivity to warmer dust located in the interior of the cloud complex.

Maps of $B_{\rm POS}$ constructed in this work have two important considerations. First, the magnetic field strength is underestimated in areas where the turbulence correlation length ($\delta$) is not resolved. Specifically, this is clearly a problem for the bar in the 214~\micron\ data. This particular shortcoming can be fixed only by increasing the angular resolution of the polarimetric data. However, the application of a statistical correction factor in such areas could be studied using simulated polarimetric data. Second, the quality of $B_{\rm POS}$ maps is dependent on (resolution and accuracy) the quality of the maps of mass density and velocity dispersion. On the other hand, the largest uncertainty source for the $B_{\rm POS}$ maps presented here, likely comes from the $\ratio$ parameter. In this work, we assume a fractional uncertainty in $B_{\rm POS}$ of 50\%\citep{Ostriker2001}.

\subsection{Line-of-Sight (LOS) Component}\label{sec:los_field}

The DCF technique as described above enables one to produce a map of the magnitude of the POS component of the magnetic field. Because the total magnetic field is three-dimensional, it is desirable to also take into account any significant contribution of the LOS component of the field for the end goal of estimating $M/\Phi$ over the cloud. Here a novel approach is explored for estimating the angle of the field relative to the LOS that utilizes the local dispersion $\mathcal{S}$ that is described above. This rough estimate of the magnitude of the LOS component of the magnetic field, $B_{LOS}$ that is described here is used as a map-based correction to estimate the total magnetic field from the POS component as determined by the DCF technique above. 

\begin{figure}[!h]
    \centering
    \includegraphics[width=3.2in]{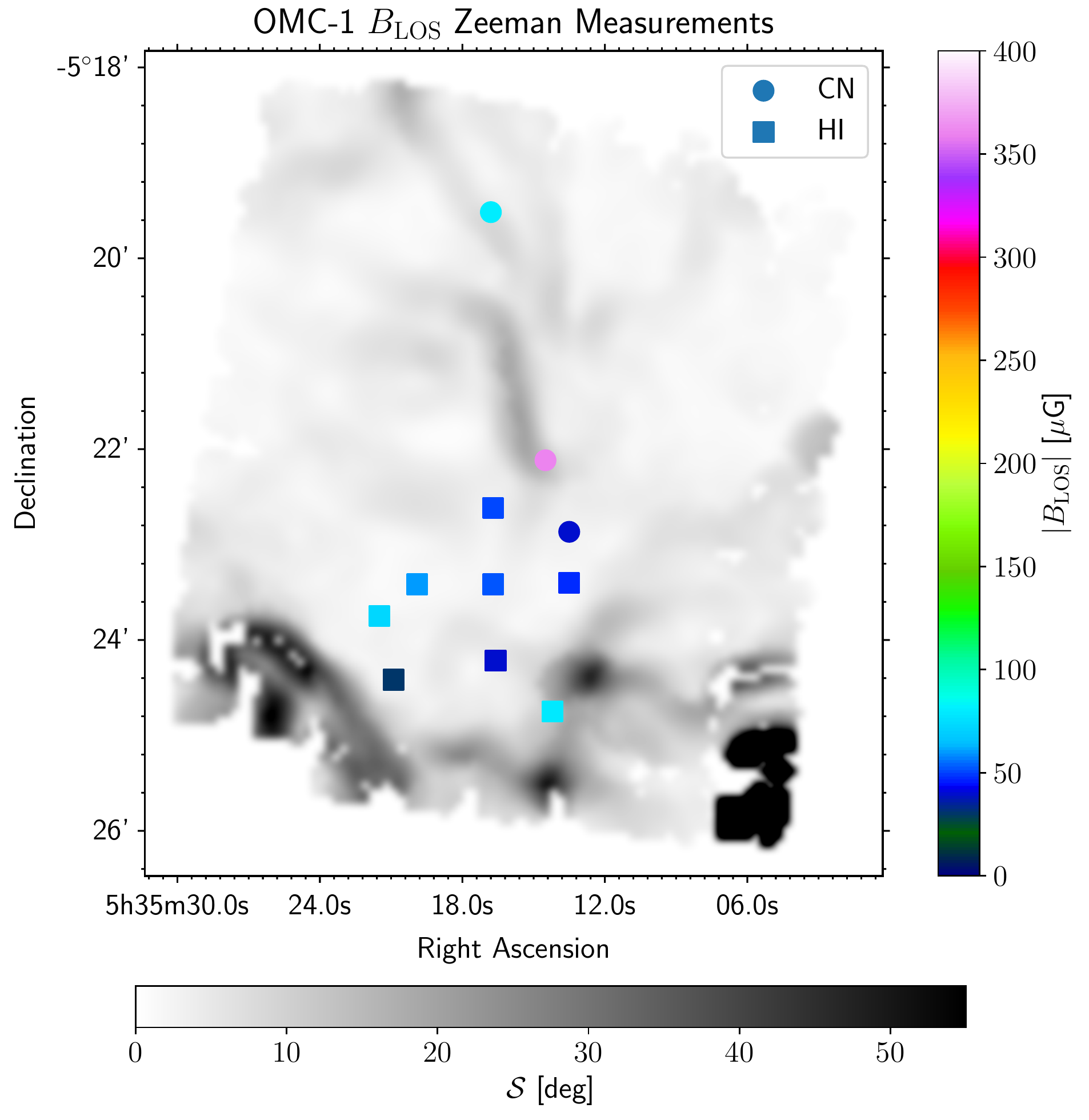}
    \caption{Map of dispersion values $\mathcal{S}$ for HAWC+ 214 $\mu$m data. Symbols in color correspond to Zeeman measurements of the line-of-sight magnetic field, $B_{\rm LOS}$. Squares and circles correspond to the Zeeman-splitting data of HI and CN lines, correspondingly.}
    \label{fig:Blos_S_map}
\end{figure}

\citet{Hensley2019} demonstrates that for the diffuse ISM, the dispersion of polarization angles $\mathcal{S}$ is modulated by the angle between that the magnetic field forms with the LOS direction, $\varphi$. They propose the relation $\sin^{2}(\varphi) \propto \mathcal{S}^{n}$ and found values of $n$ between $-0.478$ and $-0.528$ (for different ranges of $N$(HI) using the Planck 353 GHz all-sky data). Because $n < 0$, values of $\mathcal{S}$ near zero are associated with $\varphi$ close to $\pi/2$ -- the magnetic field is close to the plane-of-sky. On the other hand, large values of $\mathcal{S}$ then indicates that the magnetic field is oriented closer to the line-of-sight direction ($\varphi \rightarrow$ 0). Therefore, $\mathcal{S}$ is assumed to be a rough tracer of $B_{\rm LOS}$ across the cloud. This relationship can be calibrated by examining Zeeman measurements of the magnetic field in OMC-1, which directly measure $B_{\rm LOS}$. Figure \ref{fig:Blos_S_map} shows a map of $\mathcal{S}$ values for the 214~\micron\ data with available measurements of $B_{\rm LOS}$ for OMC-1 superposed. $B_{\rm LOS}$ values were obtained from the fitting of Zeeman splitting lines of 21-cm HI \citep[squares, ][]{Troland2016} and CN measurements \citep[circles, ][]{Crutcher1996}. Visually, low values of $B_{\rm LOS}$ seem spatially aligned with low values of dispersion, while one measure of strong $B_{\rm LOS}$ is located near the BN/KL object, where the dispersion is large. 

Here, this general principle is applied to the OMC-1 data as follows.  Caveates and justification of assumptions are considered at the end of this section.
In order to determine the LOS component in this work, first the $\varphi-\mathcal{S}$ relation is expressed as \citep{Hensley2019}

\begin{equation}
\sin^{2}(\varphi) = a\mathcal{S}^{n},
\label{eq:S_angle}
\end{equation}

\noindent
with $\varphi$ being the angle between the magnetic field direction and the LOS. The coefficient $a$ and exponent $n$ are experimentally-determined parameters. Following this convention, the magnitude of the components $B_{\rm POS}$ and $B_{\rm LOS}$ are therefore related as

\begin{equation}
\tan(\varphi) = \frac{B_{\rm POS}}{B_{\rm LOS}}.
\label{eq:blos_comp}
\end{equation}
Combining Eqs. \ref{eq:S_angle} and \ref{eq:blos_comp}, $B_{\rm LOS}$ can be calculated as
\begin{equation}
B_{\rm LOS} = B_{\rm POS}\sqrt{ \frac{1-a\mathcal{S}^{n}}{a\mathcal{S}^{n}} }.
\label{eq:blos}
\end{equation}

The use of Eq. \ref{eq:blos} to estimate the LOS component is limited to values of $a\mathcal{S}^{n} < 1$. Depending on the specific values of $a$ and $n$, a cutoff value in angle dispersion $\mathcal{S}_{\rm c}\equiv a^{-1/n}$ is imposed ($n<0$). Once the technique is calibrated and the values of $n$ and $a$ are found, regions where $\mathcal{S}<\mathcal{S}_\mathrm{c}$ are assumed to have negligible contribution to the total magnetic field from the LOS component.

\begin{figure}[!h]
    \centering
    \includegraphics[width=3.3in]{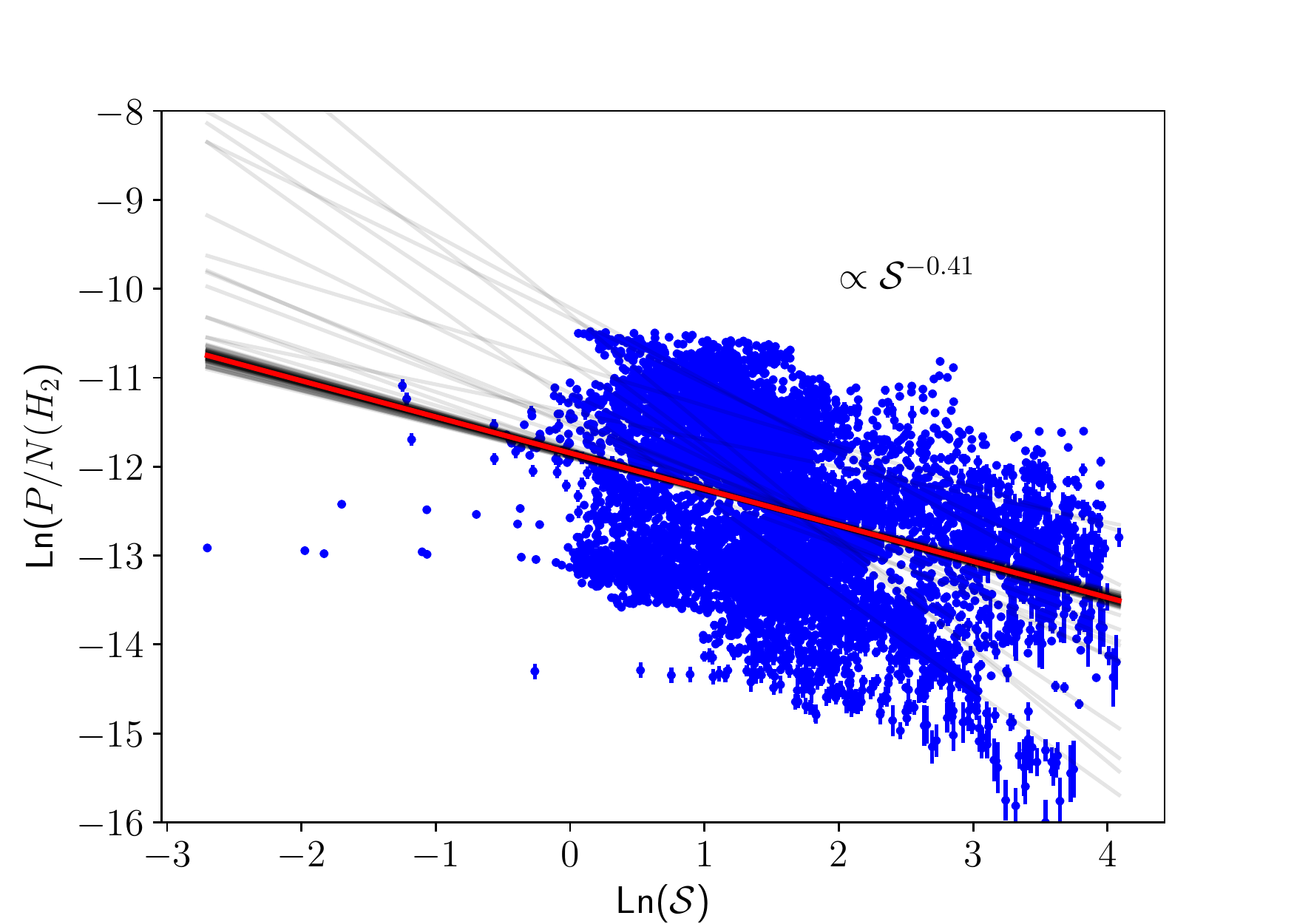}
    \includegraphics[width=3.2in]{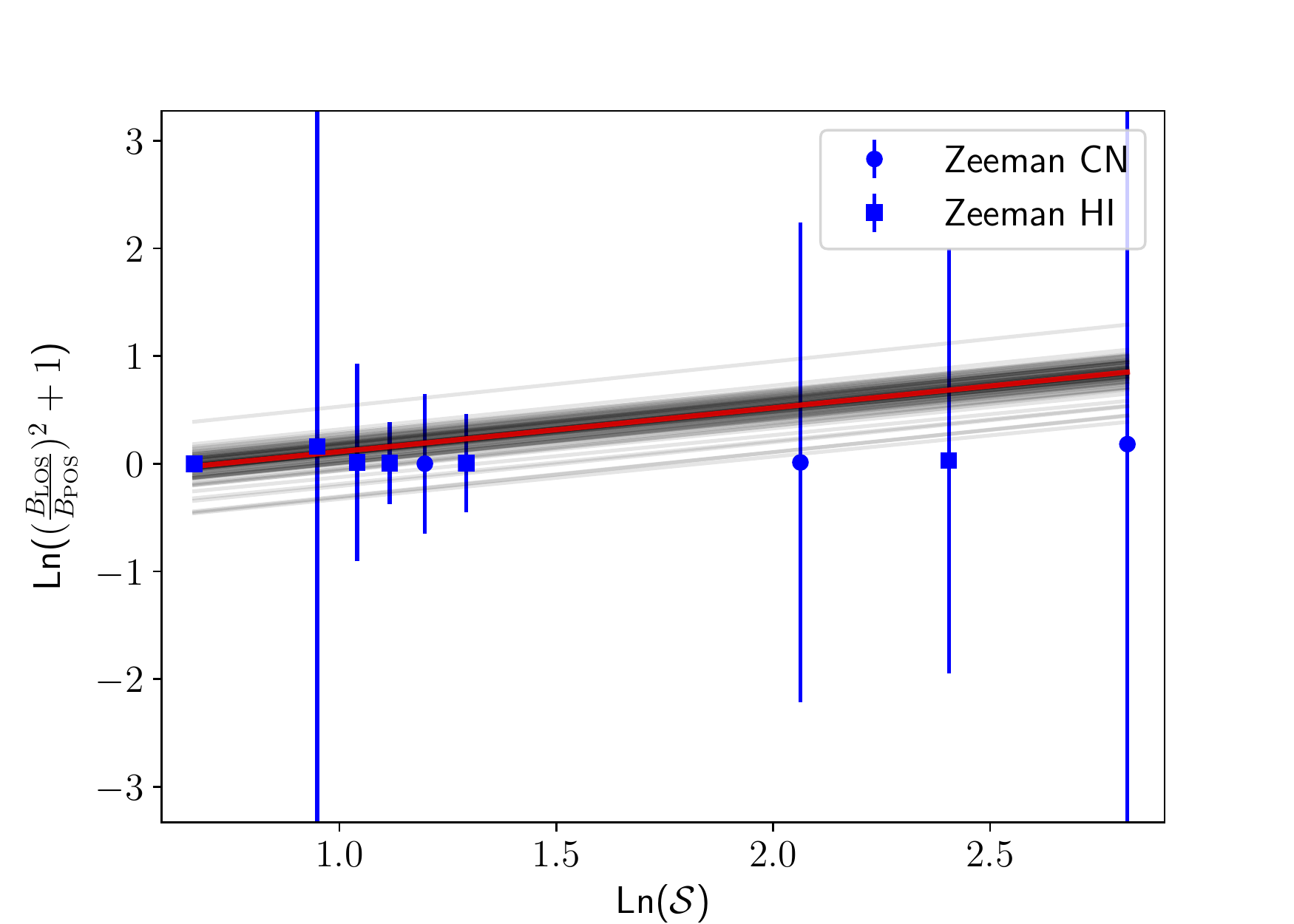}
    \caption{{\it Left:} Scatter plot of polarized flux $P$ divided by the column density ($H_{2}$) as a function of the dispersion $\mathcal{S}$ for HAWC+ 214 $\mu$m data. The red line corresponds to the best-fit linear model while grey lines correspond to different realizations of the MCMC solver and provide the uncertainty for the best-fit model. {\it Right:} Line-of-sight field values from Zeeman CN and HI measurements are used to plot $\ln{\left[\left({B_{\rm LOS}}/{B_{\rm POS}}\right)^2+1\right]}$ as a function of $\mathcal{S}$ and modeled according to Eq. {\ref{eq:linfit_a}}. In this plot only the intercept is fitted using the Zeeman measurements; the slope for the model from the fit on the left ($P/N(H_{2})$ v. $\mathcal{S}$) is used for the slope for the model on the right hand figure.  As in Figure~\ref{fig:Blos_S_map}, CN Zeeman measurements are plotted with circles while the squares correspond to HI Zeeman measurements.}
    \label{fig:Blos_vs_S}
\end{figure}

The power-law relation in Eq. \ref{eq:S_angle} is also consistent with the empirical anti-correlation between $\mathcal{S}$ and the polarization fraction $p$ \citep{Fissel2016,Planck2018}. Thus, following \citet{Hensley2019}, the value of $n$ can be determined using the ratio of the polarized intensity to the column density, $P/N(H_2)$, as a function of the dispersion $\mathcal{S}$ (see their Eq. 21). Figure \ref{fig:Blos_vs_S}~({\it Left}) displays the data for the 214~\micron\ observations. The exponent $n$ is identified as the best-fit value of the slope of a linear fit to these data. It can be seen that the best linear fit (red line) has a negative slope with a relative small uncertainty, given by the $1\sigma$ value of the posterior MCMC distribution. 

On the other hand, Eq.~\ref{eq:blos} can be linearized to solve for the calibration constant $a$, 
\begin{equation}
    \ln{\left[\left(\frac{B_{\rm LOS}}{B_{\rm POS}}\right)^2+1\right]}=-\ln{a}-n\ln{\mathcal{S}}.
\label{eq:linfit_a}
\end{equation}

In this form, the intercept ($-\ln{a}$) sets the calibration for the power law relationship. To estimate this, a set of Zeeman measurements is utilized as a transfer standard. To determine ($-\ln{a}$), the value of $n$ is fixed to that found above and thus Eq.~\ref{eq:blos} is fit for the intercept only using the Zeeman values for $B_{\rm LOS}$ and the corresponding $B_{\rm POS}$ and $\ln{\mathcal{S}}$ from the analysis above for each points that a Zeeman measurement exists (See Fig.~\ref{fig:Blos_vs_S},{\it Left}, for a graphical depiction of these locations.) can be utilized.  Figure \ref{fig:Blos_vs_S}~({\it Right}) shows the fit for the intercept. The uncertainties in the ordinate variable are obtained from those of $B_{\rm LOS}$ and $B_{\rm POS}$, properly propagated (the uncertainty in $B_{\rm POS}$ is assumed to be 50\% of its value.) These uncertainties are then inflated to force the $\chi^2$ value of the fit to unity. Values of $n$, $a$, and $\mathcal{S}_{\rm c}$ for all four HAWC+ wavelengths are summarized in Table \ref{tbl:lm_params}. 

Although the HI Zeeman measurements are not necessarily co-located with the dust grains along the line-of-sight, they are still a potentially good indicator of the LOS field strength if the field does not significantly vary along the LOS.  Including these data points along with the CN measurements does not have significant effect on the values of $a$ resulting from the fits. However, the inclusion of HI data points improves the uncertainty of $a$ by making the posterior distributions narrower.

Values of $n$ are negative and range from $\sim$ -0.7 to -0.3, in agreement with \citet{Hensley2019}. However, no clear trend is observed with the FIR wavelength. Values of the coefficient $a$, on the other hand, seem to roughly increase with wavelength from 1.34 at 214 \micron\ to 2.77 at 53 \micron.

\begin{table}[!h]
    \centering
    \begin{tabular}{cccc}
        Wavelength [$\mu$m] & $n$ & $a$ & $\mathcal{S}_{\rm c}$ [$^{\rm o}$]\\
        \hline\hline
        53 & $-0.68^{+0.01}_{-0.01}$ & $2.77^{+0.68}_{-0.61}$ & 4.47 \\
        89 & $-0.34^{+0.01}_{-0.05}$ & $1.42^{+0.14}_{-0.12}$ & 2.80 \\
        154 & $-0.52^{+0.01}_{-0.01}$ & $1.39^{+0.39}_{-0.32}$ & 1.88 \\
        214 & $-0.41^{+0.01}_{-0.02}$ & $1.34^{+0.14}_{-0.13}$ & 2.04\\\hline
    \end{tabular}
    \caption{Parameters of linear fit performed to find values of coefficient $a$ according to Eq. \ref{eq:linfit_a} for all four HAWC+ bands. Values of $n$ and $a$ correspond to the exponent and coefficient that characterize the power-law of Eq. \ref{eq:S_angle} and that are necessary for calculating values of $B_{\rm LOS}$. The parameter $\mathcal{S}_{\rm c}$ is the minimum dispersion value for which the magnetic field has negligible component in the LOS direction.}
    \label{tbl:lm_params}
\end{table}

\begin{figure}
    \centering
    \includegraphics[width=3.5in]{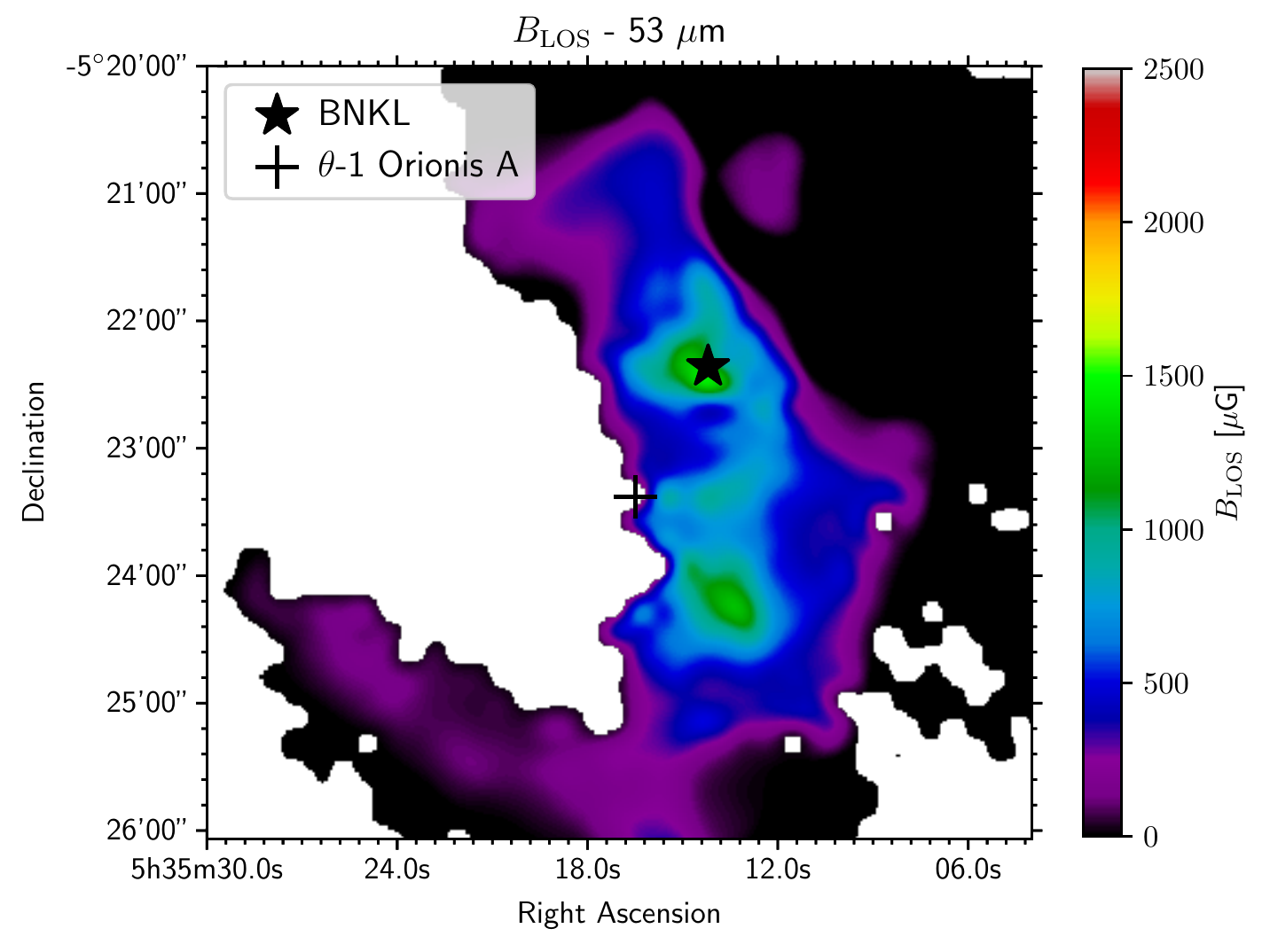}
    \includegraphics[width=3.5in]{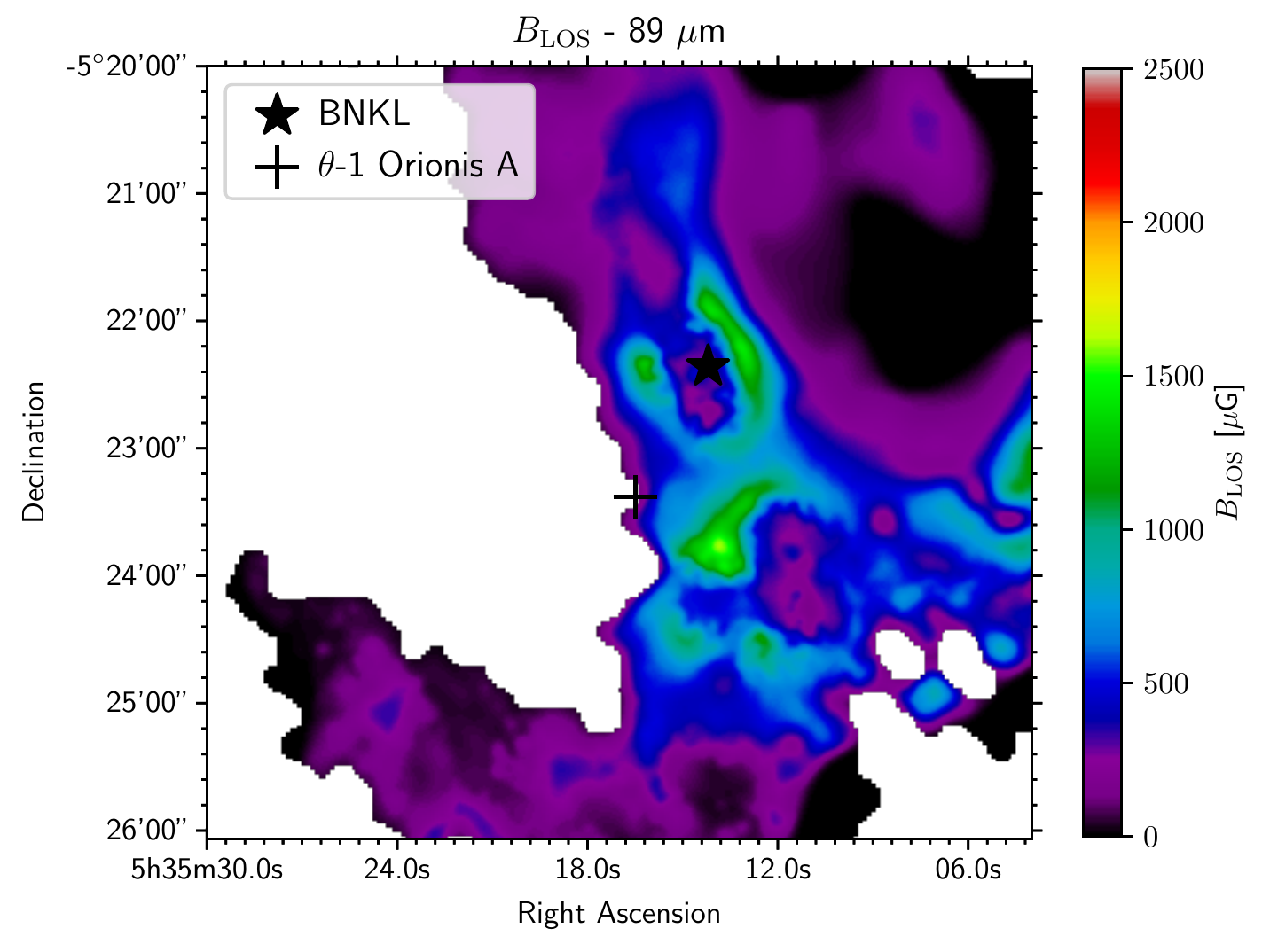}
    \includegraphics[width=3.5in]{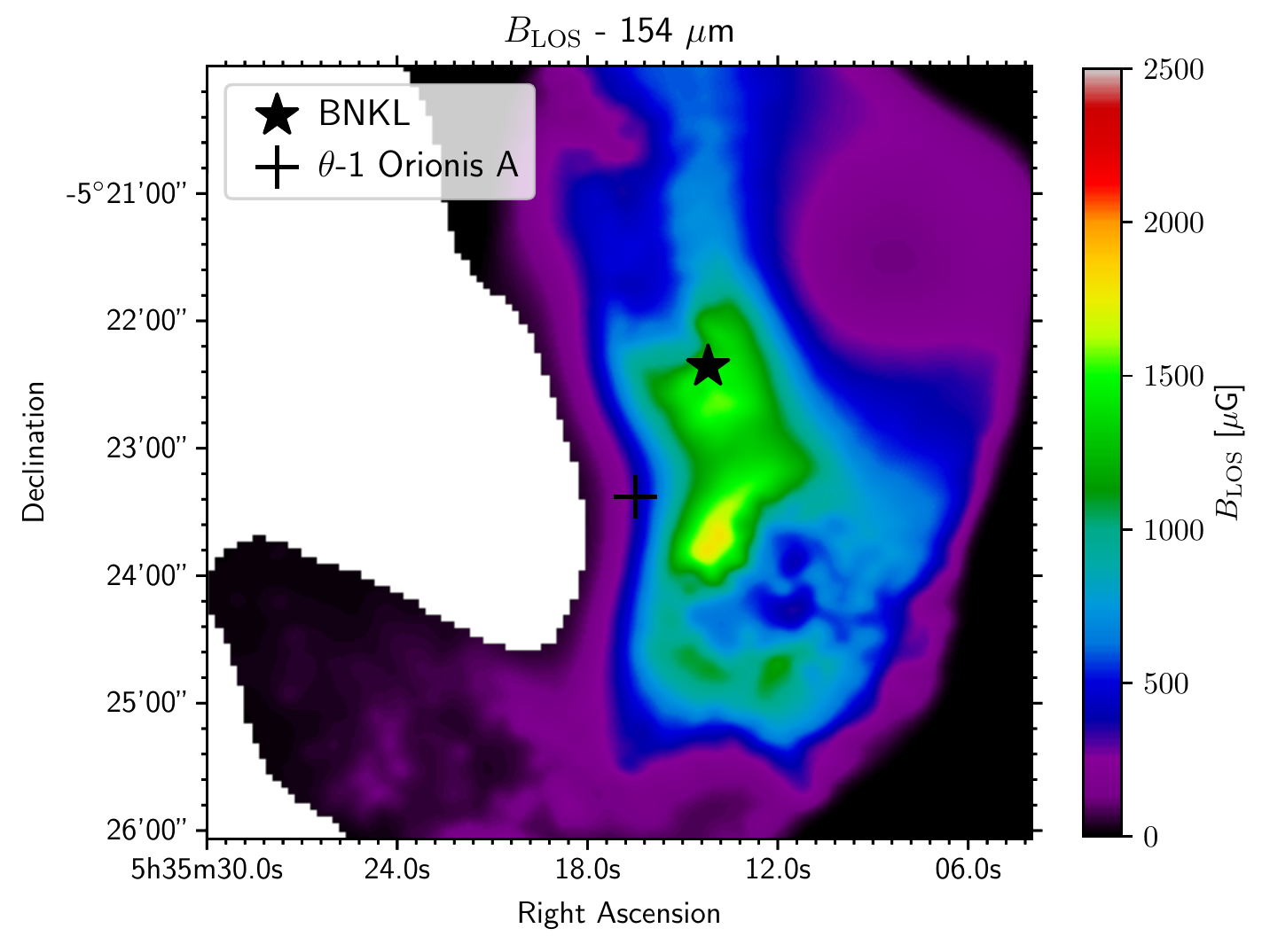}
    \includegraphics[width=3.5in]{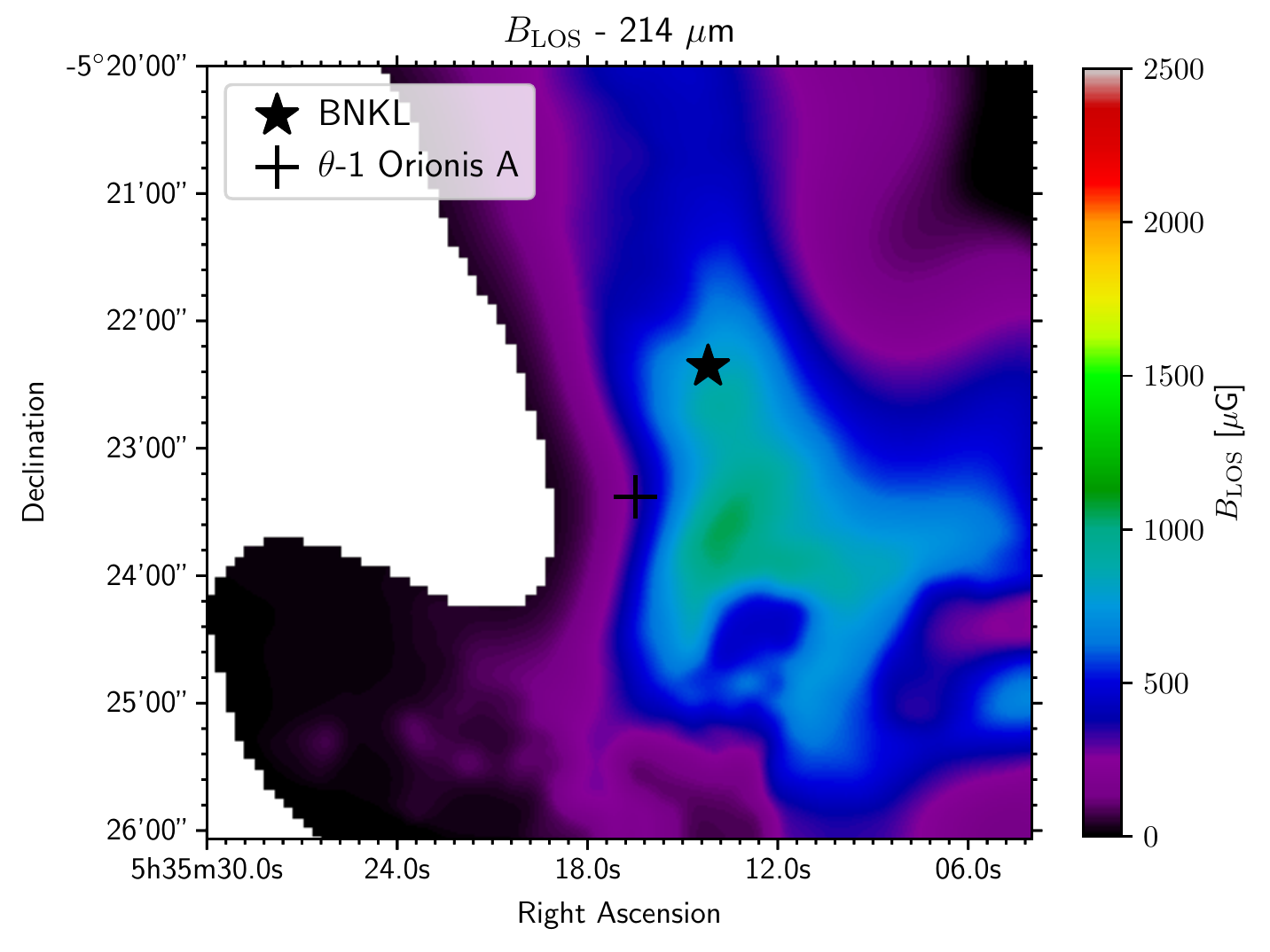}
    \caption{Maps of the line-of-sight (LOS) field strength for 53 $\micron$ ({\it Top left}), 89 $\micron$ ({\it Top right}), 154 $\micron$ ({\it Bottom left}), and 214 $\micron$ ({\it Bottom right}). For reference, the locations of the BN/KL object (star) and the Trapezium cluster (cross) are included as well. Angular resolution in each map matches those of the $B_{\rm POS}$ maps. LOS strengths can only be calculated for values of $\mathcal{S} > \mathcal{S}_{\rm c} \simeq$ 4$^{\rm o}$, 3$^{\rm o}$, 2$^{\rm o}$, and 2$^{\rm o}$ for the 53, 89, 154, and 214 $\micron$ data, respectively.}
    \label{fig:Blos_maps}
\end{figure}

The maps of $\mathcal{S}$ shown in Figure \ref{fig:dispersion} are used to obtain estimates of $B_{\rm LOS}$ across the field of view of OMC-1. The resulting maps are displayed in Figure \ref{fig:Blos_maps}, utilizing Eq. \ref{eq:blos} along with the values in Table \ref{tbl:lm_params}. As previously mentioned, values of $B_{\rm LOS}$ can only be estimated for values of dispersion $\mathcal{S} > \mathcal{S}_{\rm c}$, therefore for locations with values of $\mathcal{S} < \mathcal{S}_{\rm c}$ the $B_{\rm LOS}$ value is set to zero, and the total fiedl is taken to be $B_{\rm POS}$. Strong LOS magnetic fields ($\gtrsim$1000 $\mu G)$ appear near the BN/KL object, where either the dispersion $\mathcal{S}$ and/or mass density, $\rho$, are observed to be large. On the other hand, in the OMC-1 bar, although the dispersion $\mathcal{S}$ is observed to be large, the density is low resulting in low values of LOS magnetic field.

This approach to calculating $B_{\rm LOS}$, although promising, is in an early stage of development. The use of $\mathcal{S}$ as a proxy for the inclination angle of the field depends on other factors that contribute to $\mathcal{S}$ being subdominant. As mentioned in Section~\ref{sec:LocDisp}, other physical quantities that can affect $\mathcal{S}$ include variations in grain alignment efficiency and variation of magnetic field structure within the volume of the beam through the cloud.  \citet{Chuss2019} found that the inverse relation between $p$ and $I$ did not require loss of grain alignment efficiency in dense regions of OMC-1. In addition, from our DCF fit parameters, it is found that the turbulence scale is resolved over most of the cloud. Therefore, it may be reasonable to assume that at least in OMC-1, the dispersion is dominated by the geometry of the magnetic field.  However, future studies of additional clouds along with numerical models will be required to understand this in sufficient detail to quantify the uncertainties of this technique.  Additionally, more precise Zeeman measurements would strengthen the calibration of the technique.

\subsection{Total Magnetic Field Strength and Mass-Magnetic Flux Criticality}\label{sec:m_phi}

The balance between gravitational collapse and magnetic pressure support of ionized material in molecular clouds, can be quantified by the ratio $M/\Phi$, where $M$ is the total mass and $\Phi$ is the magnetic flux. Following \citet{Crutcher2004}, the value of $M/\Phi$ divided by the critical ratio $(M/\Phi)_{\rm c}$ can be calculated as

\begin{equation}
    \lambda = \frac{M/\Phi}{(M/\Phi)_{\rm c}} = 7.6\times 10^{-21} \left(\frac{N(H_{2})}{B_{\rm Total}}\right)
    \label{eq:m_phi}
\end{equation}

\noindent
where $N(H_{2})$ is the column density in cm$^{-2}$ and the total magnetic field strength in $\mu$G, $B_{\rm Total}$,  can be written as

\begin{equation}
    B_{\rm Total} = (B_{\rm POS}^{2} + B_{\rm LOS}^{2})^{1/2} = \frac{B_{\rm POS}}{\left({a\mathcal{S}^n}\right)^{1/2}}.
\label{eq:btot}
\end{equation}
Based on this ratio, a volume of mass inside the magnetized molecular cloud can be established to be subcritical ($\lambda <$ 1) or supercritical ($\lambda >$ 1).

\begin{figure}
    \centering
    \includegraphics[width=3.2in]{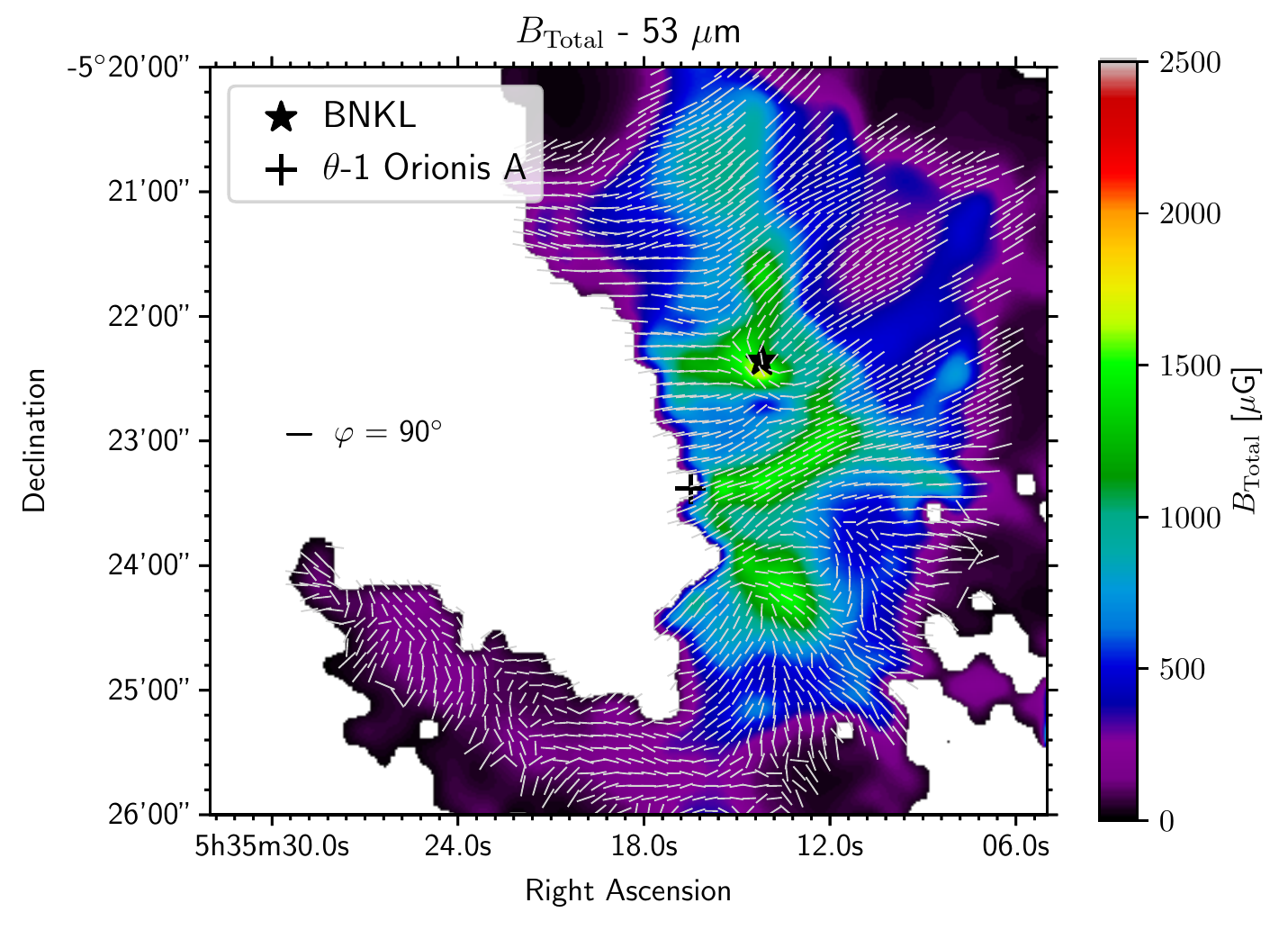}
    \includegraphics[width=3.2in]{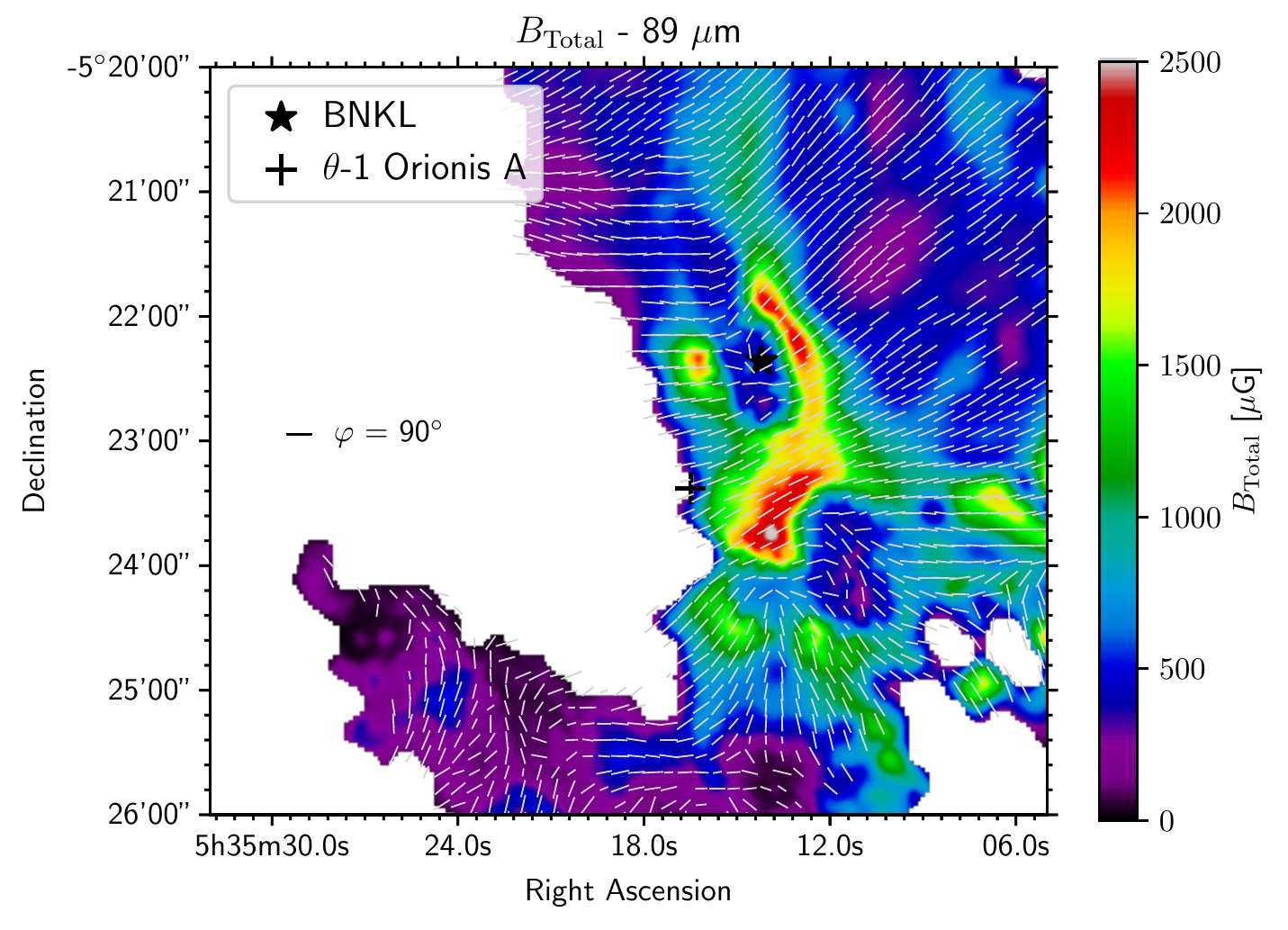}
    \includegraphics[width=3.2in]{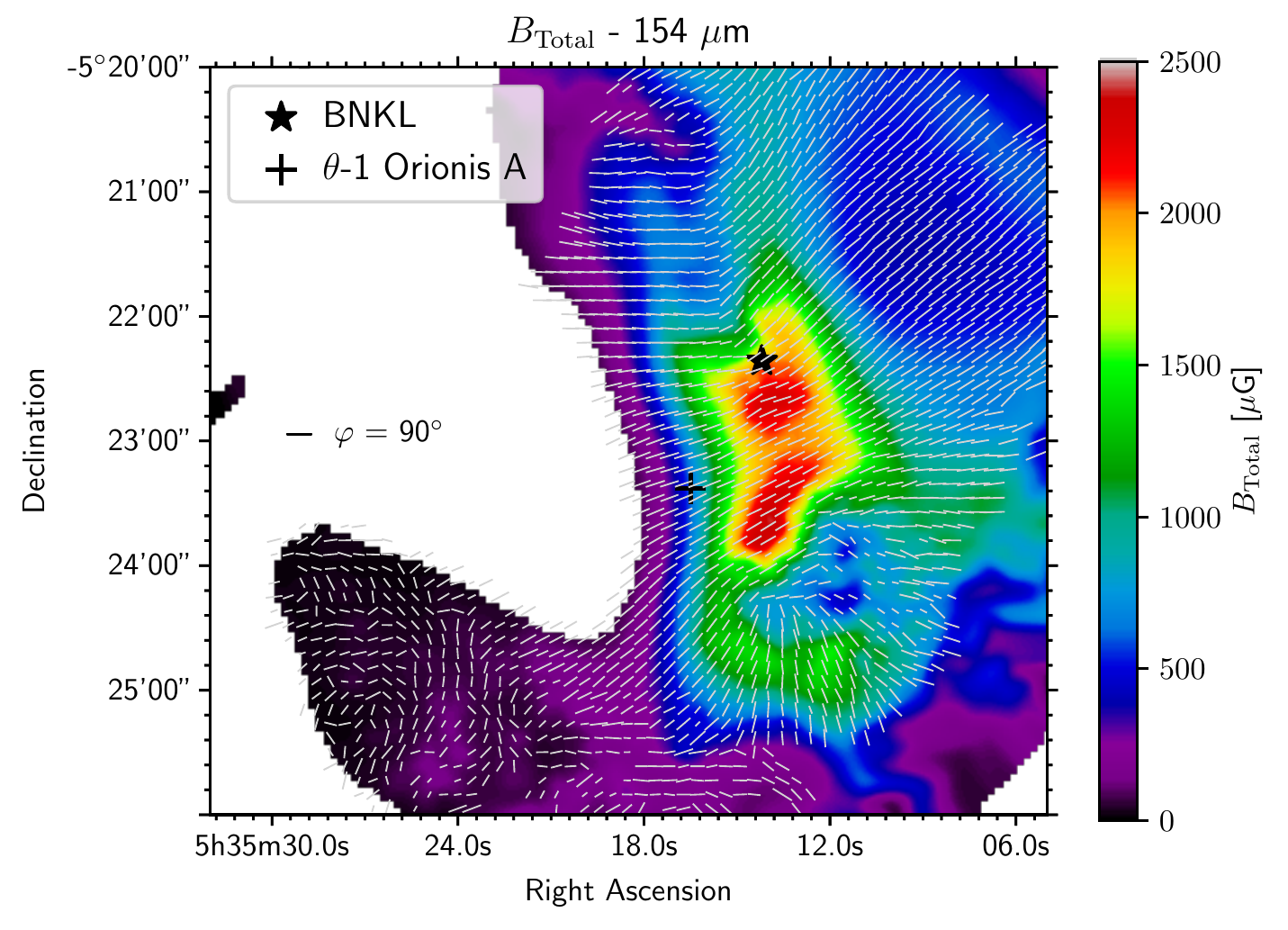}
    \includegraphics[width=3.2in]{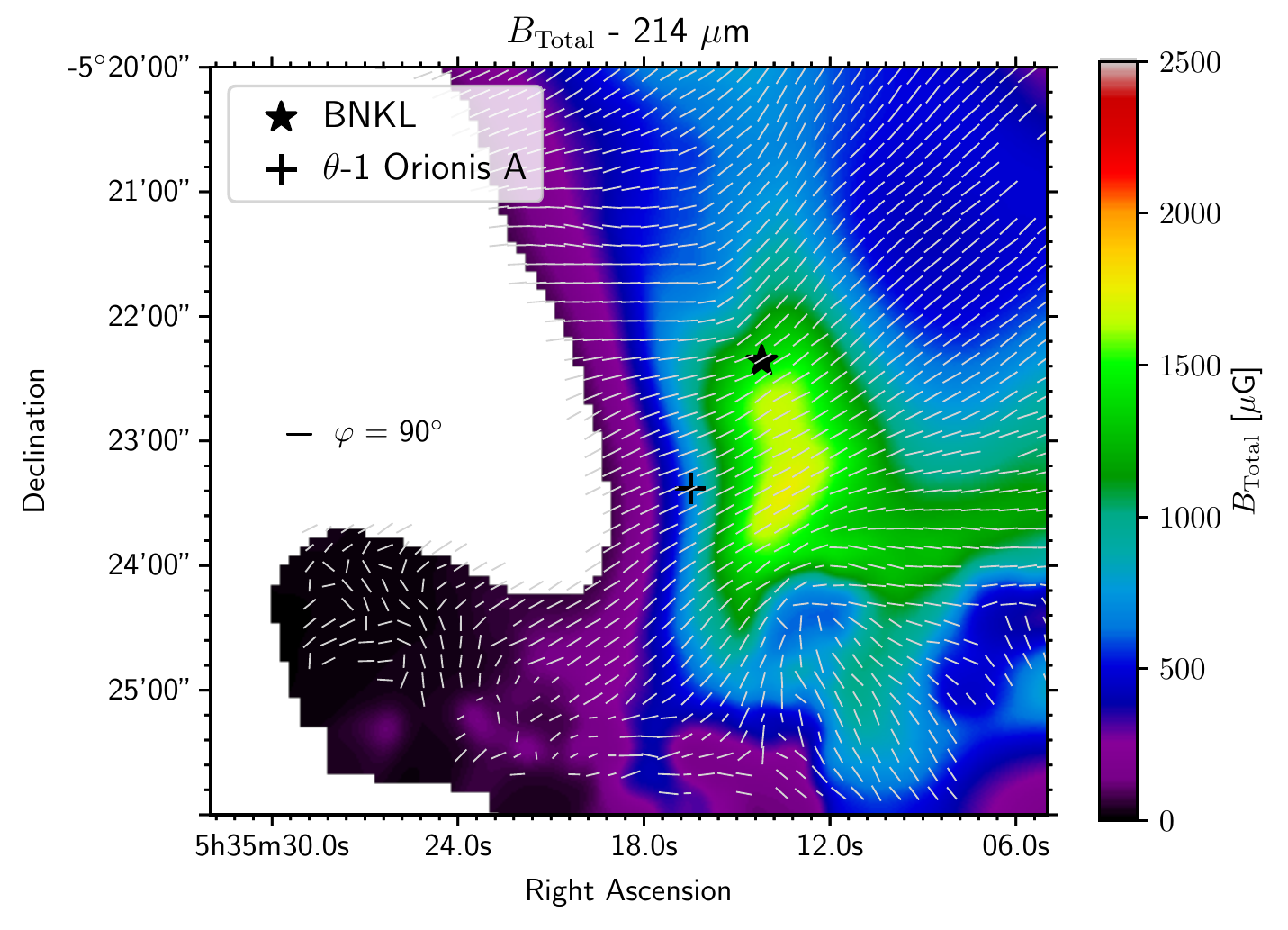}
    \caption{Total magnetic field strength in across the OMC-1 region for all four HAWC+ wavelengths (53~\micron, {\it Top left}; 89~\micron, {\it Top right}; 154~\micron, {\it Bottom left}; 214~\micron, {\it Bottom right}.) Rotated polarization vectors -- which display the POS field direction -- are multiplied by the factor $\sin(\varphi)$ to show the locations in which the inclination angle $\varphi$ can be estimated. Vectors with longer lengths are closer to the POS direction. The length of a vector lying entirely in the POS is shown for reference. The angular resolution of these maps matches those of the $B_{\rm POS}$ maps in Figure \ref{fig:B_maps}.}
    \label{fig:btot_maps}
\end{figure}

To evaluate $\lambda$, first $B_{\rm Total}$ is estimated using the results presented in the previous section and the $N(H_{2})$ map that is presented in section \ref{sec:H2_vel}. These maps are displayed in Figure \ref{fig:btot_maps}. Rotated polarization vectors are also displayed to visualize the POS magnetic field direction. In this case the polarization vectors are multiplied by a factor of $\sin(\varphi)$ in order to provide a rough indication of 3-dimensional field geometry. Vectors with shorter length in Figure \ref{fig:btot_maps} correspond to positions where $B_{\rm LOS}$ dominates. Polarization vectors with longer lengths indicate locations for which $B_{\rm Total}$ is predominantly in the POS direction.

\begin{figure}
    \centering
    \includegraphics[width=3.3in]{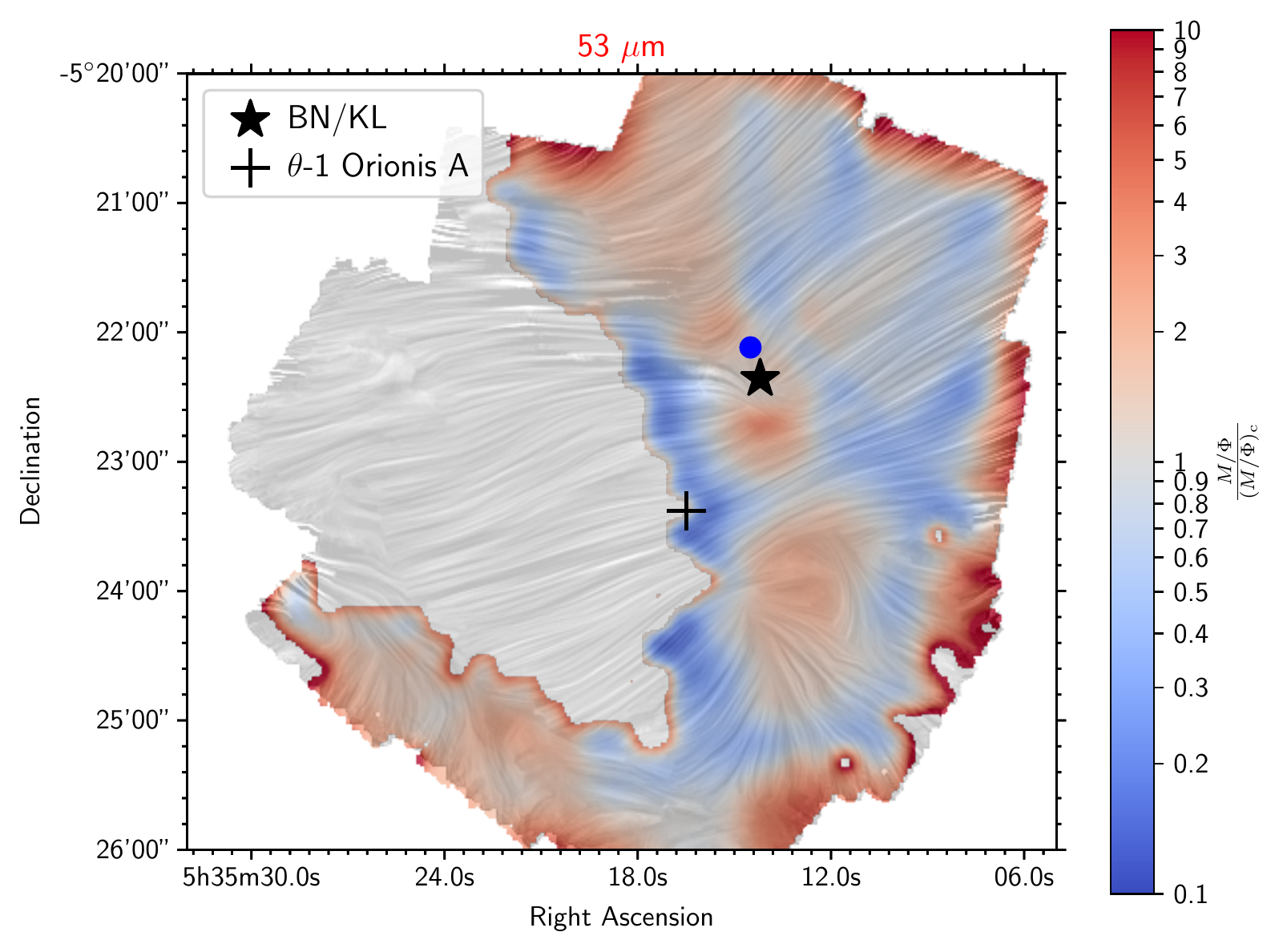}
    \includegraphics[width=3.3in]{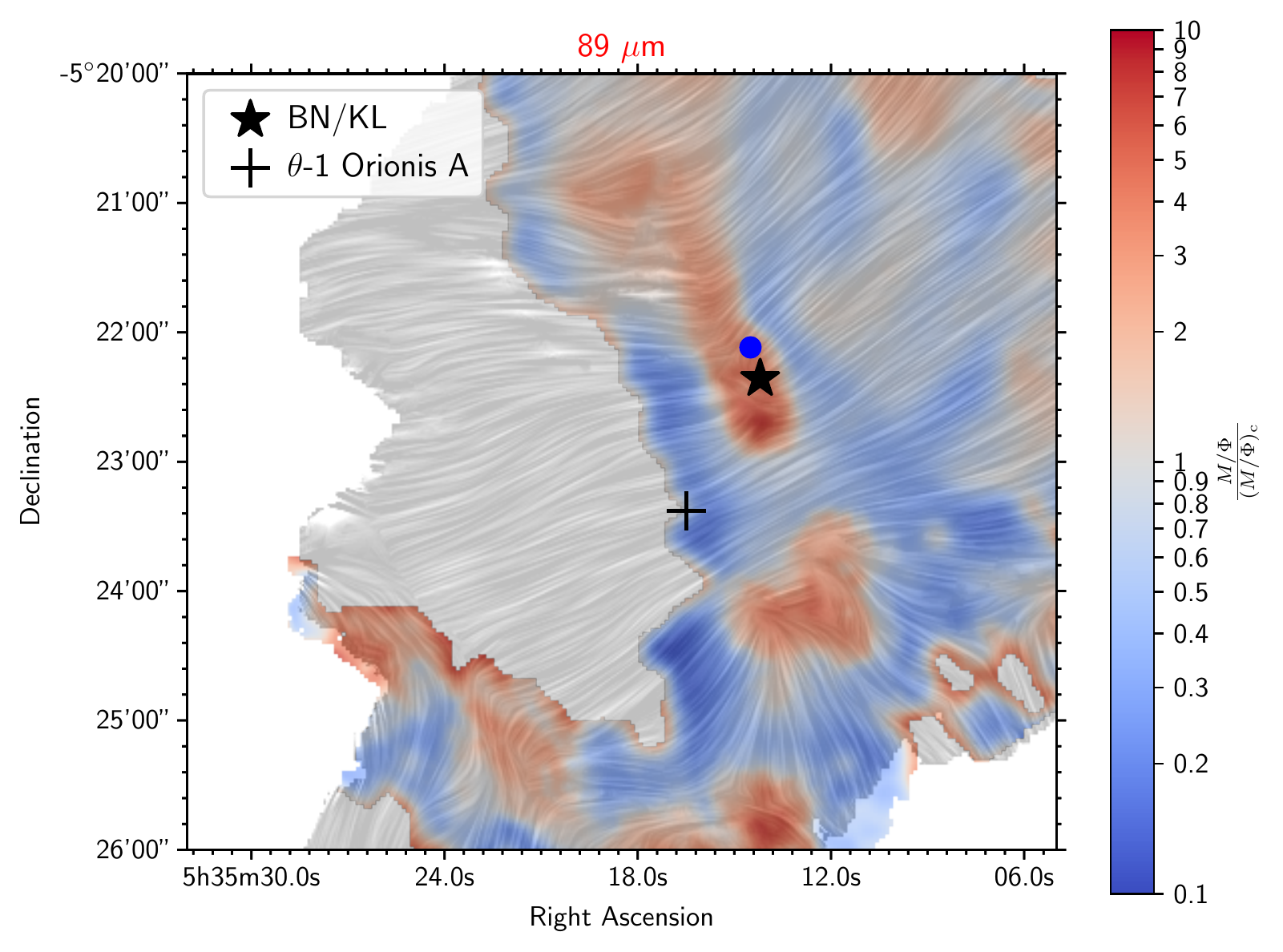}
    \includegraphics[width=3.3in]{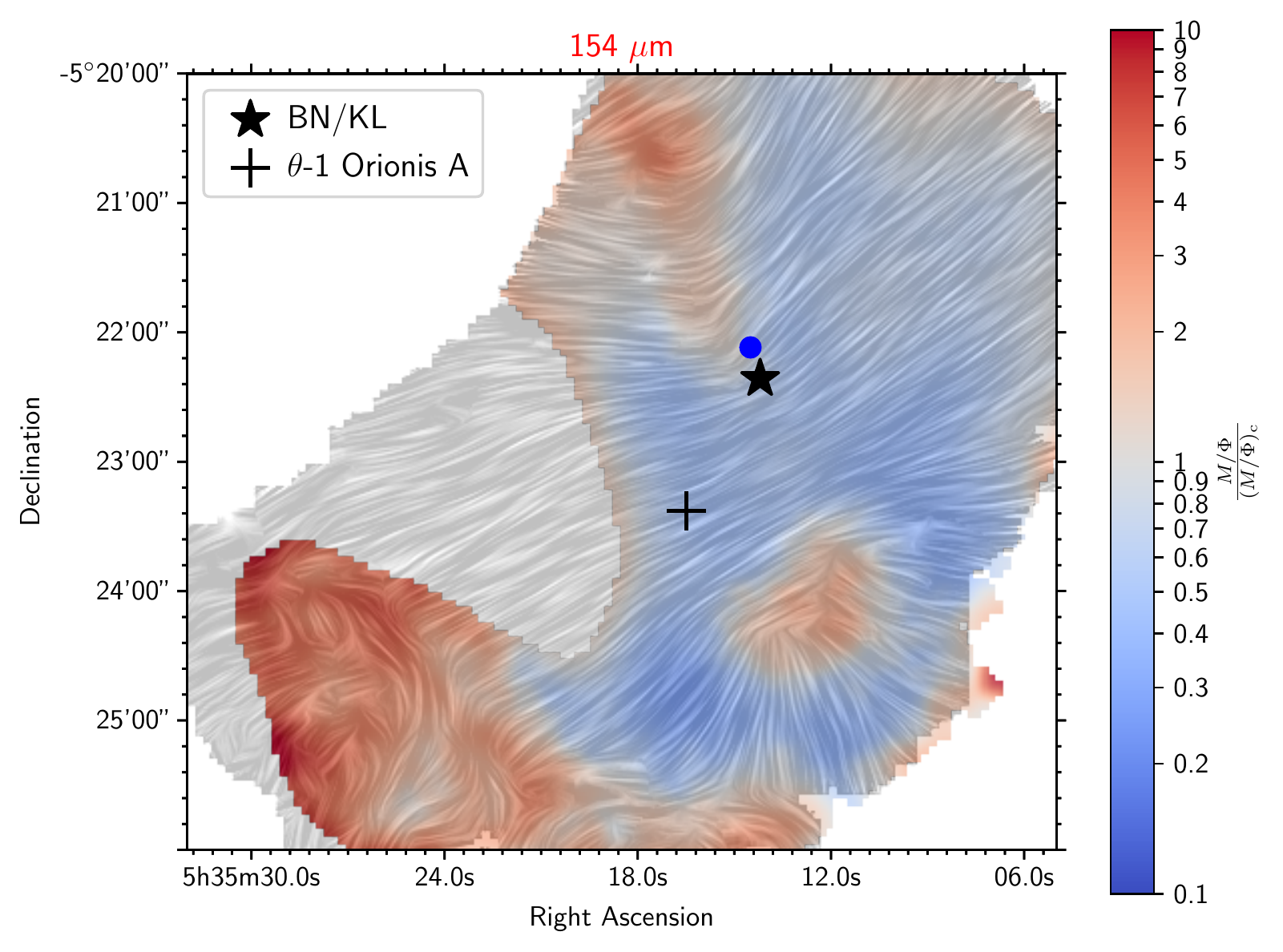}
    \includegraphics[width=3.3in]{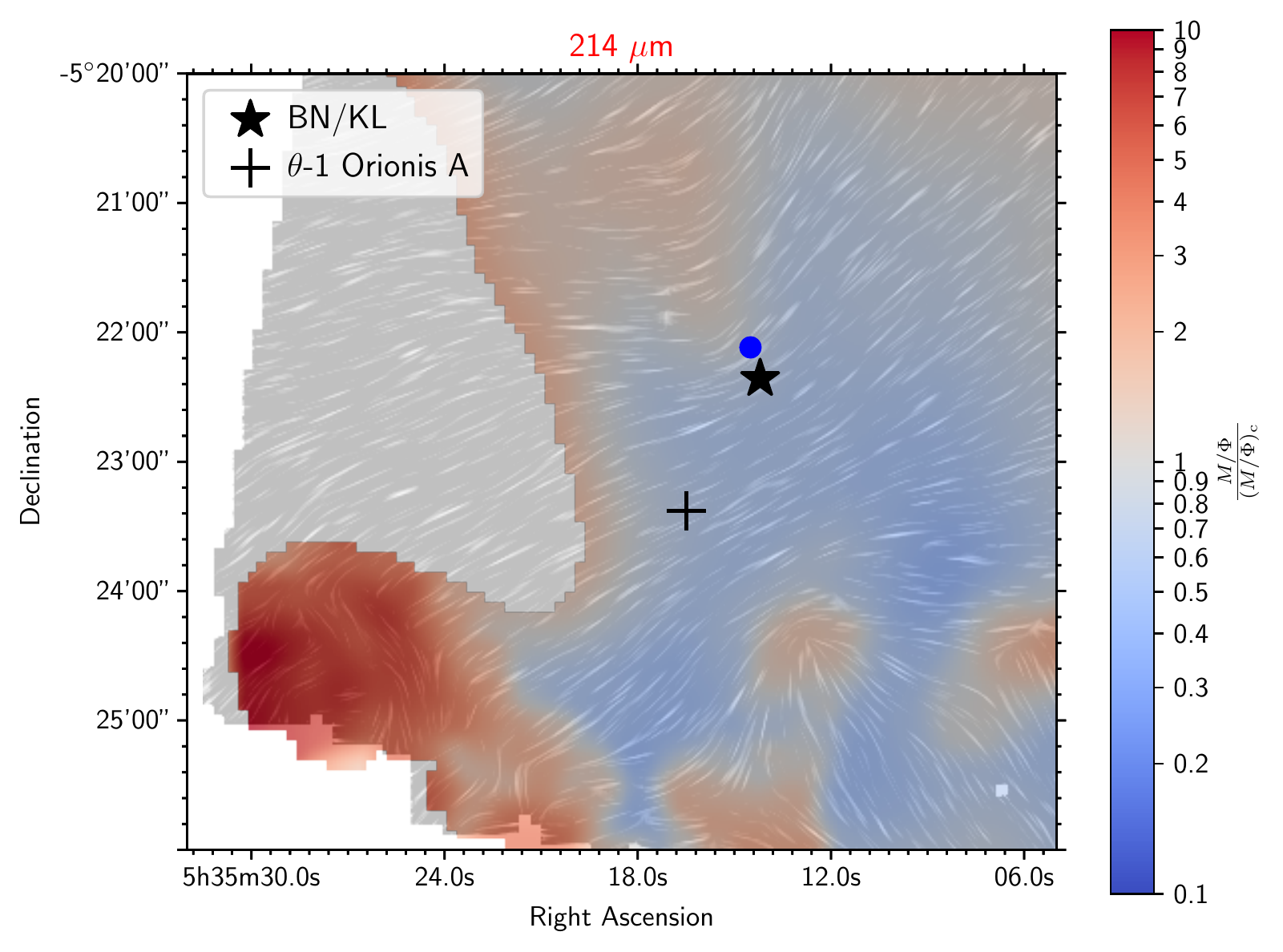}
    \caption{Mass-to-magnetic-flux ratio, $M/\Phi$, normalized by the critical mass-to-magnetic-flux ratio, $(M/\Phi)_{\rm c}$,  for the OMC-1 region, according to Eq. \ref{eq:m_phi}. $M$ is obtained from column density $N(H_{2})$ map and the total magnetic field  $B_{\rm{Total}}$ is calculated using $B_{\rm{POS}}$ from Figure \ref{fig:B_maps} and $B_{\rm{LOS}}$ derived from Figure \ref{fig:Blos_maps}. Gray color corresponds to the unity value, signaling the separation between the two different regimes: sub (blue) and supercritical (red). Regions with values $< 1$ are subcritical and gravitational collapse will likely not occur. For reference, the locations of the BN/KL object (star) and the Trapezium cluster (cross) are included as well. Blue circle just above the BN/KL location correspond to the location of $B_{\rm LOS} = 360\,\mu$G (see previous section).}
    \label{fig:m_phi_maps}
\end{figure}

The resulting maps of $\lambda$ for all wavelengths are presented in Figure \ref{fig:m_phi_maps}. Values range from $\sim$0.1 to $\sim$10 and further spatial structure appears at smaller wavelengths given the increased angular resolution. In the maps of Figure \ref{fig:m_phi_maps}, the $\lambda$ values are shown with a diverging color bar for easier interpretation. Gray indicates those locations where $\lambda$ is around unity, shades of red correspond to $\lambda > 1$ (super-critical) and shades of blue correspond to $\lambda < 1$ (sub-critical). Super-critical regions appear spatially aligned with the highest-density ($N(H_{2}) \gtrsim 10^{23}$ cm$^{-2}$) filament in OMC-1 ({\it e.g.}, Figure \ref{fig:den_vel_maps}, {\it Left}) specially in the 53 and 89~\micron\ maps. In each map, sub-critical regions (blue) are observed to be co-spatial with lower-density ($N(H_{2}) \lesssim 10^{23}$ cm$^{-2}$) regions of relatively strong magnetic field strengths ($\sim$ 500 $\mu$G), such as the filamentary structures located northwest of the BN/KL object. For these filamentary structures, the total magnetic field seems to comes entirely from the POS component, $B_{\rm LOS} \ll B_{\rm POS}$. These findings suggest that the results from Planck \citep{PlanckXXXV2016}, which finds sub-critical clouds with intermediate range of column density to be mostly perpendicular with the magnetic field orientation, are valid at sub-parsec spatial scales. This is consistent with the findings of \citet{Pillai2020} in Serpens South.

\subsection{Wavelength Dependence of Maps}

In Figures~\ref{fig:B_maps}, \ref{fig:btot_maps}, and \ref{fig:m_phi_maps}, differences are evident between maps at different wavelengths.  There are several potential explanations for such differences.
The resolution of the 53~\micron\ and 89~\micron\ maps are approximately that of the velocity dispersion map (32\arcsec\ and 33\arcsec\, respectively). The 154 and 214~\micron\ maps have resolutions set by the dispersion function kernels (58\arcsec\ and 77\arcsec\, respectively). The coarser resolutions at the longer wavelengths likely account for some of the discrepancy between images. More specificically, because the same kernel size was used for each map, the angular size of the kernel scales with wavelength. For the construction of the 53 \micron\ $B_{\rm POS}$ map, the $\ratio$ parameter map (which resolution is 20.7$^{\prime\prime}$) is smoothed to the resolution of the velocity dispersion map.  For other wavelengths, the velocity map is smoothed to the resolution of the $\ratio$ parameter map. This difference in procedure can potentially account for some of the differences between the 53 and 89~\micron\  $B_{\rm POS}$ maps.

In addition, different wavelengths are likely to selectively probe different regions of the cloud along the line-of-sight.  This is especially likely to be true in the dense region around BN/KL, where the polarization angle also differs between wavelengths.  For example, near BN/KL, the long wavelength polarimetry is likely tracing the cooler outer part of the cloud, while the short wavelength data preferentially sample the warm dust that corresponds to the BN/KL and the associated explosion \citep{Chuss2019}.  Infrared polarimetry support this picture \citep{Dyck1973}.  The magnetic field direction implied by polarimetry by absorption is consistent with the HAWC+ 154 and 214~\micron\ data, indicating that the majority of the absorption is being done by the cool, dense part of the cloud.  

Another caveat here is that the estimated density map that is used for the DCF technique assumes that a single temperature thermal model for the dust is sufficient. Although this is likely to be true over parts of the cloud, it is difficult to apply this to the BN/KL region in particular. As argued above from the polarimetry perspective, this region is likely to contain regions of various temperatures and densities along the line of sight.  Future work should consider the three-dimensional distribution of dust and magnetic field geometry in working towards magnetic field measurements of higher fidelity. Specifically, estimates of the density and velocity dispersion maps that are functions of wavelength are desired. 

Finally, it should be noted that despite these caveats, the $M/\Phi$ maps look similar in each of the four wavelengths.  This relative consistency may be partially due to the fact that this ratio is slightly less sensitive to mass density than the magnetic field maps themselves. 

\section{Summary} \label{sec:summary}

The relationship between magnetic field strength and dispersion of polarization angles was investigated in the Orion Molecular Cloud 1 (OMC-1) using far-infrared (FIR) dust polarization observations from SOFIA/HAWC+. Maps of plane-of-sky (POS) and the line-of-sight (LOS) magnetic field strength were constructed by quantifying the spatial variation of dispersion of polarization vectors.

\begin{enumerate}

\item Maps of the POS field strength in OMC-1 have been produced by means of the Davis-Chandrasekhar-Fermi (DCF) method in combination with the analysis of a two-point structure function of the angle difference ($\dispfunct$). $B_{\rm POS}$ was estimated for each pixel by calculating $\dispfunct$ over a circular kernel centered at such pixel. Maps of POS field show strengths ranging $\sim$ 0 - 2000 $\mu$G with strong fields observed in areas where mass density and velocity dispersion are large ({\it i.e.}, the BN/KL object) and the dispersion in polarization angles (as measured by the ratio of turbulent-to-ordered magnetic energies) is small.

\item For $B_{\rm LOS}$, the local dispersion parameter $\mathcal{S}$ -- a root-mean-square (RMS) value of angle dispersion over a circular region of $\sim$32$^{\prime\prime}$--  in combination with Zeeman measurements of $B_{\rm LOS}$ provided an exploratory path to infer the strength of the total field from that of the POS component by estimating the field inclination at each point in the $B_{\rm POS}$ map. This allows a coarse estimate of the LOS magnetic field strength across the map to complement the map of the POS component. This approach depends on many important considerations and must be tested in larger FIR polarimetric data sets, observations and simulated data.

\item The estimation of both POS and LOS field strengths, allows one to produce a map of $M/\Phi$, the mass-to-magnetic flux critical ratio by improving the estimates of the total magnetic field strength. The inferred maps of $M/\Phi$ show consistency with early results that establish higher-density filamentary structures in clouds perpendicular with their ambient magnetic field \citep{PlanckXXXV2016}.

\end{enumerate}

Keeping in mind the limitations of the derived maps, the results of this work show the potential for testing the specific scenarios of clumps and star formation. For example \citet{Crutcher2009} proposed that the quantity $\mathcal{R} $ -- the ratio of $M/\Phi$ between the core and the envelope -- can be used to distinguish between ambipolar diffusion and turbulence-regulated star formation.


\acknowledgments

Based on observations made with the NASA/DLR Stratospheric Observatory for Infrared Astronomy (SOFIA). SOFIA is jointly operated by the Universities Space Research Association, Inc. (USRA), under NASA contract NAS2-97001, and the Deutsches SOFIA Institut (DSI) under DLR contract 50 OK 0901 to the University of Stuttgart. Financial support for this work was provided by NASA through awards \#SOF 05-0038 and \#SOF 05-0018 issued by USRA. DC would like to thank S. Clark, B. Hensley, D.A. Harper, and I. Stephens for useful discussions. Parts of the analysis were performed using the Clusty Computing Facility in the Villanova Department of Astrophysics and Planetary Science. We thank Andrej Pr\v{s}a for his support in leading and maintaining this resource. Portions of this work were carried out at the Jet Propulsion Laboratory, operated by the California Institute of Technology under a contract with NASA.



\software{ \texttt{python, Ipython} \citep{Perez2007}, \texttt{numpy} \citep{vanderWalt2011}, \texttt{scipy} \citep{Jones2001} \texttt{matplotlib} \citep{Hunter2007}, \texttt{emcee} \citep{Foreman-Mackey2013}, \texttt{corner} \citep{Foreman-Mackey2016}, \texttt{astropy} \citep{astropy:2013, astropy:2018}, LIC code (ported from publically-available IDL source by Diego Falceta-Gon\c{c}alves), \texttt{joblib}}.

\bibliography{bib}{}

\begin{thebibliography}{}
\expandafter\ifx\csname natexlab\endcsname\relax\def\natexlab#1{#1}\fi
\providecommand{\url}[1]{\href{#1}{#1}}

\bibitem[{{Astropy Collaboration} {et~al.}(2013){Astropy Collaboration},
  {Robitaille}, {Tollerud}, {Greenfield}, {Droettboom}, {Bray}, {Aldcroft},
  {Davis}, {Ginsburg}, {Price-Whelan}, {Kerzendorf}, {Conley}, {Crighton},
  {Barbary}, {Muna}, {Ferguson}, {Grollier}, {Parikh}, {Nair}, {Unther},
  {Deil}, {Woillez}, {Conseil}, {Kramer}, {Turner}, {Singer}, {Fox}, {Weaver},
  {Zabalza}, {Edwards}, {Azalee Bostroem}, {Burke}, {Casey}, {Crawford},
  {Dencheva}, {Ely}, {Jenness}, {Labrie}, {Lim}, {Pierfederici}, {Pontzen},
  {Ptak}, {Refsdal}, {Servillat}, \& {Streicher}}]{astropy:2013}
{Astropy Collaboration}, {Robitaille}, T.~P., {Tollerud}, E.~J., {et~al.} 2013,
  \aap, 558, A33

\bibitem[{{Bally} {et~al.}(2017){Bally}, {Ginsburg}, {Arce}, {Eisner},
  {Youngblood}, {Zapata}, \& {Zinnecker}}]{Bally2017}
{Bally}, J., {Ginsburg}, A., {Arce}, H., {et~al.} 2017, \apj, 837, 60

\bibitem[{Cabral \& Leedom(1993)}]{Cabral1993}
Cabral, B., \& Leedom, L.~C. 1993, in Proceedings of the 20th annual conference
  on Computer graphics and interactive techniques, ACM, 263--270

\bibitem[{{Chandrasekhar} \& {Fermi}(1953)}]{Chandrasekhar1953}
{Chandrasekhar}, S., \& {Fermi}, E. 1953, \apj, 118, 113

\bibitem[{{Chen} {et~al.}(2019){Chen}, {King}, {Li}, {Fissel}, \&
  {Mazzei}}]{Chen2019}
{Chen}, C.-Y., {King}, P.~K., {Li}, Z.-Y., {Fissel}, L.~M., \& {Mazzei}, R.~R.
  2019, \mnras, 485, 3499

\bibitem[{{Chuss} {et~al.}(2019){Chuss}, {Andersson}, {Bally}, {Dotson},
  {Dowell}, {Guerra}, {Harper}, {Houde}, {Jones}, {Lazarian}, {Lopez
  Rodriguez}, {Michail}, {Morris}, {Novak}, {Siah}, {Staguhn}, {Vaillancourt},
  {Volpert}, {Werner}, {Wollack}, {Benford}, {Berthoud}, {Cox}, {Crutcher},
  {Dale}, {Fissel}, {Goldsmith}, {Hamilton}, {Hanany}, {Henning}, {Looney},
  {Moseley}, {Santos}, {Stephens}, {Tassis}, {Trinh}, {Van Camp},
  {Ward-Thompson}, \& {HAWC + Science Team}}]{Chuss2019}
{Chuss}, D.~T., {Andersson}, B.~G., {Bally}, J., {et~al.} 2019, \apj, 872, 187

\bibitem[{{Clark} \& {Hensley}(2019)}]{Clark2019}
{Clark}, S.~E., \& {Hensley}, B.~S. 2019, \apj, 887, 136

\bibitem[{{Crutcher}(2004)}]{Crutcher2004}
{Crutcher}, R.~M. 2004, \apss, 292, 225

\bibitem[{Crutcher {et~al.}(2009)Crutcher, Hakobian, \& Troland}]{Crutcher2009}
Crutcher, R.~M., Hakobian, N., \& Troland, T.~H. 2009, The Astrophysical
  Journal, 692, 844.
\newblock \url{https://doi.org/10.1088\%2F0004-637x\%2F692\%2F1\%2F844}

\bibitem[{Crutcher \& Kemball(2019)}]{Crutcher2019}
Crutcher, R.~M., \& Kemball, A.~J. 2019, Frontiers in Astronomy and Space
  Sciences, 6, 66.
\newblock \url{https://www.frontiersin.org/articles/10.3389/fspas.2019.00066}

\bibitem[{{Crutcher} {et~al.}(1996){Crutcher}, {Troland}, {Lazareff}, \&
  {Kazes}}]{Crutcher1996}
{Crutcher}, R.~M., {Troland}, T.~H., {Lazareff}, B., \& {Kazes}, I. 1996, \apj,
  456, 217

\bibitem[{{Davis}(1951)}]{Davis1951}
{Davis}, L. 1951, PhRv, 81, 890

\bibitem[{{Dyck} {et~al.}(1973){Dyck}, {Capps}, {Forrest}, \&
  {Gillett}}]{Dyck1973}
{Dyck}, H.~M., {Capps}, R.~W., {Forrest}, W.~J., \& {Gillett}, F.~C. 1973,
  \apjl, 183, L99

\bibitem[{{Falceta-Gon{\c{c}}alves} {et~al.}(2008){Falceta-Gon{\c{c}}alves},
  {Lazarian}, \& {Kowal}}]{Falceta2008}
{Falceta-Gon{\c{c}}alves}, D., {Lazarian}, A., \& {Kowal}, G. 2008, \apj, 679,
  537

\bibitem[{{Federrath} \& {Klessen}(2012)}]{Federrath2012}
{Federrath}, C., \& {Klessen}, R.~S. 2012, \apj, 761, 156

\bibitem[{Federrath {et~al.}(2016)Federrath, Rathborne, Longmore, Kruijssen,
  Bally, Contreras, Crocker, Garay, Jackson, Testi, \& Walsh}]{Federrath2016}
Federrath, C., Rathborne, J.~M., Longmore, S.~N., {et~al.} 2016, The
  Astrophysical Journal, 832, 143.
\newblock \url{https://doi.org/10.3847\%2F0004-637x\%2F832\%2F2\%2F143}

\bibitem[{{Fissel} {et~al.}(2016){Fissel}, {Ade}, {Angil{\`e}}, {Ashton},
  {Benton}, {Devlin}, {Dober}, {Fukui}, {Galitzki}, {Gandilo}, {Klein},
  {Korotkov}, {Li}, {Martin}, {Matthews}, {Moncelsi}, {Nakamura},
  {Netterfield}, {Novak}, {Pascale}, {Poidevin}, {Santos}, {Savini}, {Scott},
  {Shariff}, {Diego Soler}, {Thomas}, {Tucker}, {Tucker}, \&
  {Ward-Thompson}}]{Fissel2016}
{Fissel}, L.~M., {Ade}, P.~A.~R., {Angil{\`e}}, F.~E., {et~al.} 2016, \apj,
  824, 134

\bibitem[{Foreman-Mackey(2016)}]{Foreman-Mackey2016}
Foreman-Mackey, D. 2016, JOSS, 24, doi:10.21105/joss.00024

\bibitem[{Foreman-Mackey {et~al.}(2013)Foreman-Mackey, Hogg, Lang, \&
  Goodman}]{Foreman-Mackey2013}
Foreman-Mackey, D., Hogg, D.~W., Lang, D., \& Goodman, J. 2013, PASP, 125, 306.
\newblock \url{http://stacks.iop.org/1538-3873/125/i=925/a=306}

\bibitem[{{Friesen} {et~al.}(2017){Friesen}, {Pineda}, {co-PIs}, {Rosolowsky},
  {Alves}, {Chac{\'o}n-Tanarro}, {How-Huan Chen}, {Chun-Yuan Chen}, {Di
  Francesco}, {Keown}, {Kirk}, {Punanova}, {Seo}, {Shirley}, {Ginsburg},
  {Hall}, {Offner}, {Singh}, {Arce}, {Caselli}, {Goodman}, {Martin}, {Matzner},
  {Myers}, {Redaelli}, \& {GAS Collaboration}}]{Friesen2017}
{Friesen}, R.~K., {Pineda}, J.~E., {co-PIs}, {et~al.} 2017, \apj, 843, 63

\bibitem[{Harper {et~al.}(2018)Harper, Runyan, Dowell, Wirth, Amato, Ames,
  Amiri, Banks, Bartels, Benford, Berthoud, Buchanan, Casey, Chapman, Chuss,
  Derro, Dotson, Evans, Fixsen, Gatley, Guerra, Halpern, Hamilton, Hamlin,
  Hansen, Heimsath, Hermida, Hilton, Hirsch, Hollister, Hostetter, Irwin,
  Jhabvala, Jhabvala, Kastner, Kovács, Loewenstein, Looney, Lopez-Rodriguez,
  Maher, Michail, Miller, Moseley, Novak, Pernic, Rennick, Rhody, Sandberg,
  Sandford, Santos, Shafer, Sharp, Shirron, Siah, Silverberg, Sparr, Spotz,
  Staguhn, Toorian, Towey, Tuttle, Vaillancourt, Voellmer, Volpert, i~Wang, \&
  Wollack}]{Harper2018}
Harper, D.~A., Runyan, M.~C., Dowell, C.~D., {et~al.} 2018, JAI, 7, 1840008

\bibitem[{{Hensley} {et~al.}(2019){Hensley}, {Zhang}, \& {Bock}}]{Hensley2019}
{Hensley}, B.~S., {Zhang}, C., \& {Bock}, J.~J. 2019, arXiv e-prints,
  arXiv:1909.07394

\bibitem[{{Hildebrand} {et~al.}(2009){Hildebrand}, {Kirby}, {Dotson}, {Houde},
  \& {Vaillancourt}}]{Hildebrand2009}
{Hildebrand}, R.~H., {Kirby}, L., {Dotson}, J.~L., {Houde}, M., \&
  {Vaillancourt}, J.~E. 2009, \apj, 696, 567

\bibitem[{{Houde} {et~al.}(2004){Houde}, {Dowell}, {Hildebrand}, {Dotson},
  {Vaillancourt}, {Phillips}, {Peng}, \& {Bastien}}]{Houde2004}
{Houde}, M., {Dowell}, C.~D., {Hildebrand}, R.~H., {et~al.} 2004, \apj, 604,
  717

\bibitem[{{Houde} {et~al.}(2013){Houde}, {Fletcher}, {Beck}, {Hildebrand },
  {Vaillancourt}, \& {Stil}}]{Houde2013}
{Houde}, M., {Fletcher}, A., {Beck}, R., {et~al.} 2013, \apj, 766, 49

\bibitem[{{Houde} {et~al.}(2011){Houde}, {Rao}, {Vaillancourt}, \&
  {Hildebrand}}]{Houde2011}
{Houde}, M., {Rao}, R., {Vaillancourt}, J.~E., \& {Hildebrand}, R.~H. 2011,
  \apj, 733, 109

\bibitem[{{Houde} {et~al.}(2009){Houde}, {Vaillancourt}, {Hildebrand},
  {Chitsazzadeh}, \& {Kirby}}]{Houde2009}
{Houde}, M., {Vaillancourt}, J.~E., {Hildebrand}, R.~H., {Chitsazzadeh}, S., \&
  {Kirby}, L. 2009, \apj, 706, 1504

\bibitem[{Hunter(2007)}]{Hunter2007}
Hunter, J.~D. 2007, CSE, 9

\bibitem[{Jones {et~al.}(2001)Jones, Oliphant, \& et~al.}]{Jones2001}
Jones, E., Oliphant, T., \& et~al., P.~P. 2001, SciPy: Open Source Scientific
  Tools for Python, , .
\newblock \url{http://www.scipy.org/}

\bibitem[{{Kounkel} {et~al.}(2017){Kounkel}, {Hartmann}, {Loinard},
  {Ortiz-Le{\'o}n}, {Mioduszewski}, {Rodr{\'\i}guez}, {Dzib}, {Torres}, {Pech},
  {Galli}, {Rivera}, {Boden}, {Evans}, {Brice{\~n}o}, \& {Tobin}}]{Kounkel2017}
{Kounkel}, M., {Hartmann}, L., {Loinard}, L., {et~al.} 2017, \apj, 834, 142

\bibitem[{{Mouschovias}(1976)}]{Mouschovias1976}
{Mouschovias}, T.~C. 1976, \apj, 207, 141

\bibitem[{{Ostriker} {et~al.}(2001){Ostriker}, {Stone}, \&
  {Gammie}}]{Ostriker2001}
{Ostriker}, E.~C., {Stone}, J.~M., \& {Gammie}, C.~F. 2001, \apj, 546, 980

\bibitem[{P\'{e}rez \& Granger(2007)}]{Perez2007}
P\'{e}rez, F., \& Granger, B.~E. 2007, CSE, 9, doi:10.1109/MCSE.2007.53

\bibitem[{{Pillai} {et~al.}(2020){Pillai}, {Clemens}, {Reissl}, {Myers},
  {Kauffmann}, {Lopez-Rodriguez}, {Alves}, {Franco}, {Henshaw}, {Menten},
  {Nakamura}, {Seifried}, {Sugitani}, \& {Wiesemeyer}}]{Pillai2020}
{Pillai}, T. G.~S., {Clemens}, D.~P., {Reissl}, S., {et~al.} 2020, Nature
  Astronomy, arXiv:2009.14100

\bibitem[{{Planck Collaboration} {et~al.}(2016){Planck Collaboration}, {Ade},
  {Aghanim}, {Alves}, {Arnaud}, {Arzoumanian}, {Ashdown}, {Aumont},
  {Baccigalupi}, {Band ay}, {Barreiro}, {Bartolo}, {Battaner}, {Benabed},
  {Beno{\^\i}t}, {Benoit-L{\'e}vy}, {Bernard}, {Bersanelli}, {Bielewicz},
  {Bock}, {Bonavera}, {Bond}, {Borrill}, {Bouchet}, {Boulanger}, {Bracco},
  {Burigana}, {Calabrese}, {Cardoso}, {Catalano}, {Chiang}, {Christensen},
  {Colombo}, {Combet}, {Couchot}, {Crill}, {Curto}, {Cuttaia}, {Danese},
  {Davies}, {Davis}, {de Bernardis}, {de Rosa}, {de Zotti}, {Delabrouille},
  {Dickinson}, {Diego}, {Dole}, {Donzelli}, {Dor{\'e}}, {Douspis}, {Ducout},
  {Dupac}, {Efstathiou}, {Elsner}, {En{\ss}lin}, {Eriksen},
  {Falceta-Gon{\c{c}}alves}, {Falgarone}, {Ferri{\`e}re}, {Finelli}, {Forni},
  {Frailis}, {Fraisse}, {Franceschi}, {Frejsel}, {Galeotta}, {Galli}, {Ganga},
  {Ghosh}, {Giard}, {Gjerl{\o}w}, {Gonz{\'a}lez-Nuevo}, {G{\'o}rski},
  {Gregorio}, {Gruppuso}, {Gudmundsson}, {Guillet}, {Harrison}, {Helou},
  {Hennebelle}, {Henrot-Versill{\'e}}, {Hern{\'a}ndez-Monteagudo}, {Herranz},
  {Hildebrand t}, {Hivon}, {Holmes}, {Hornstrup}, {Huffenberger}, {Hurier},
  {Jaffe}, {Jaffe}, {Jones}, {Juvela}, {Keih{\"a}nen}, {Keskitalo}, {Kisner},
  {Knoche}, {Kunz}, {Kurki-Suonio}, {Lagache}, {Lamarre}, {Lasenby},
  {Lattanzi}, {Lawrence}, {Leonardi}, {Levrier}, {Liguori}, {Lilje},
  {Linden-V{\o}rnle}, {L{\'o}pez-Caniego}, {Lubin}, {Mac{\'\i}as-P{\'e}rez},
  {Maino}, {Mandolesi}, {Mangilli}, {Maris}, {Martin},
  {Mart{\'\i}nez-Gonz{\'a}lez}, {Masi}, {Matarrese}, {Melchiorri}, {Mendes},
  {Mennella}, {Migliaccio}, {Miville-Desch{\^e}nes}, {Moneti}, {Montier},
  {Morgante}, {Mortlock}, {Munshi}, {Murphy}, {Naselsky}, {Nati},
  {Netterfield}, {Noviello}, {Novikov}, {Novikov}, {Oppermann}, {Oxborrow},
  {Pagano}, {Pajot}, {Paladini}, {Paoletti}, {Pasian}, {Perotto}, {Pettorino},
  {Piacentini}, {Piat}, {Pierpaoli}, {Pietrobon}, {Plaszczynski},
  {Pointecouteau}, {Polenta}, {Ponthieu}, {Pratt}, {Prunet}, {Puget}, {Rachen},
  {Reinecke}, {Remazeilles}, {Renault}, {Renzi}, {Ristorcelli}, {Rocha},
  {Rossetti}, {Roudier}, {Rubi{\~n}o-Mart{\'\i}n}, {Rusholme}, {Sandri},
  {Santos}, {Savelainen}, {Savini}, {Scott}, {Soler}, {Stolyarov}, {Sudiwala},
  {Sutton}, {Suur-Uski}, {Sygnet}, {Tauber}, {Terenzi}, {Toffolatti}, {Tomasi},
  {Tristram}, {Tucci}, {Umana}, {Valenziano}, {Valiviita}, {Van Tent},
  {Vielva}, {Villa}, {Wade}, {Wandelt}, {Wehus}, {Ysard}, {Yvon}, \&
  {Zonca}}]{PlanckXXXV2016}
{Planck Collaboration}, {Ade}, P.~A.~R., {Aghanim}, N., {et~al.} 2016, \aap,
  586, A138

\bibitem[{{Planck Collaboration} {et~al.}(2018){Planck Collaboration},
  {Aghanim}, {Akrami}, {Alves}, {Ashdown}, {Aumont}, {Baccigalupi},
  {Ballardini}, {Banday}, {Barreiro}, {Bartolo}, {Basak}, {Benabed}, {Bernard},
  {Bersanelli}, {Bielewicz}, {Bock}, {Bond}, {Borrill}, {Bouchet}, {Boulanger},
  {Bracco}, {Bucher}, {Burigana}, {Calabrese}, {Cardoso}, {Carron}, {Chary},
  {Chiang}, {Colombo}, {Combet}, {Crill}, {Cuttaia}, {de Bernardis}, {de
  Zotti}, {Delabrouille}, {Delouis}, {Di Valentino}, {Dickinson}, {Diego},
  {Dor{\'e}}, {Douspis}, {Ducout}, {Dupac}, {Efstathiou}, {Elsner},
  {En{\ss}lin}, {Eriksen}, {Fantaye}, {Fernandez-Cobos}, {Ferri{\`e}re},
  {Forastieri}, {Frailis}, {Fraisse}, {Franceschi}, {Frolov}, {Galeotta},
  {Galli}, {Ganga}, {G{\'e}nova-Santos}, {Gerbino}, {Ghosh},
  {Gonz{\'a}lez-Nuevo}, {G{\'o}rski}, {Gratton}, {Green}, {Gruppuso},
  {Gudmundsson}, {Guillet}, {Handley}, {Hansen}, {Helou}, {Herranz}, {Hivon},
  {Huang}, {Jaffe}, {Jones}, {Keih{\"a}nen}, {Keskitalo}, {Kiiveri}, {Kim},
  {Krachmalnicoff}, {Kunz}, {Kurki-Suonio}, {Lagache}, {Lamarre}, {Lasenby},
  {Lattanzi}, {Lawrence}, {Le Jeune}, {Levrier}, {Liguori}, {Lilje},
  {Lindholm}, {L{\'o}pez-Caniego}, {Lubin}, {Ma}, {Mac{\'\i}as-P{\'e}rez},
  {Maggio}, {Maino}, {Mandolesi}, {Mangilli}, {Marcos-Caballero}, {Maris},
  {Martin}, {Mart{\'\i}nez-Gonz{\'a}lez}, {Matarrese}, {Mauri}, {McEwen},
  {Melchiorri}, {Mennella}, {Migliaccio}, {Miville-Desch{\^e}nes}, {Molinari},
  {Moneti}, {Montier}, {Morgante}, {Moss}, {Natoli}, {Pagano}, {Paoletti},
  {Patanchon}, {Perrotta}, {Pettorino}, {Piacentini}, {Polastri}, {Polenta},
  {Puget}, {Rachen}, {Reinecke}, {Remazeilles}, {Renzi}, {Ristorcelli},
  {Rocha}, {Rosset}, {Roudier}, {Rubi{\~n}o-Mart{\'\i}n}, {Ruiz-Granados},
  {Salvati}, {Sandri}, {Savelainen}, {Scott}, {Sirignano}, {Sunyaev},
  {Suur-Uski}, {Tauber}, {Tavagnacco}, {Tenti}, {Toffolatti}, {Tomasi},
  {Trombetti}, {Valiviita}, {Van Tent}, {Vielva}, {Villa}, {Vittorio},
  {Wandelt}, {Wehus}, {Zacchei}, \& {Zonca}}]{Planck2018}
{Planck Collaboration}, {Aghanim}, N., {Akrami}, Y., {et~al.} 2018, ArXiv
  e-prints, arXiv:1807.06212

\bibitem[{{Price} \& {Bate}(2008)}]{Price2008}
{Price}, D.~J., \& {Bate}, M.~R. 2008, \mnras, 385, 1820

\bibitem[{{Price-Whelan} {et~al.}(2018){Price-Whelan}, {Sip{\'{o}}cz},
  {G{\"u}nther}, {Lim}, {Crawford}, {Conseil}, {Shupe}, {Craig}, {Dencheva},
  {Ginsburg}, {VanderPlas}, {Bradley}, {P{'e}rez-Su{'a}rez}, {de Val-Borro},
  {Paper Contributors}, {Aldcroft}, {Cruz}, {Robitaille}, {Tollerud},
  {Coordination Committee}, {Ardelean}, {Babej}, {Bach}, {Bachetti}, {Bakanov},
  {Bamford}, {Barentsen}, {Barmby}, {Baumbach}, {Berry}, {Biscani}, {Boquien},
  {Bostroem}, {Bouma}, {Brammer}, {Bray}, {Breytenbach}, {Buddelmeijer},
  {Burke}, {Calderone}, {Cano Rodr{'i}guez}, {Cara}, {Cardoso}, {Cheedella},
  {Copin}, {Corrales}, {Crichton}, {D{ extquoteright}Avella}, {Deil},
  {Depagne}, {Dietrich}, {Donath}, {Droettboom}, {Earl}, {Erben}, {Fabbro},
  {Ferreira}, {Finethy}, {Fox}, {Garrison}, {Gibbons}, {Goldstein}, {Gommers},
  {Greco}, {Greenfield}, {Groener}, {Grollier}, {Hagen}, {Hirst}, {Homeier},
  {Horton}, {Hosseinzadeh}, {Hu}, {Hunkeler}, {Ivezi{'c}}, {Jain}, {Jenness},
  {Kanarek}, {Kendrew}, {Kern}, {Kerzendorf}, {Khvalko}, {King}, {Kirkby},
  {Kulkarni}, {Kumar}, {Lee}, {Lenz}, {Littlefair}, {Ma}, {Macleod},
  {Mastropietro}, {McCully}, {Montagnac}, {Morris}, {Mueller}, {Mumford},
  {Muna}, {Murphy}, {Nelson}, {Nguyen}, {Ninan}, {N{"o}the}, {Ogaz}, {Oh},
  {Parejko}, {Parley}, {Pascual}, {Patil}, {Patil}, {Plunkett}, {Prochaska},
  {Rastogi}, {Reddy Janga}, {Sabater}, {Sakurikar}, {Seifert}, {Sherbert},
  {Sherwood-Taylor}, {Shih}, {Sick}, {Silbiger}, {Singanamalla}, {Singer},
  {Sladen}, {Sooley}, {Sornarajah}, {Streicher}, {Teuben}, {Thomas},
  {Tremblay}, {Turner}, {Terr{'o}n}, {van Kerkwijk}, {de la Vega}, {Watkins},
  {Weaver}, {Whitmore}, {Woillez}, {Zabalza}, \& {Contributors}}]{astropy:2018}
{Price-Whelan}, A.~M., {Sip{\'{o}}cz}, B.~M., {G{\"u}nther}, H.~M., {et~al.}
  2018, \aj, 156, 123

\bibitem[{{Schleuning}(1998)}]{Schleuning1998}
{Schleuning}, D.~A. 1998, \apj, 493, 811

\bibitem[{{Serkowski}(1974)}]{Serkowski1974}
{Serkowski}, K. 1974, Methods of Experimental Physics, 12, 361

\bibitem[{{Tahani} {et~al.}(2019){Tahani}, {Plume}, {Brown}, {Soler}, \&
  {Kainulainen}}]{Tahani2019}
{Tahani}, M., {Plume}, R., {Brown}, J.~C., {Soler}, J.~D., \& {Kainulainen}, J.
  2019, \aap, 632, A68

\bibitem[{{Tang} {et~al.}(2010){Tang}, {Ho}, {Koch}, \& {Rao}}]{Tang2010}
{Tang}, Y.-W., {Ho}, P.~T.~P., {Koch}, P.~M., \& {Rao}, R. 2010, \apj, 717,
  1262

\bibitem[{Temi {et~al.}(2018)Temi, Hoffman, Ennico, \& Le}]{SOFIA}
Temi, P., Hoffman, D., Ennico, K., \& Le, J. 2018, Journal of Astronomical
  Instrumentation, 07, 1840011.
\newblock \url{https://doi.org/10.1142/S2251171718400111}

\bibitem[{{Troland} {et~al.}(2016){Troland}, {Goss}, {Brogan}, {Crutcher}, \&
  {Roberts}}]{Troland2016}
{Troland}, T.~H., {Goss}, W.~M., {Brogan}, C.~L., {Crutcher}, R.~M., \&
  {Roberts}, D.~A. 2016, \apj, 825, 2

\bibitem[{{Vall{\'e}e} \& {Bastien}(1999)}]{Vallee1999}
{Vall{\'e}e}, J.~P., \& {Bastien}, P. 1999, \apj, 526, 819

\bibitem[{{van der Walt} {et~al.}(2011){van der Walt}, Colbert, \&
  Varoquaux}]{vanderWalt2011}
{van der Walt}, S., Colbert, S.~C., \& Varoquaux, G. 2011, CSE, 13,
  doi:10.1109/MCSE.2011.37

\bibitem[{{Ward-Thompson} {et~al.}(2017){Ward-Thompson}, {Pattle}, {Bastien},
  {Furuya}, {Kwon}, {Lai}, {Qiu}, {Berry}, {Choi}, {Coud{\'e}}, {Di Francesco},
  {Hoang}, {Franzmann}, {Friberg}, {Graves}, {Greaves}, {Houde}, {Johnstone},
  {Kirk}, {Koch}, {Kwon}, {Lee}, {Li}, {Matthews}, {Mottram}, {Parsons}, {Pon},
  {Rao}, {Rawlings}, {Shinnaga}, {Sadavoy}, {van Loo}, {Aso}, {Byun},
  {Eswaraiah}, {Chen}, {Chen}, {Chen}, {Ching}, {Cho}, {Chrysostomou}, {Chung},
  {Doi}, {Drabek-Maunder}, {Eyres}, {Fiege}, {Friesen}, {Fuller}, {Gledhill},
  {Griffin}, {Gu}, {Hasegawa}, {Hatchell}, {Hayashi}, {Holland}, {Inoue},
  {Inutsuka}, {Iwasaki}, {Jeong}, {Kang}, {Kang}, {Kang}, {Kawabata}, {Kemper},
  {Kim}, {Kim}, {Kim}, {Kim}, {Kim}, {Kim}, {Lacaille}, {Lee}, {Lee}, {Li},
  {Li}, {Liu}, {Liu}, {Liu}, {Liu}, {Lyo}, {Mairs}, {Matsumura},
  {Moriarty-Schieven}, {Nakamura}, {Nakanishi}, {Ohashi}, {Onaka}, {Peretto},
  {Pyo}, {Qian}, {Retter}, {Richer}, {Rigby}, {Robitaille}, {Savini}, {Scaife},
  {Soam}, {Tamura}, {Tang}, {Tomisaka}, {Wang}, {Wang}, {Whitworth}, {Yen},
  {Yoo}, {Yuan}, {Zhang}, {Zhang}, {Zhou}, {Zhu}, {Andr{\'e}}, {Dowell},
  {Falle}, \& {Tsukamoto}}]{Ward-Thompson2017}
{Ward-Thompson}, D., {Pattle}, K., {Bastien}, P., {et~al.} 2017, \apj, 842, 66

\end{thebibliography}
\bibliographystyle{aasjournal}

\appendix
\section{DCF Parameters Maps for $\delta$ constant}\label{sec:appendixA}

\begin{figure}[!h]
    \centering
    \includegraphics[width=3.3in]{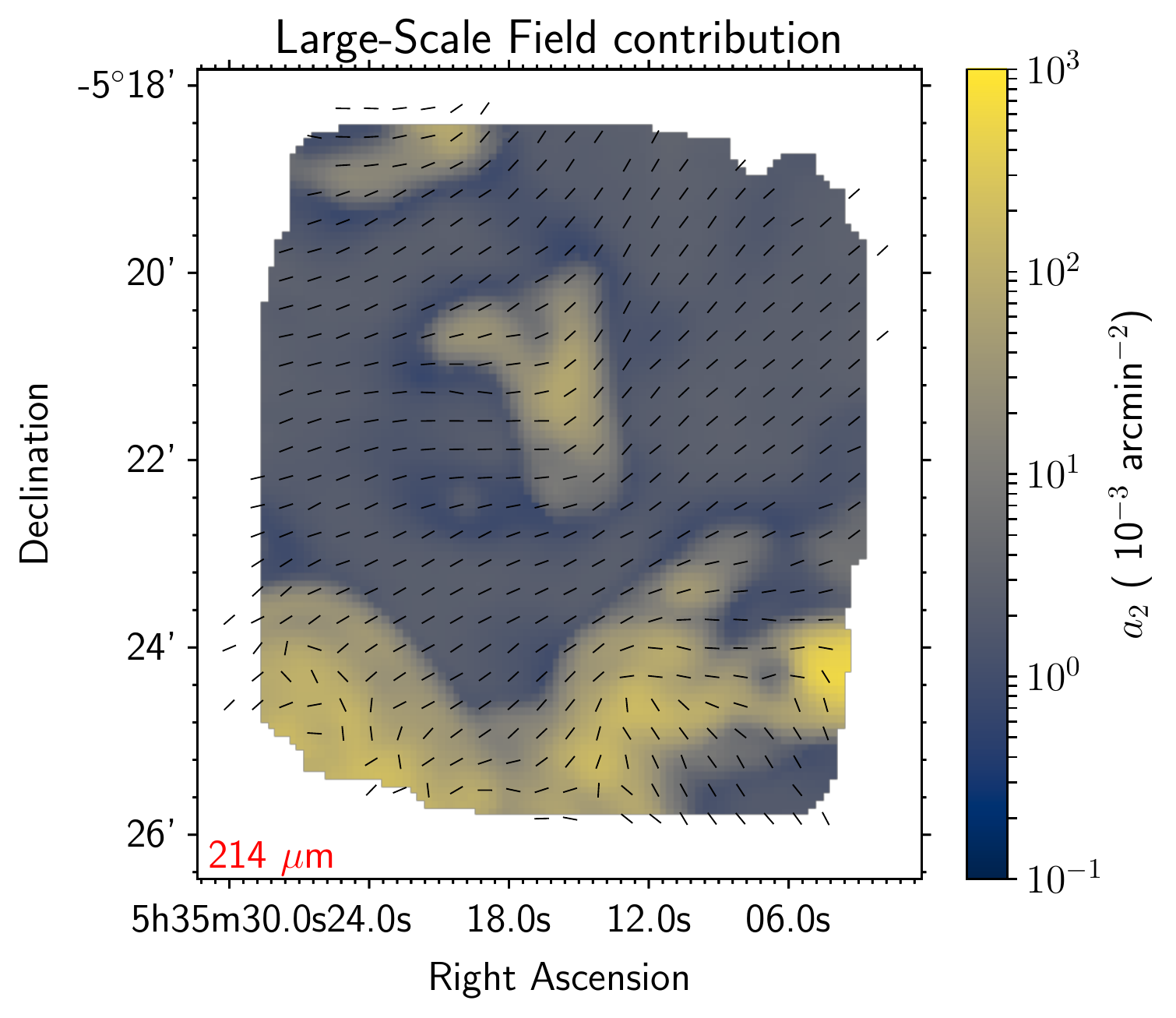}
    \includegraphics[width=3.2in]{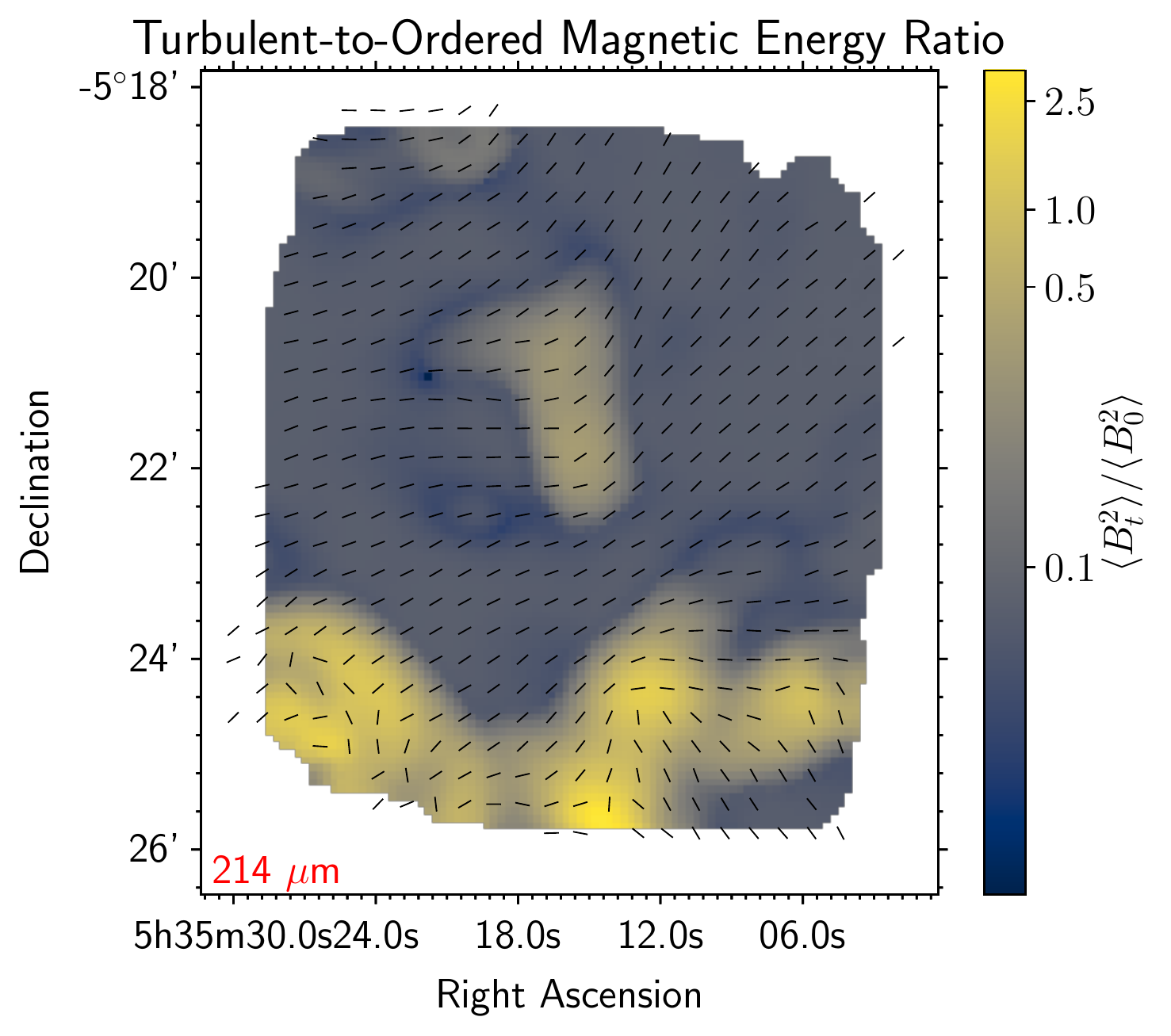}
    \caption{Maps of parameters $a_{2}$ ({\it Left}) and  $\ratio$ ({\it Right}) for 214-$\micron$ data using $w_{\rm opt}$ = 9 pixels. For calculating these maps, the parameter $\delta$ -- the turbulence correlation lengths -- is kept fixed to a whole-map value, $\delta_{0} = $27.0$^{\prime\prime}$. The spatial distribution of the parameters in this case are very similar to that when the parameter $\delta$ is also determined by the MCMC solver.}
    \label{fig:DCF_maps_delta_fixed}
\end{figure}

Figure \ref{fig:DCF_maps_delta_fixed} displays the $a_{2}$ ({\it Left}) and $\ratio$ ({\it Right}) maps when the turbulence correlation length $\delta = $ 27.0$^{\prime\prime}$ is assumed constant over the whole field of view. Compared to maps in Figure \ref{fig:param_maps_delta}(Left and Middle), the solutions for $a_{2}$ do not seem affected by the $\delta$-value, which is expected since the those parameters describe different spatial scales. On the other hand, values of $\ratio$ are lower in Figure \ref{fig:DCF_maps_delta_fixed}. However, the spatial distributions of $a_{2}$ and $\ratio$ in this case ($\delta$ constant) are very close to those of the $\delta$-variable case. Therefore, this is evidence that such spatial distribution is not the result of covariance among the parameters.

\section{MCMC Posterior Distributions and Chains Convergence} \label{sec:appendixB}

\begin{figure}[!h]
    \centering
    \includegraphics[width=3.3in]{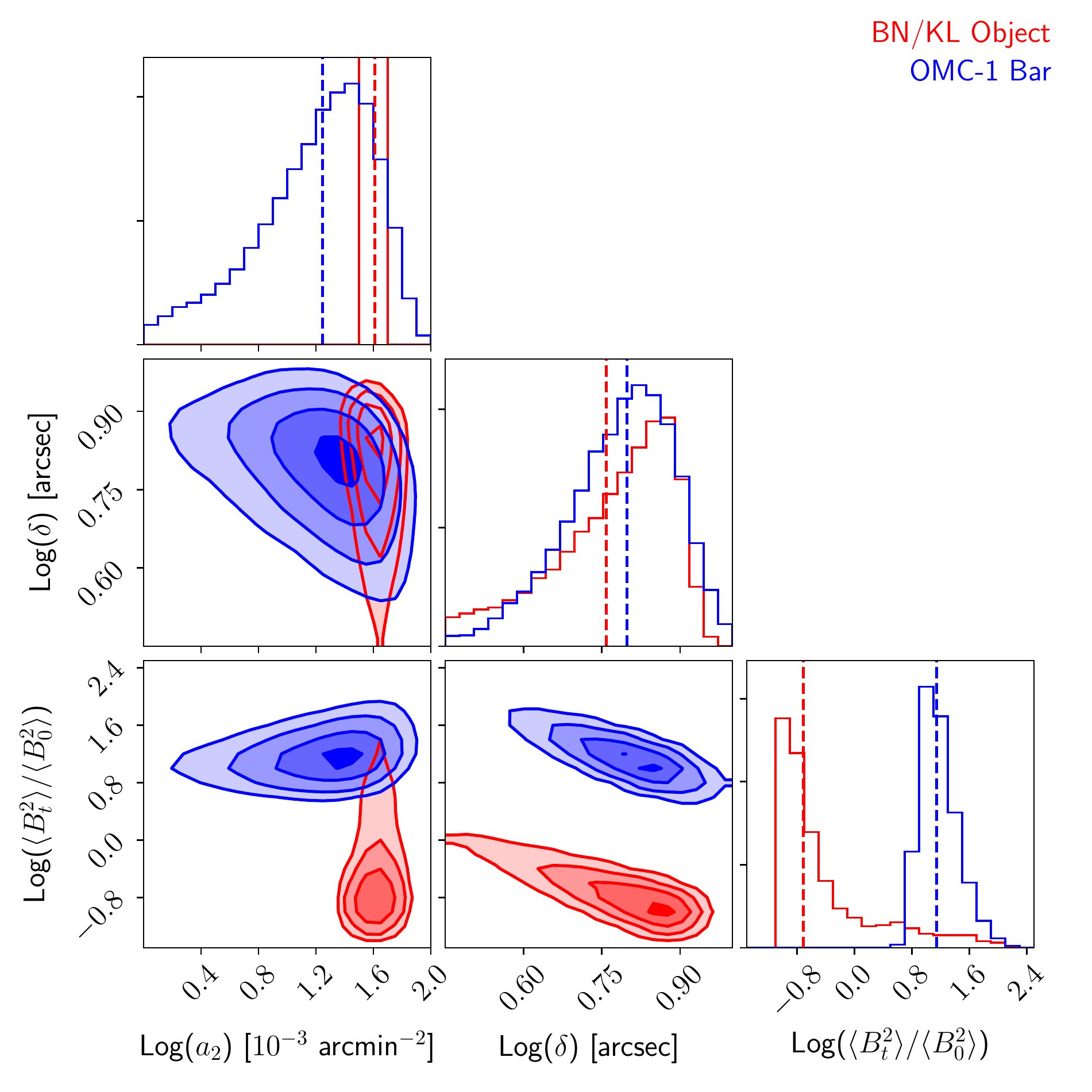}
    \includegraphics[width=3.3in]{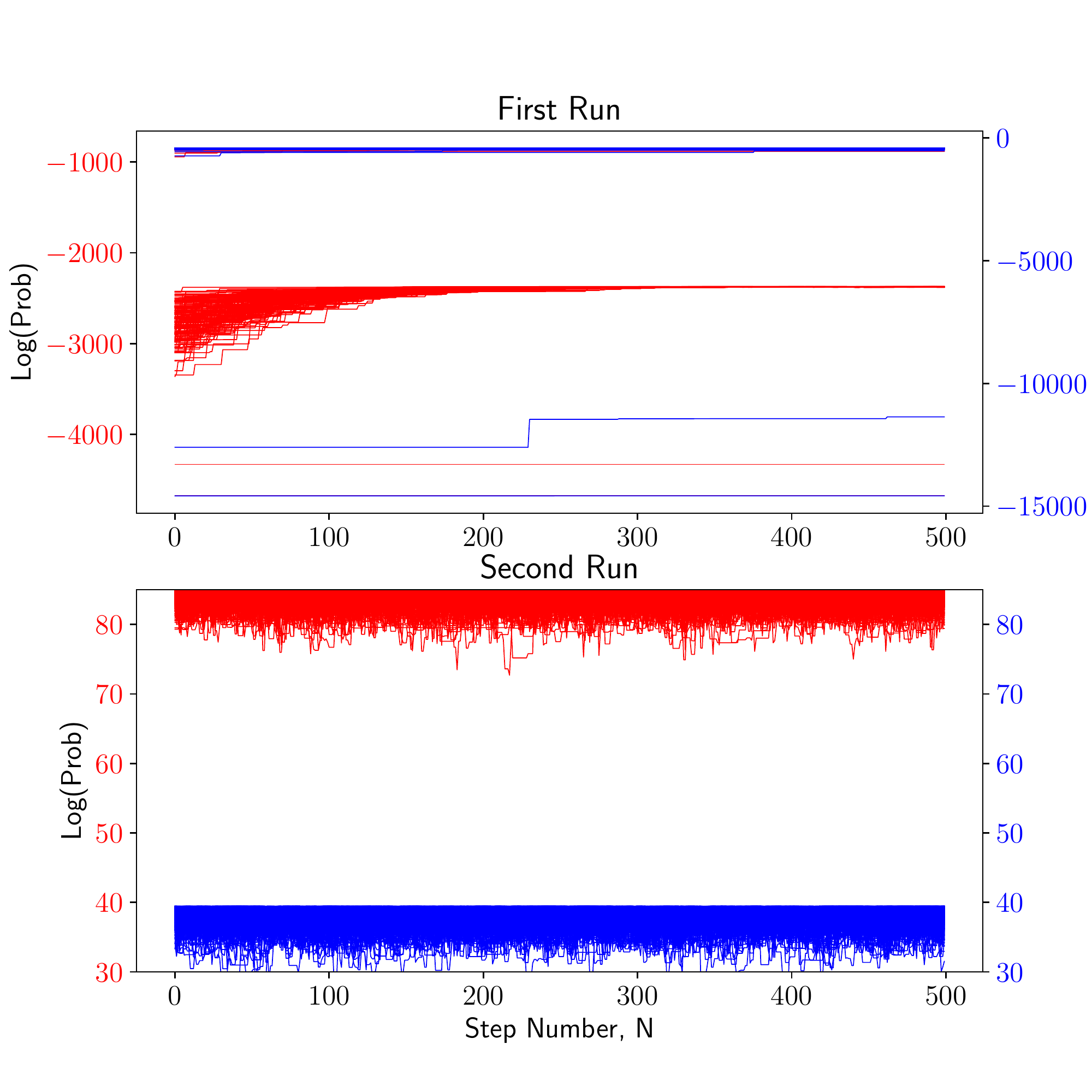}
    \caption{Posterior distributions ({\it Left}) and maximum likelihood ({\it Right}) of the chains for the MCMC fitting of dispersion functions calculated at the BN/KL (red) and the bar (blue) positions, using the 214-$\mu$m data (Figure \ref{fig:sample_DFs},{\it Left}). Unimodal distributions for parameters (log of) $a_{0}$, $\delta$, and $\ratio$ are constructed with the chains from the second MCMC run, which have already converged into the global solution.}
    \label{fig:MCMC_corner_plots}
\end{figure}

The MCMC process for fitting the measured dispersion function to Eq. \ref{eq:disp_model} is routinely performed twice for every pixel, each run with 500 steps. In the first run, a global solution is estimated. Then, the second run uses the median values from the global solution as initial guess. Figure \ref{fig:MCMC_corner_plots}({\it Left}) displays the posterior distributions for parameters (log of) $a_{2}$, $\delta$, and $\ratio$ for the MCMC fitting performed at the BN/KL position (red) and the OMC-1 bar (blue), using the 214-$\mu$m data (see Figure \ref{fig:sample_DFs}). At both locations, posterior distributions appear unimodal and their shapes suggest that values are not severely limited by the imposed bounds ({\it i.e.} $a_{2}, \ratio> 0$ and $0 < \delta < \delta_{max}$). A simple way to test the quality of the resulting posterior distributions is by inspecting the convergence of the MCMC chains. This is shown in Figure \ref{fig:MCMC_corner_plots}({\it Right}) with the values of the maximum likelihood, $log(Prob)$, as a function of steps for the same OMC-1 locations. It can be seen that during the first MCMC run all chains have converged  ({\it i.e.} $Log(Prob)$ stabilizes) well before 500 steps and therefore a global solution is found. During the the second run, the chains appear already converged since the variation of $Log(Prob)$ is small for all step values. More importantly, the variance of the chains for the second run is approximately 1\% of their mean. Values in maps of parameters such as those in Figures \ref{fig:param_maps_delta} and \ref{fig:DCF_maps_delta_fixed} are always calculated from posterior distributions obtained from the second run. 

\end{document}